\documentclass[preprint,journal]{vgtc}            

\onlineid{1018}

\preprinttext{%
  \parbox{0.9\textwidth}{%
    $\copyright$ 2026 IEEE.
    This is the author's version of an article that has been accepted
    for publication in
    IEEE Transactions on Visualization and Computer Graphics.
  }%
}

\vgtccategory{Research}

\vgtcpapertype{algorithm/technique}

\title{Structuring Line Ensembles with Path-Integrated Fidelity and Structural Inconsistency Fields}
\author{%
  \authororcid{Yumeng Xue}{0000-0002-8195-517X},
  \authororcid{Patrick Paetzold}{0000-0002-1315-4602},
  \authororcid{Bin Chen}{0009-0005-5047-214X},
  \authororcid{Yunhai Wang}{0000-0003-0059-6580},
  \authororcid{Christophe Hurter}{0000-0003-4318-6717},
  and \authororcid{Oliver Deussen}{0000-0001-5803-2185}
}

\authorfooter{
  \item
  	Y. Xue, P. Paetzold, B. Chen, and O. Deussen are with the University of Konstanz, Germany.
  	E-mail: \{ﬁrstname.lastname\}@uni-konstanz.de.
  \item
  	Y. Wang is with Renmin University of China, China.
  	E-mail: wang.yh@ruc.edu.cn.
  \item C. Hurter is with ENAC, Université de Toulouse, France, and with IPAL Singapore International Research Laboratory on Artificial Intelligence, Singapore.
  	E-mail: christophe.hurter@enac.fr.
  \item  O. Deussen and Y. Wang are joint corresponding authors.
}

\abstract{%
When visualizing large-scale line ensembles, trajectory continuity and visual scalability are inherently antagonistic. Trajectory-centric renderings preserve path information but rapidly degenerate into clutter as line density increases and mutual occlusion dominates. In contrast, field-based density representations enhance visibility while sacrificing structural coherence: density reflects accumulation rather than agreement, such that scalar aggregation alone cannot discriminate between consistent and conflicting configurations. Rather than replacing density-based views, we complement them with a path-integrated trajectory-fidelity measure that quantifies the agreement of each trajectory with a surrounding tensor field. 
By projecting this passage-centered structural support back into image space, we obtain what we call a Structural Inconsistency Field, which localizes regions where dense patterns correspond to coherent structure versus disagreement, outliers, or connectivity-induced ambiguity.
Dynamic leave-one-out correction reduces self-bias in the path integral. Efficient fixed-grid updates combined with prefix-sum evaluation enable interactive analysis and iterative extraction of coherent structures. 
Synthetic benchmarks, scalability analyses, and real-world case studies demonstrate that, when paired with conventional density views, the proposed method disambiguates dense line patterns by exposing spatially localized coherence and structural breakdown that remain concealed in density-only representations.
}


\teaser{
    \centering
    \begin{subfigure}{0.36\textwidth}
        \centering
        \includegraphics[width=\linewidth]{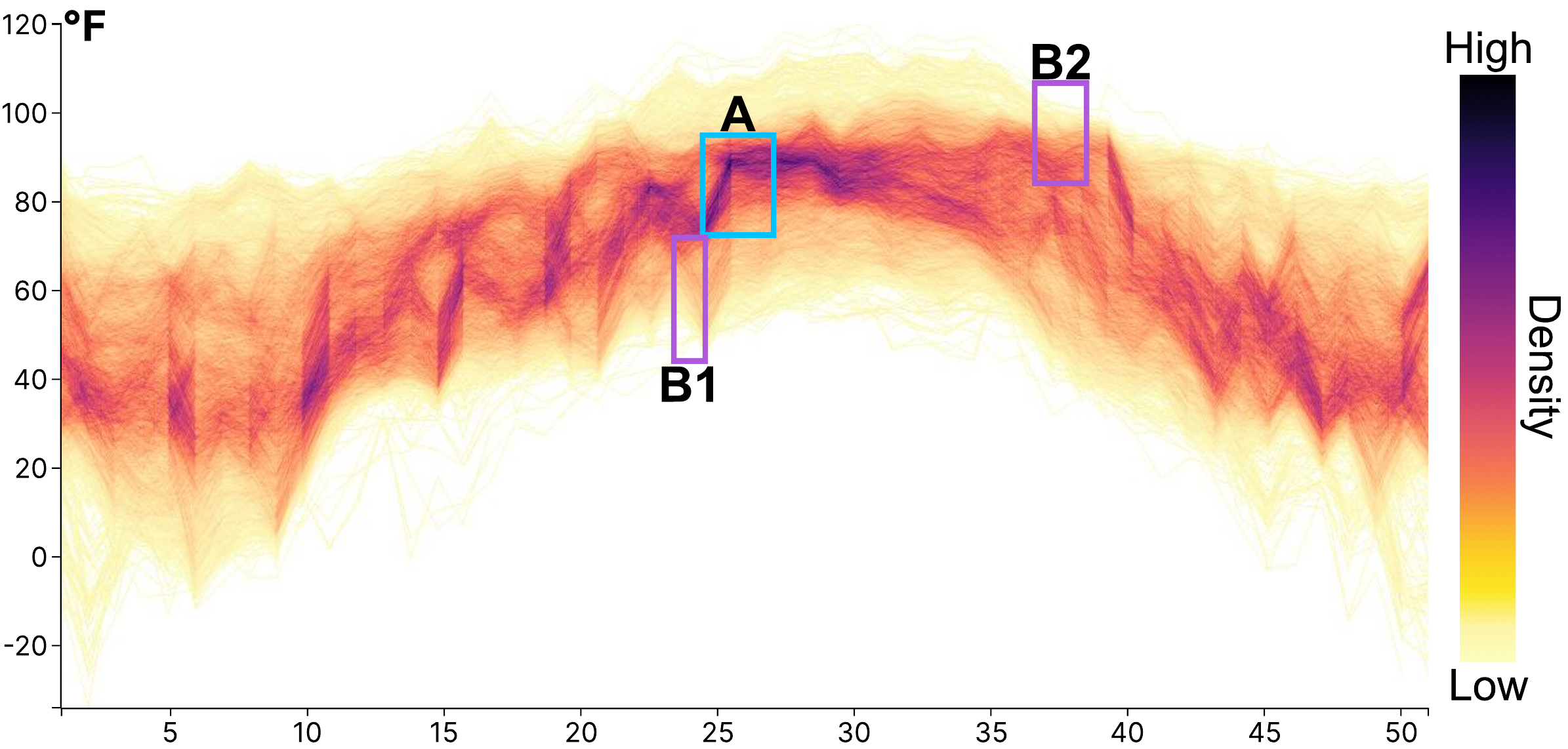}
        \caption{Conventional line density plot}
        \label{fig:teaser_density}
    \end{subfigure}
    \hfill
    \begin{subfigure}{0.36\textwidth}
        \centering
        \includegraphics[width=\linewidth]{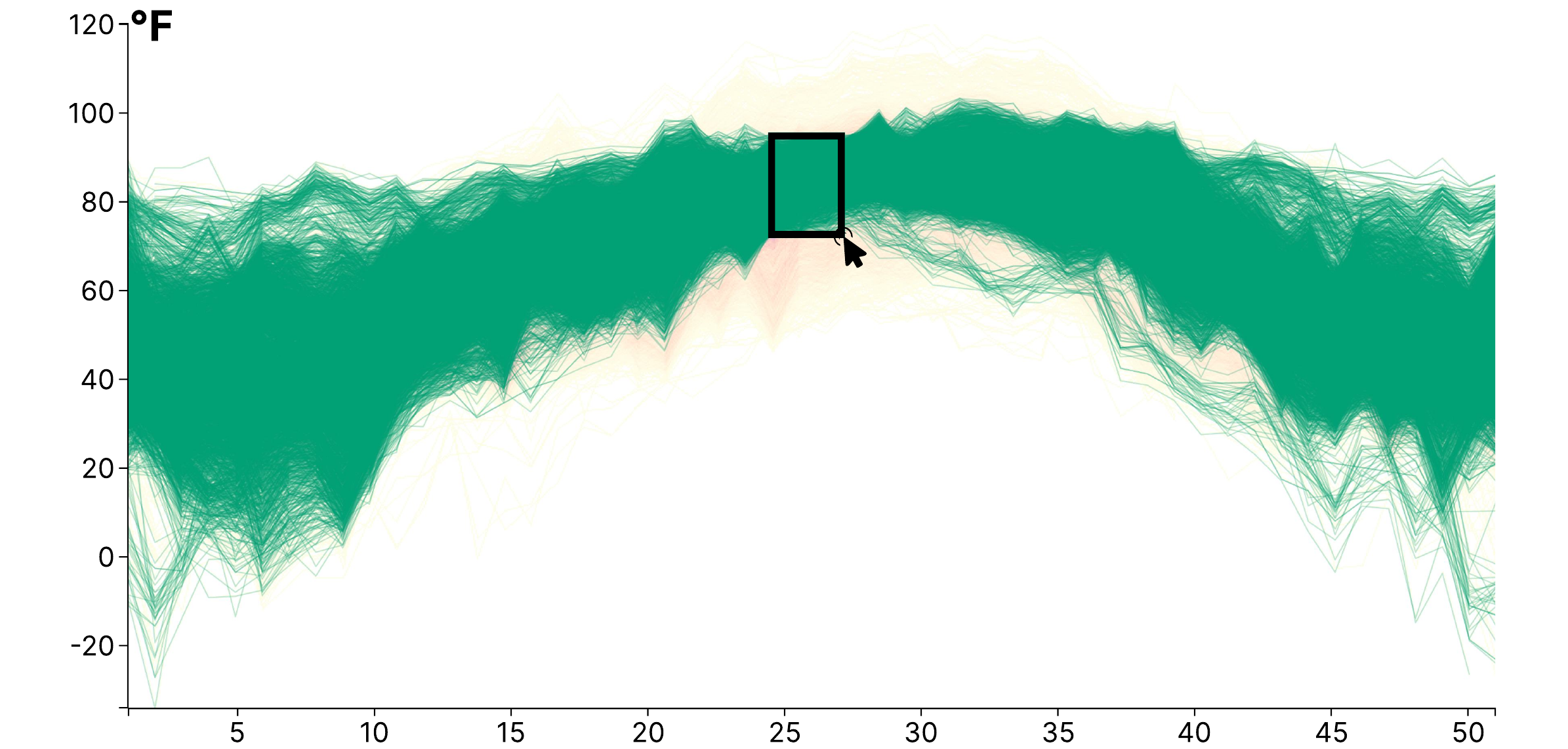}
        \caption{Density-guided query result}
        \label{fig:teaser_density_selection}
    \end{subfigure}
    \hfill
    \begin{subfigure}{0.25\textwidth}
        \centering
        \includegraphics[width=\linewidth]{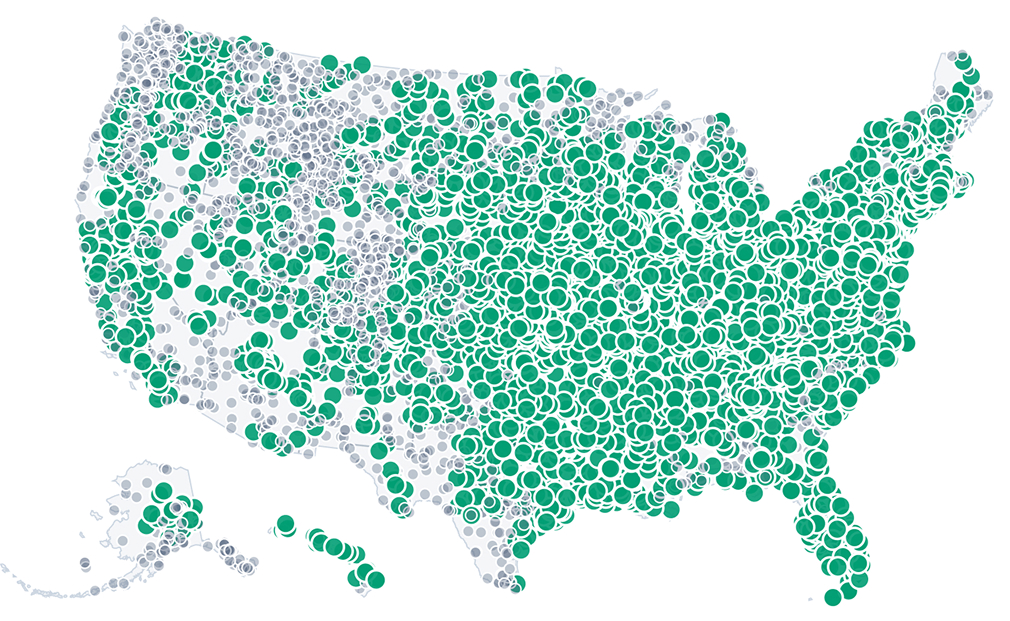}
        \caption{Geographic footprint of density-guided subset}
        \label{fig:teaser_density_map}
    \end{subfigure}

    \vspace{2pt}

    \begin{subfigure}{0.36\textwidth}
        \centering
        \includegraphics[width=\linewidth]{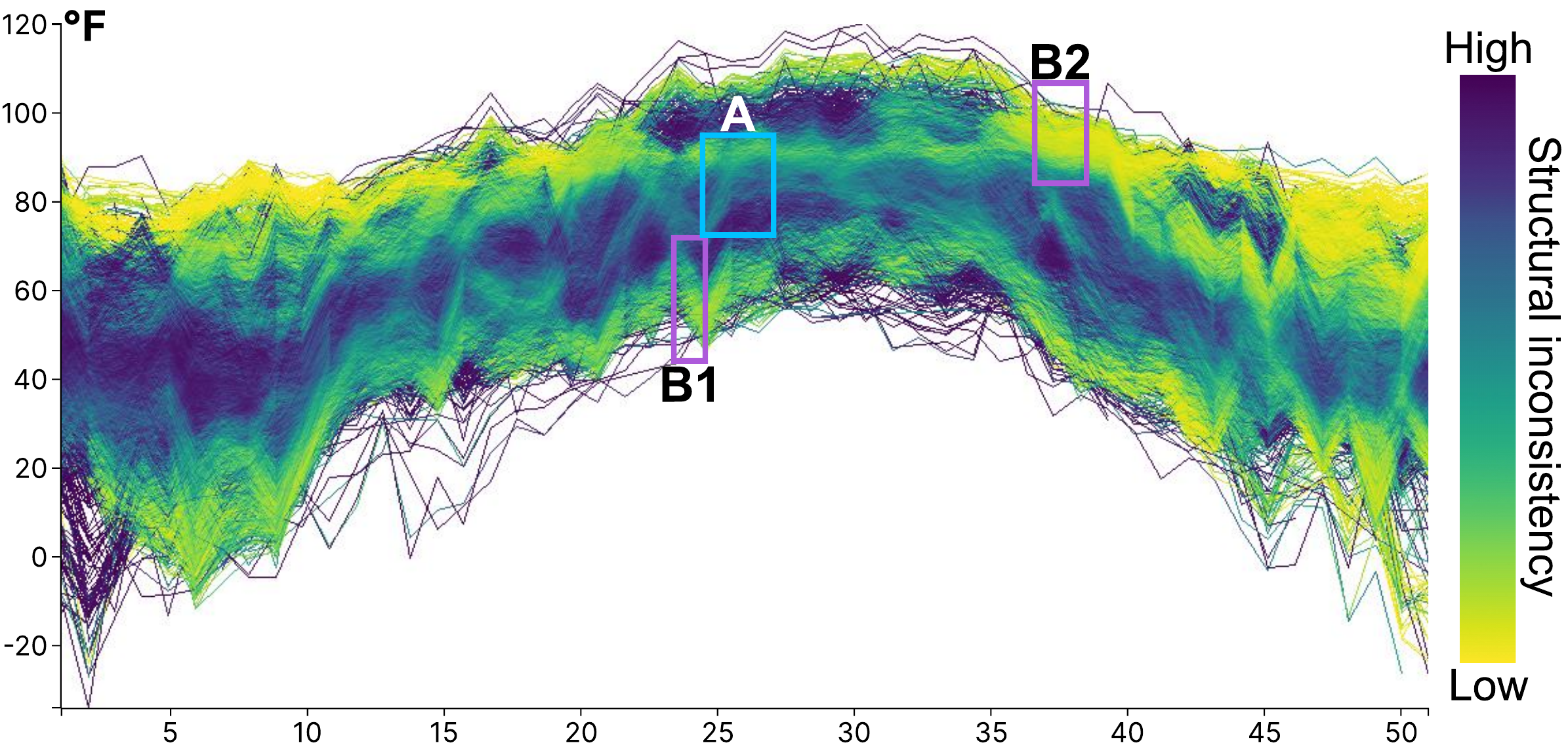}
        \caption{Structural inconsistency field}
        \label{fig:teaser_friction}
    \end{subfigure}
    \hfill
    \begin{subfigure}{0.36\textwidth}
        \centering
        \includegraphics[width=\linewidth]{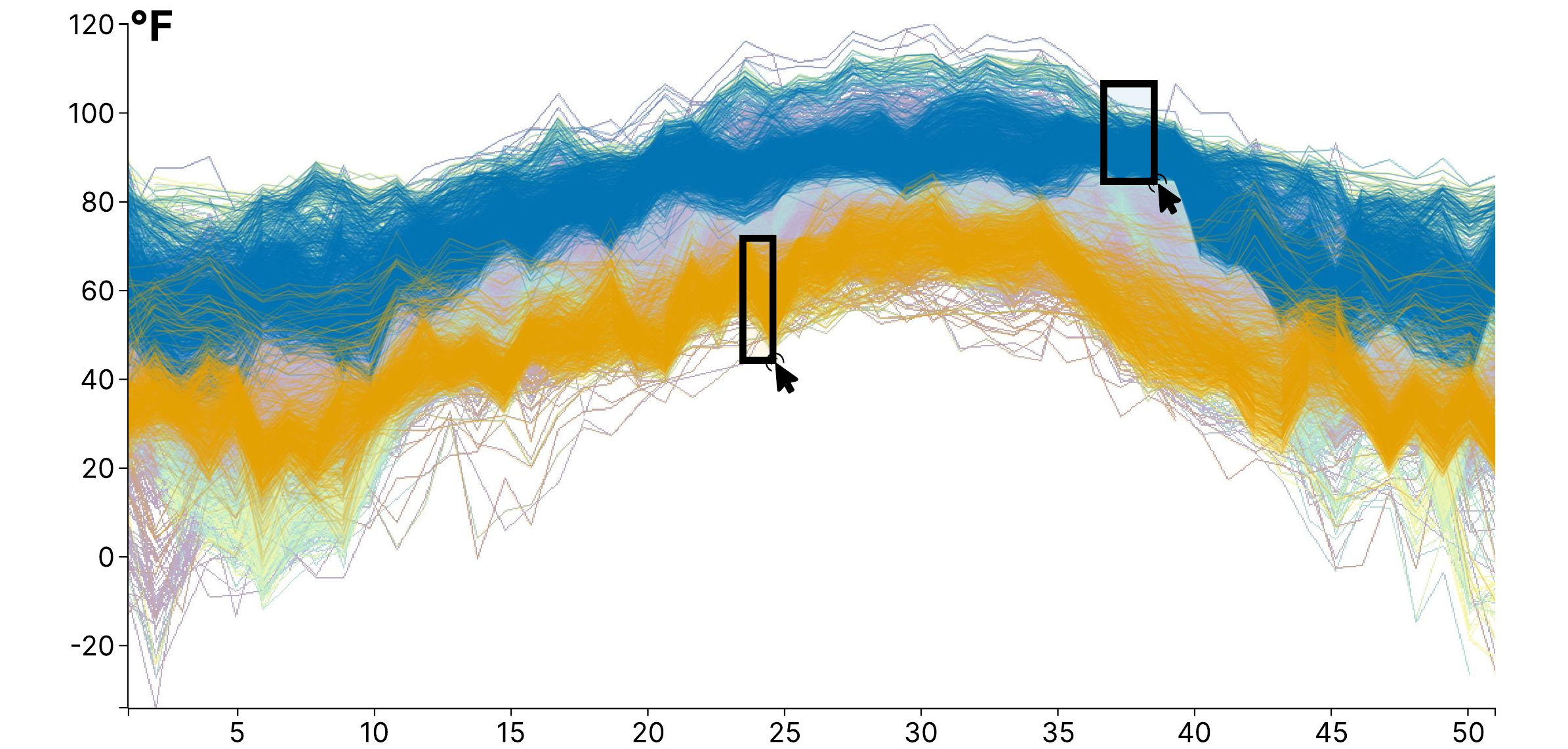}
        \caption{Structure-guided query results}
        \label{fig:teaser_friction_selection}
    \end{subfigure}
    \hfill
    \begin{subfigure}{0.25\textwidth}
        \centering
        \includegraphics[width=\linewidth]{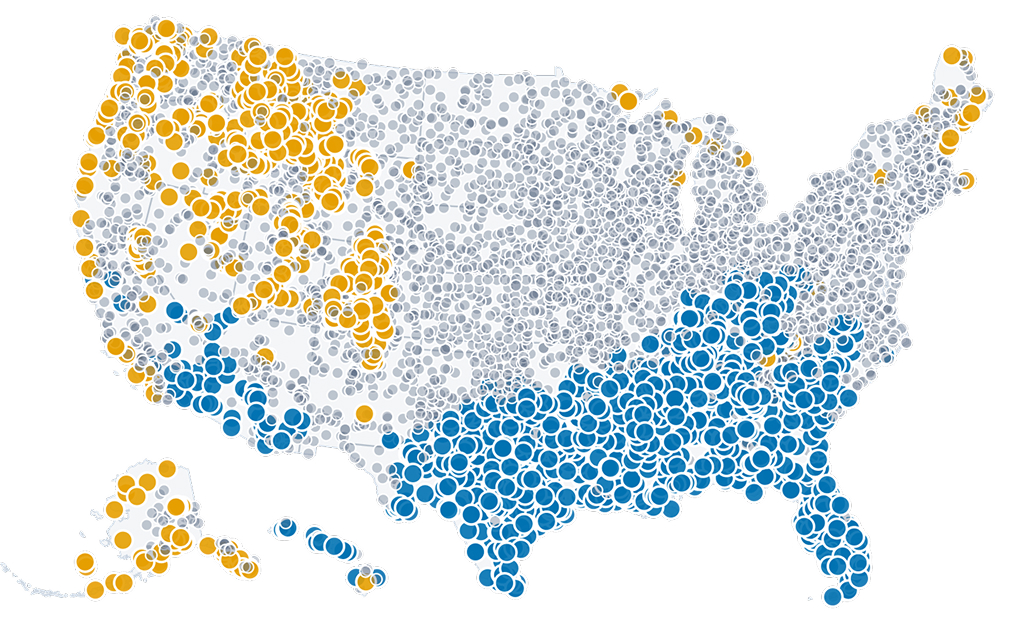}
        \caption{Geographic footprint of structure-guided subsets}
        \label{fig:teaser_friction_map}
    \end{subfigure}
    \caption{\textbf{Weekly maximum temperature series from 6,187 U.S. weather stations.} The density view (a) emphasizes hotspot A, whose query result remains broad and geographically diffuse (b,c). The structural inconsistency field (d) instead makes corridors B1 and B2 salient, yielding more coherent query results and a clearer north--south geographic separation (e,f).}
    \label{fig:teaser}
}

\renewcommand{\manuscriptnotetxt}{}

\graphicspath{{figs/}{figures/}{pictures/}{images/}{./}} 

\usepackage{tabu}                      
\usepackage{booktabs}                  
\usepackage{placeins}
\usepackage{lipsum}                    
\usepackage{mwe}                       
\usepackage{amsmath} 
\usepackage{amsfonts}

\usepackage{subcaption} 

\usepackage{mathptmx}                  

\usepackage{amsmath,amssymb,amsfonts}
\usepackage{graphicx}
\usepackage{textcomp}
\usepackage{xcolor}
\usepackage{algorithm}
\usepackage{algorithmic}
\usepackage{enumitem}

\begin{document}


\firstsection{Introduction}

\maketitle

\label{sec:intro}
Visualizing large-scale line ensembles remains a persistent challenge across domains such as flow visualization~\cite{brambilla2012illustrative}, diffusion tensor imaging (DTI)~\cite{kratz2013visualization}, and urban mobility analysis~\cite{chen2015survey}. Visualization is central to identifying global trends, coherent flow bundles, and anomalous behaviors. However, as datasets grow to millions of trajectories, directly rendering all paths---the path-based perspective---inevitably results in severe overplotting. The resulting ``hairball'' obscures both structural organization and distributional properties, thereby undermining interpretability. To alleviate this visibility bottleneck, prior work has shifted toward field-based, grid-centered representations that summarize trajectories as static spatial fields. Density-based approaches such as DenseLines~\cite{moritz2018visualizing} effectively reduce occlusion and support scalable rendering. This shift, however, introduces a complementary limitation: once trajectories are collapsed into scalar occupancy, the visualization no longer conveys \emph{how} individual lines relate or agree. Density encodes magnitude rather than structure, and scalar aggregation alone cannot resolve relational consistency. The first row of \cref{fig:teaser} makes this limitation explicit: in \cref{fig:teaser_density}, the cyan window A reveals a dominant hotspot, yet the magenta reference windows B1 and B2 that later become structure-guided query targets remain only weakly expressed in this density view. Querying hotspot A therefore yields the large, structurally mixed subset in \cref{fig:teaser_density_selection}, whose geographic footprint in \cref{fig:teaser_density_map} remains diffuse. Density therefore remains a useful overview of where trajectories accumulate, but it does not explain whether the visible patterns are structurally coherent or internally mixed.

What is needed is a structural account that preserves the scalability and overview advantages of field-based aggregation while assessing whether a visually plausible pattern is genuinely supported by the underlying trajectories. Recent work by Xue \textit{et al.}~\cite{xue2024reducing} increases the visual discriminability of ambiguous density patterns through image-space colorization; however, it remains coupled to the rendered image rather than to the generating trajectory ensemble. What is still lacking is a trajectory-grounded explanation: \textit{which trajectories support an observed pattern, and where does this support deteriorate?} The second row of \cref{fig:teaser} outlines our answer: the Structural Inconsistency Field in \cref{fig:teaser_friction} complements the density overview by making low-inconsistency (consistent) regions visually salient as regions of interest.
In contrast to the density hotspot A, the two corresponding low-inconsistency corridors B1 and B2 become salient and queryable in this structural view, even though they are only weakly expressed in density. Querying those regions yields the more regular bundles shown in \cref{fig:teaser_friction_selection}, and their spatial footprint in \cref{fig:teaser_friction_map} reveals a clearer north--south geographic separation than the density-guided alternative. Although the teaser example employs time-series data, the method targets general 2D line data that can be processed on a screen-space binning grid, including temporal curves, planar trajectories, and projected fibers.

The main construction proceeds in three steps. First, each grid cell stores not only how many trajectory samples pass through it, but also a structure tensor that summarizes their local orientation distribution. Second, each trajectory is evaluated against this tensor field by checking whether its tangent agrees with the locally supported orientation at the locations it traverses. To prevent self-support, dynamic leave-one-out (LOO) evaluation removes the queried trajectory’s own tensor contribution before this comparison. Integrating the resulting local agreements along the trajectory yields a trajectory-fidelity score. Third, each occurrence of a trajectory through a grid cell defines a passage. For each passage, local agreement is averaged over a trajectory segment centered on that cell. The Structural Inconsistency Field (SIF) at the cell is then defined as one minus the mean support over all passages through it. Consequently, low SIF values indicate stable structural support, whereas high values indicate conflict, ambiguity, or weak support. Although conceptually related to Line Integral Convolution (LIC)~\cite{cabral1993lic} through path integration, our method uses integration for analysis rather than rendering. It evaluates observed trajectories against an ensemble-derived tensor field.

The trajectory-fidelity score and the SIF provide complementary views: the former supports whole-trajectory ranking, whereas the latter localizes where path support is maintained or lost. Together, they reveal both which trajectories are structurally supported and where dense patterns reflect consensus, conflict, or diffuse clutter.

By transferring the dominant aggregation cost to fixed-grid tensor computations and evaluating centered trajectory extents via prefix sums, our method enables interactive analysis of large ensembles. Analysts can examine density and inconsistency as complementary overview representations, select a region of interest, extract the corresponding subset, and iteratively peel it away before recomputing the field. Re-evaluating the same trajectory-fidelity measure after each removal progressively exposes secondary structures that remain concealed beneath dominant patterns.

\noindent The primary contributions of this work are:
\begin{itemize}
\item \textbf{Path-Integrated Trajectory Fidelity:} A trajectory-level support metric that accumulates local structural agreement along each path, extending structure-tensor analysis beyond local orientation estimation.
\item \textbf{Dynamic Leave-One-Out (LOO) Scoring:} A counterfactual evaluation strategy that mitigates self-bias by comparing each trajectory against the surrounding ensemble rather than a field partially induced by itself.
\item \textbf{Structural Inconsistency Field:} A passage-centered aggregation scheme that projects trajectory support back into image space, localizing where dense patterns are structurally supported versus where they deteriorate.
\item \textbf{Interactive Structure-Guided Peeling:} A downstream analysis workflow leveraging the Structural Inconsistency Field and trajectory-fidelity measures for iterative subset extraction and re-evaluation.
\end{itemize}

\section{Related Work}
\label{sec:related_work}
Our work introduces a tensor-guided path-integration framework for quantifying structural consensus in dense trajectory ensembles. We review the related literature through the lens of structural consensus.
This perspective distinguishes our approach from prior methods in scalar aggregation, tensor-based feature extraction, geometric and coherence-based clustering, and perceptual enhancement techniques.

\subsection{Field Aggregation: From Density to Tensors}
Field-based, aggregation-centered methods address visual clutter by aggregating discrete trajectories into continuous fields, shifting the focus from individual trajectories to collective statistics.

\noindent\textbf{Scalar Density Estimation.}
To mitigate overdraw in massive datasets, Kernel Density Estimation (KDE) has been adapted to map trajectory data into continuous scalar fields. Willems \textit{et al.}~\cite{willems2009vessel} introduced a convolution-based density map for vessel trajectories, accumulating a scalar value $D(\mathbf{x})$ representing the integrated occupancy of lines. Lampe and Hauser~\cite{lampe2011curve} formalized this as Curve Density Estimates (CDE), utilizing a screen-space line kernel to compute probability density functions (PDFs) that reveal the spatial distribution of bundles.
To improve scalability, continuous parallel coordinates~\cite{heinrich2009continuous} derive a density model for parallel coordinates from continuous scatterplot densities, while DenseLines~\cite{moritz2018visualizing} introduced a discrete density representation of time series that supports interactive exploration at a very large scale.
Further extending this, Scheepens \textit{et al.}~\cite{scheepens2011composite, scheepens2011interactive} proposed composite density maps to encode multivariate attributes (e.g., speed, vessel type) into the density field.
These methods excel at revealing spatial occupancy and enabling real-time exploration of large datasets. However, the aggregation is typically scalar and commutative, which discards directional variance during summation. Consequently, the same high-density region may correspond to very different structures, such as a coherent laminar bundle, a structured crossing, or isotropic noise, without the density field distinguishing among them.

\noindent\textbf{Tensor-Based Methods.}
To recover orientation information lost in scalar fields, researchers have employed structure tensors (local second-moment matrices). Widely used in image processing~\cite{weickert1999coherence} to detect edges and coherence, structure tensors summarize local orientation information without sign ambiguity.
In flow visualization, Diewald \textit{et al.}~\cite{diewald00anisotropic} applied anisotropic nonlinear diffusion to random input textures, smoothing them along vector-field integral lines while enhancing edges in the orthogonal direction to produce multiscale, flow-aligned patterns.
In DTI visualization, Kindlmann and Westin~\cite{kindlmann2006diffusion} introduced glyph packing to make continuous tensor structures more apparent through dense, data-driven glyph placement, using a particle system whose inter-glyph interactions are shaped by the local tensor field.
More recently, Jankowai \textit{et al.}~\cite{jankowai19robust} introduced a robustness-based pipeline for extracting, classifying, and simplifying degenerate points, such as wedges and trisectors, in 2D symmetric tensor fields.
These approaches have successfully leveraged tensor fields for field-based feature extraction and visualization, enabling the analysis of anisotropy and field topology. However, they primarily treat the tensor field as the final representation or extraction target rather than as a metric for evaluating the fidelity of individual trajectories. Our work employs the tensor field as a background metric against which individual trajectories are measured.

\subsection{Trajectory-Centric Analysis: Distance vs.\ Coherence}
The trajectory-centric perspective preserves the identity of individual trajectories. Existing methods can be broadly categorized into those that detect geometric similarity and those that identify flow coherence.

\noindent\textbf{Distance-Based Clustering.}
Many trajectory-clustering approaches rely on geometric proximity~\cite{chen2015survey, morris2008survey}. Algorithms employ partition-and-group strategies~\cite{lee2007trajectory} or other clustering formulations adapted to trajectory data~\cite{jain1999data, zhang2009learning} based on pairwise metrics like Hausdorff distance~\cite{Huttenlocher1993}, Chamfer distance~\cite{butt1998optimum}, or Dynamic Time Warping (DTW)~\cite{bian2018survey}. Dirichlet process mixture models~\cite{hu2013incremental} extend this to probabilistic modeling.
These methods effectively identify geometrically similar groups in moderate-sized ensembles. However, many all-pairs distance formulations scale quadratically, limiting applicability to massive datasets without aggressive downsampling. Additionally, geometric distance measures do not account for flow context: trajectories may be spatially close yet structurally incompatible within the surrounding ensemble.

\noindent\textbf{Geometric Depth \& Centrality.}
To avoid pairwise costs, statistical depth measures rank trajectories from ``median'' to ``outlier'' based on spatial centrality. Contour Boxplots~\cite{whitaker2013contour} and Curve Boxplots~\cite{mirzargar2014curve}, derived from Band Depth~\cite{lopez09depth,lopez2014simplicial}, visualize geometric spread. Related geometric summarization approaches also identify multi-granular trends in ensemble data~\cite{van2016multi}.
These techniques provide valuable summaries of geometric variation and outlier detection. However, they are defined primarily by spatial centrality or geometric grouping rather than by alignment with the dominant orientation structure: a trajectory may be spatially central yet misaligned with the surrounding bundle support. Our approach instead measures path-integrated fidelity to the collective orientation structure.

\noindent\textbf{Flow Coherence (LCS).}
In fluid dynamics, coherence is analyzed via Finite-Time Lyapunov Exponents (FTLE)~\cite{guo2016finite}. FTLE fields measure particle separation rates; ridges correspond to Lagrangian Coherent Structures (LCS) that act as transport barriers.
FTLE and LCS have become a standard framework for identifying material boundaries in continuous flows. However, they rely on integrating a dense velocity field, which is challenging to define for sparse or discrete trajectory datasets. Moreover, FTLE emphasizes separation (divergence), whereas our method focuses on alignment (consensus) derived from spatial statistics of discrete trajectories.

\subsection{Perceptual Enhancement Techniques}
When analytical reduction alone is insufficient, perceptual techniques improve legibility through illumination, texture, deformation, or opacity optimization.

\noindent\textbf{Illumination and Texture.}
To enhance shape perception in density representations, illumination models are applied. Standard Phong shading~\cite{phongshading} and exaggerated shading~\cite{rusinkiewicz06exaggerated} use surface normals for 3D cues. Chen \textit{et al.}~\cite{chen2024visualization} introduced Visualization-Driven Illumination (VIDP) for density plots. For lines, Eichelbaum \textit{et al.}~\cite{eichelbaum13lineao} proposed LineAO for screen-space ambient occlusion. More recently, Xue \textit{et al.}~\cite{xue2025enhancing} presented a pixel-based illumination method that enhances line-density structure while preserving color encoding.
Texture-based methods like Line Integral Convolution (LIC)~\cite{cabral1993lic} and Image-Based Flow Visualization (IBFV)~\cite{vanwijk2002ibfv} advect noise to reveal field orientation.
Illustrative vector-field rendering further enhances flow structures through stylized visual cues; for example, Chen \textit{et al.}~\cite{chen2011illustrative} proposed an illustrative visualization framework for 3D vector fields.
These techniques effectively provide perceptual cues for depth and orientation. However, they improve rendering and interpretability of the displayed field without, by themselves, defining a trajectory-level measure of structural consensus or a quantitative field for evaluating local trajectory agreement.

\noindent\textbf{Edge Bundling.}
To reduce screen-space clutter, edge bundling algorithms deform lines to group spatially proximate paths. Techniques range from hierarchical approaches~\cite{holten2006hierarchical} to force-directed models~\cite{holten2009force, hurter12graph} and geometry-based constraints~\cite{zwan2016cubu, zeng2019route}. Hurter \textit{et al.}~\cite{hurter2018functional} further formulated bundled simplification through functional decomposition, representing curve sets with spline bases and principal component functions to support statistically controlled bundling and unbundling. Recent work has addressed ambiguity more explicitly: Lhuillier \textit{et al.}~\cite{lhuillier2017state} review it as a central issue in bundling, Wallinger \textit{et al.}~\cite{wallinger2021edge} propose Edge-Path bundling to reduce independent edge ambiguities, and a follow-up accelerates Edge-Path bundling through graph spanners~\cite{wallinger2023faster}.
Bundling produces aesthetically clear visualizations and effectively reduces clutter in many contexts. However, bundling intentionally alters the trajectory geometry, which can be problematic in applications that require faithful path shapes or precise spatial attribution (e.g., traffic safety analysis).

\noindent\textbf{Perceptual Optimization.}
As a non-destructive alternative to geometric simplification or deformation, perceptual optimization improves visibility by adjusting rendering attributes such as opacity and color.
Günther \textit{et al.}~\cite{guenther2013opacity} address visibility in dense 3D line fields through opacity optimization. Complementary color-design approaches operate at a different level: Palettailor~\cite{lu2020palettailor} optimizes discriminable categorical palettes, while Xue \textit{et al.}~\cite{xue2024reducing} use image-space colorization to reduce ambiguities in line-based density plots.
These methods improve visibility through rendering, opacity, or color design rather than by defining a trajectory-aware structural metric. Accordingly, they are complementary perceptual baselines rather than direct counterparts to methods that score individual trajectories or quantify local structural agreement. In contrast, our tensor-guided peeling uses structural consensus to prioritize trajectories and progressively reveal the coherent skeleton.

Taken together, existing approaches either aggregate trajectories into fields that obscure trajectory-level support from the surrounding orientation structure or preserve trajectory identity at the cost of scalability and contextual alignment. Importantly, in ambiguous configurations such as crossings versus coherent bundles, density, depth, and separation-based measures may still be insufficient to disambiguate local structural support. Our tensor-guided path-integral formulation provides a complementary criterion by evaluating trajectory alignment against an ensemble-derived tensor field, enabling distinctions that are difficult to obtain from prior representations alone.

\section{Background: From Scalar Mass to Vector Order}
\label{sec:background}

Our framework rests on the transition from discrete trajectory-centric representations to continuous field-based aggregations. This section serves one specific purpose: it defines the local field quantities that are subsequently used by the path-integrated metric. We therefore focus the discussion of the background on two main aspects only: a zero-order mass field that measures where trajectories accumulate, and a second-order tensor field that describes how they are locally oriented. The methodological question of how these local quantities are evaluated along trajectories is deferred to the next section.

\subsection{The Zero-Order Moment: Line Density (CDE)}
\label{sec:bg_cde}

Standard point-based kernel density estimation (KDE) is ill-suited for trajectory data.
We therefore adopt curve density estimation (CDE)~\cite{lampe2011curve}, which defines scalar occupancy by convolving curve geometry with a spatial kernel along arc length. In this paper, the role of CDE is purely foundational: it provides the zero-order mass term from which our tensor field is built. Since this construction is inherited from prior work, we keep the formal continuous definition and illustrative diagram in the appendix (\cref{appendix:cde,fig:app_cde}) and retain here only the discrete form used downstream.

For efficient fixed-grid computation, each segment is rasterized into a set of pixels $P_{\text{segment}}$. Following the same bin-based approximation used in prior line-density work~\cite{xue2025enhancing}, we assign a constant segment weight to the covered pixels:
\begin{equation}
    D(\mathbf{x}) \approx \sum_{\text{segments}} \omega_s \sum_{\mathbf{p} \in P_{\text{segment}}} K_h(\mathbf{x} - \mathbf{p})
    \label{eq:discrete_cde}
\end{equation}
where $K_h$ is the spatial smoothing kernel with bandwidth $h$, $\mathbf{p}$ denotes the center of each rasterized pixel, and $\omega_s$ denotes the constant arc-length factor assigned to the current rasterized segment in this fixed-grid approximation.
The resulting field $D(\mathbf{x})$ measures scalar occupancy (e.g., \cref{fig:teaser_density}). This zero-order moment is lifted in the next subsection to a second-order tensor field. It supplies the mass term while the tensor adds directional information.

\begin{figure}[tb]
    \centering
    \begin{subfigure}[b]{0.34\linewidth}
        \centering
        \includegraphics[width=\linewidth]{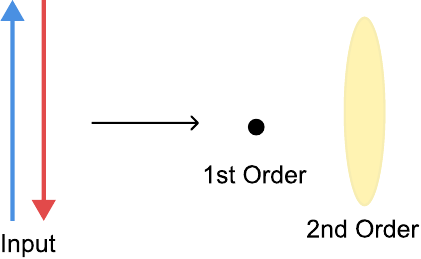}
        \caption{Sign Ambiguity}
        \label{fig:sign_ambiguity} 
    \end{subfigure}
    \hfill
    \begin{subfigure}[b]{0.25\linewidth}
        \centering
        \includegraphics[width=\linewidth]{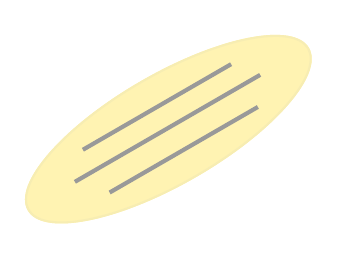}
        \caption{Rank-1: Coherence}
        \label{fig:rank1}
    \end{subfigure}
    \hfill
    \begin{subfigure}[b]{0.34\linewidth}
        \centering
        \includegraphics[width=\linewidth]{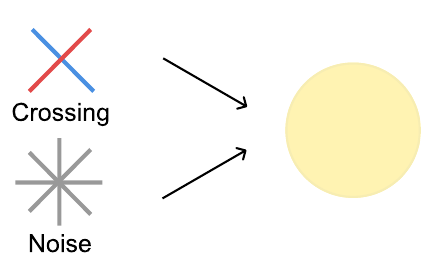}
        \caption{Rank-2: Ambiguity}
        \label{fig:rank2}
    \end{subfigure}
    \vspace{-2mm}
    \caption{\textbf{Geometric interpretation of the Structure Tensor.} (a) For bi-directional flow, the first-order vector mean cancels out (circle), while the second-order structure tensor preserves orientation (ellipse). (b) High anisotropy ($\lambda_1 \gg \lambda_2$) indicates coherent laminar flow. (c) Low anisotropy ($\lambda_1 \approx \lambda_2$) is semantically blind, failing to distinguish between structured crossings and unstructured noise.}
    \label{fig:tensor_concept}
     \vspace{-4mm}
\end{figure}

\subsection{The Second-Order Moment: Structure Tensor Field}
\label{sec:bg_structure_tensor}
To augment scalar occupancy with local orientation while retaining a field-based representation, naïve first-order vector averaging is inadequate, as opposing directions cancel in bidirectional corridors and crossings, producing spurious voids in high-density regions. We therefore employ the second-order moment, i.e., the \textbf{structure tensor}, a local second-moment matrix~\cite{weickert1999coherence} that can be interpreted as a covariance-like summary of local orientation statistics, albeit without mean-centering, which encodes orientation via tangent outer products and avoids sign-cancellation artifacts (\cref{fig:sign_ambiguity}).

Building directly on \cref{eq:discrete_cde}, we keep the same rasterized kernel mass contributions but lift each scalar contribution into a second-order outer product using the local normalized tangent. We therefore define the tensor field over the same fixed grid as
\begin{equation}
    \mathbf{J}(\mathbf{x}) \approx \sum_{\text{segments}} \omega_s \sum_{\mathbf{p} \in P_{\text{segment}}} K_h(\mathbf{x} - \mathbf{p}) \cdot (\mathbf{v}_{\mathbf{p}} \mathbf{v}_{\mathbf{p}}^T)
    \label{eq:discrete_tensor}
\end{equation}
Since each term $\mathbf{v}_{\mathbf{p}}\mathbf{v}_{\mathbf{p}}^T$ is symmetric positive semi-definite and has unit trace, $\mathbf{J}(\mathbf{x})$ is likewise symmetric positive semi-definite, and taking the trace of \cref{eq:discrete_tensor} directly recovers \cref{eq:discrete_cde}. In this sense, the structure tensor is a strict extension of density: it retains occupancy while additionally encoding the local orientation distribution.

The eigenstructure of $\mathbf{J}(\mathbf{x})$ yields the local geometric interpretation summarized in \cref{fig:tensor_concept}. When $\lambda_1 \gg \lambda_2$, the neighborhood is highly anisotropic and forms a coherent laminar bundle aligned with the principal eigenvector. When $\lambda_1 \approx \lambda_2$, the tensor is locally isotropic, indicating either a structured crossing or unstructured clutter; in both cases, the field is semantically ambiguous. This delineates the limit of the background: density and the structure tensor characterize locally supported mass and orientation, but do not indicate whether such support remains coherent along an entire trajectory. The next section converts these local field quantities into a path-integrated criterion for structural agreement.

\section{Methodology}
\label{sec:methodology}
The methodological question is now how to turn these local statistics into a criterion that respects trajectory identity. This reflects the central grid-based/path-based tension of our problem: the field provides scalable context, but the trajectories carry the semantics we ultimately want to judge.

Our solution is to use the tensor field not as the final visualization, but as the environment against which trajectories are evaluated. The construction is analogous to LIC in that path integration remains the core operator, but the roles are reversed: the field is no longer the object being visualized, but the contextual medium against which trajectories are judged. The methodology therefore proceeds in four steps: we construct the tensor field, define a local structural-support quantity, integrate that quantity along trajectories (with leave-one-out correction), and then aggregate passage-centered path support back into space as a structural inconsistency field. \cref{sec:system} explains how these quantities are exposed through a lightweight visual analytics interface.

\subsection{Structure Tensor Field Construction}
\label{sec:method_field}

\begin{figure}[t]
    \centering
    \begin{subfigure}[b]{0.32\linewidth}
        \centering
        \includegraphics[width=0.8\linewidth]{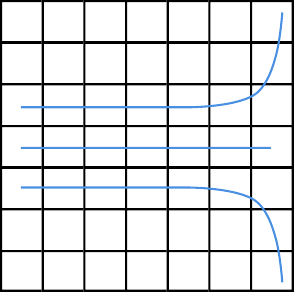}
        \caption{Input Trajectories}
        \label{fig:structure_tensor_lines} 
    \end{subfigure}
    \hfill
    \begin{subfigure}[b]{0.32\linewidth}
        \centering
        \includegraphics[width=0.8\linewidth]{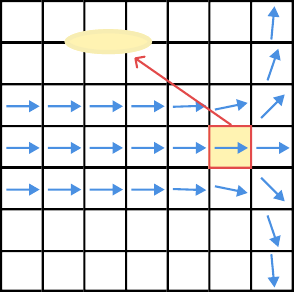}
        \caption{Local Accumulation ($1 \times 1$)}
        \label{fig:structure_tensor_1}
    \end{subfigure}
    \hfill
    \begin{subfigure}[b]{0.32\linewidth}
        \centering
        \includegraphics[width=0.8\linewidth]{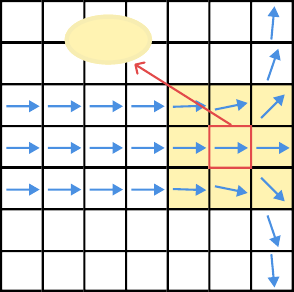}
        \caption{Neighborhood Aggregation ($3 \times 3$)}
        \label{fig:structure_tensor_3}
    \end{subfigure}
    \vspace{-2mm}
    \caption{\textbf{Construction of the structure tensor field.} (a) Trajectories are rasterized onto the analysis grid. (b) With a minimal kernel ($1\times1$), the tensor records only the instantaneous tangent. (c) A wider spatial kernel ($3\times3$) aggregates neighborhood statistics and reveals crossings or divergences through increased isotropy.}
    \label{fig:structure_tensor}
    \vspace{-3mm}
\end{figure}

The first step is to build the field-based reference representation against which trajectories will later be scored. Given the structure-tensor background in \cref{sec:bg_structure_tensor}, we discretize the domain $\Omega$ into a grid of tensors so that every raster sample along a trajectory can query the orientation statistics of its surrounding environment.

\noindent\textbf{Accumulation and Kernel Density.}
We rasterize each trajectory into tangent-carrying samples and splat them to nearby bins with nonnegative kernel weights; the exact discrete weighting convention is shown in \cref{appendix:method_impl}. Following the pixel-based discretization paradigm visualized in \cref{fig:structure_tensor}, we construct the field via a splatting approach. For each spatial bin $b$, the aggregate Structure Tensor $\mathbf{J}_{\text{total}}(b)$ is computed as the weighted superposition of all rasterized trajectory samples whose kernel support overlaps $b$:

\begin{equation}
    \mathbf{J}_{\text{total}}(b) = \sum_s \omega_s(b) \cdot (\mathbf{v}_s \mathbf{v}_s^T)
    \label{eq:tensor_accumulation}
\end{equation}
where $\mathbf{v}_s$ is the normalized tangent at sample $s$ and $\omega_s(b)$ is the corresponding spatial kernel weight. 

\noindent\textbf{The Role of Integration Scale.}
The choice of the kernel support $\omega$ determines the structural resolution of the field, as illustrated by the contrast between \cref{fig:structure_tensor_1} and \cref{fig:structure_tensor_3}. A minimal kernel ($1 \times 1$) records only the instantaneous tangent at the sampled pixel and therefore remains sensitive to rasterization artifacts and local undersampling. Expanding the kernel support turns the field into a local statistical accumulator: nearby segments contribute jointly, crossings and divergences increase the secondary eigenvalue, and the tensor transitions from a nearly singular local orientation estimate to a neighborhood-level description of anisotropy. This scale-dependent aggregation is the practical mechanism by which the field acquires the local complexity cues later exploited by the path-integrated score.

\subsection{Path Integration and Structural Agreement}
\label{sec:method_metric}

\begin{figure}[t]
    \centering
    \begin{subfigure}[b]{0.32\linewidth}
        \centering
        \includegraphics[width=0.8\linewidth]{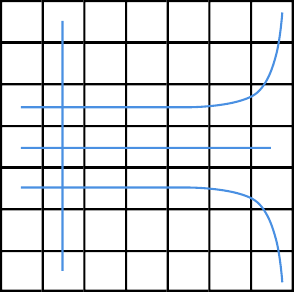}
        \caption{Raw Trajectories}
        \label{fig:metric_step1_raw}
    \end{subfigure}
    \hfill
    \begin{subfigure}[b]{0.32\linewidth}
        \centering
        \includegraphics[width=0.8\linewidth]{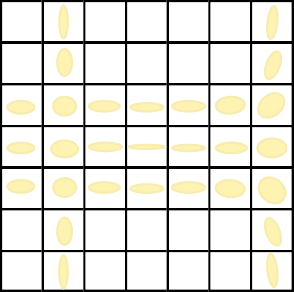}
        \caption{Tensor Field Context}
        \label{fig:metric_step2_tensor}
    \end{subfigure}
    \hfill
    \begin{subfigure}[b]{0.32\linewidth}
        \centering
        \includegraphics[width=0.8\linewidth]{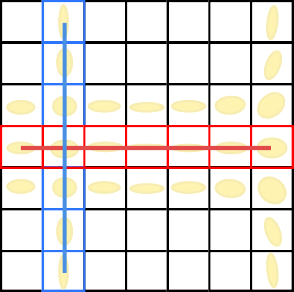}
        \caption{Path Integration Probes}
        \label{fig:metric_step3_execute}
    \end{subfigure}
    \vspace{-2mm}
    \caption{\textbf{Path integration with respect to the tensor field.} (a) Input trajectories. (b) The associated tensor field. (c) Each trajectory is evaluated via integration of tangent–field alignment along its path.}
    \label{fig:metric_process}
    \vspace{-4mm}
\end{figure}


Once the field has been constructed, the next question is how to judge an individual trajectory against it. A purely local score is not sufficient for this purpose: a trajectory may align with the field at one location yet contradict the dominant structure over the remainder of its path. We therefore define the evaluation through \textbf{path integration}, visually summarized in \cref{fig:metric_process}. Just as a matched filter responds maximally when a signal matches a template, our metric responds maximally when a trajectory's geometry aligns with the background tensor field over an extended path segment. Importantly, the quantity being measured is support by the surrounding ensemble, not the intrinsic smoothness or physical plausibility of an isolated trajectory.

\noindent\textbf{Local Structural Support.}
The process begins by establishing the context from the raw data (\cref{fig:metric_step1_raw}) into a tensor field (\cref{fig:metric_step2_tensor}). Then, as illustrated in \cref{fig:metric_step3_execute}, for any specific location $\mathbf{x}$ along a trajectory, the alignment between its tangent $\mathbf{v}_i$ and the local environment $\mathbf{J}(\mathbf{x})$ is quantified by the trace-normalized quadratic form:

\begin{equation}
    E_{\text{local}}(\mathbf{x}, \mathbf{v}_i) = \frac{\mathbf{v}_i^T \cdot \mathbf{J}(\mathbf{x}) \cdot \mathbf{v}_i}{\text{tr}(\mathbf{J}(\mathbf{x})) + \epsilon}
    \label{eq:local_consistency}
\end{equation}
Here, the denominator $\text{tr}(\mathbf{J})$ approximately normalizes out local traffic volume away from sparse regions, while $\epsilon$ prevents division by zero in sparse regions.
Structurally, this measures the fraction of local energy that aligns with the trajectory's orientation. In practice, $\mathbf{J}(\mathbf{x})$ is queried from the rasterized tensor grid at the bin containing $\mathbf{x}$, with optional interpolation between neighboring bins in the implementation. Since $\mathbf{J}(\mathbf{x}) \succeq 0$ and $\|\mathbf{v}_i\|=1$, we have $0 \le \mathbf{v}_i^T\mathbf{J}(\mathbf{x})\mathbf{v}_i \le \lambda_{\max}(\mathbf{J}(\mathbf{x})) \le \operatorname{tr}(\mathbf{J}(\mathbf{x}))$, and therefore $0 \le E_{\text{local}}(\mathbf{x},\mathbf{v}_i) < 1$ (or $E_{\text{local}}\le 1$ when $\epsilon$ is ignored). Because the tensor is formed from $\mathbf{v}\mathbf{v}^T$, this score measures undirected orientation support rather than signed directional consistency.
\begin{itemize}
    \item \textbf{Consensus ($E \to 1$):} The trajectory moves parallel to the principal axis $\mathbf{e}_1$ in a highly anisotropic region (e.g., the horizontal red path aligning with horizontal tensors in \cref{fig:metric_step3_execute}).
    \item \textbf{Conflict / Inconsistency ($E \to 0$):} The trajectory cuts orthogonally across a bundle in a strongly anisotropic region (e.g., where the horizontal red probe crosses the vertical blue bundle in \cref{fig:metric_step3_execute}).
    \item \textbf{Ambiguity ($E \approx 0.5$ in our 2D screen-space setting):} In locally isotropic regions, $\mathbf{J}(\mathbf{x}) \approx \lambda \mathbf{I}_2$, so $E_{\text{local}} \approx \lambda/(2\lambda+\epsilon) \approx 0.5$ for any unit tangent. Such regions, therefore, express weak directional support from the environment rather than strong orthogonal conflict.
\end{itemize}

\noindent\textbf{Path-Integrated Fidelity.}
The local score resolves only pointwise agreement. To compare complete trajectories, we will integrate this support signal along arc length and normalize by trajectory length; however, because the final trajectory score is defined against an unbiased environment, we postpone its formal definition until after leave-one-out correction.

\subsection{Dynamic Leave-One-Out Filtering}
\label{sec:method_loo}

Before using these trajectory scores either for ranking or for spatial aggregation, we must address a critical bias which we call the \textbf{Self-Fulfilling Prophecy}: a strong outlier contributes to the local tensor field, thereby biasing the local orientation pattern toward itself and artificially inflating its own structural-support estimate. Leave-one-out is used here as a standard counterfactual evaluation principle, widely used in density estimation and cross-validation~\cite{bowman1984alternative}: a trajectory should be judged by support from the surrounding ensemble, not by a field partially defined by itself. The motivation for leave-one-out is therefore not a computational refinement but a semantic one: without it, the metric can become self-confirming in exactly the cases where it should be most skeptical. In very dense regions, this correction has little numerical effect because one trajectory contributes only a small fraction of the local tensor mass. It becomes more important in sparse regions, for isolated outliers, and after interactive peeling, where the remaining subset can be small enough that self-support would otherwise dominate the local estimate.

\begin{figure}[t]
    \centering
    \begin{subfigure}[b]{0.32\linewidth}
        \centering
        \includegraphics[width=0.8\linewidth]{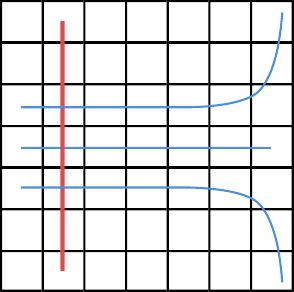}
        \caption{Candidate Trajectory (Red)}
        \label{fig:loo_step1}
    \end{subfigure}
    \hfill
    \begin{subfigure}[b]{0.32\linewidth}
        \centering
        \includegraphics[width=0.8\linewidth]{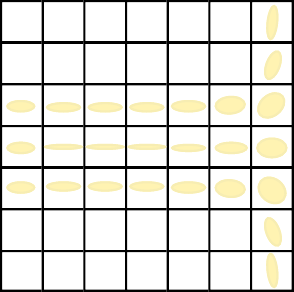}
        \caption{Unbiased Environment $\mathbf{J}_{\text{env}}$}
        \label{fig:loo_step2}
    \end{subfigure}
    \hfill
    \begin{subfigure}[b]{0.32\linewidth}
        \centering
        \includegraphics[width=0.8\linewidth]{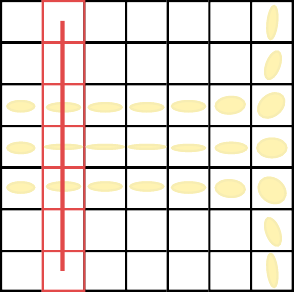}
        \caption{Structural Inconsistency Test}
        \label{fig:loo_step3}
    \end{subfigure}
    \vspace{-2mm}
    \caption{\textbf{Dynamic leave-one-out (LOO).} (a) A candidate trajectory (in red) is selected for evaluation. (b) Its contribution is removed to form the environment tensor $\mathbf{J}_{\text{env}}$. (c) The candidate is then evaluated against this unbiased environment.}
    \label{fig:loo_mechanism}
    \vspace{-4mm}
\end{figure}

To avoid self-bias during evaluation, we employ a \textbf{Dynamic Leave-One-Out (LOO)} strategy, illustrated in \cref{fig:loo_mechanism}. The term dynamic refers to the adaptation required by our tensor/path-integration setting: the counterfactual environment is recomputed for each queried trajectory rather than fixed once for the entire dataset. For each trajectory $L_i$, we compute a trajectory-conditioned environment by subtracting its own contribution from the global tensor field:
\[
    \mathbf{J}_{\text{env}}^{(i)}(b)=\mathbf{J}_{\text{total}}(b)-\mathbf{C}_i(b).
\]
Here, $\mathbf{C}_i(b)$ denotes the sum of the sample-induced tensor contributions originating from $L_i$ in bin $b$, so $\mathbf{J}_{\text{env}}^{(i)}$ reflects the tensor induced by all remaining trajectories.
If removing $L_i$ leaves no remaining tensor mass, the resulting zero support indicates an absence of independent evidence rather than directional conflict.
We then evaluate $L_i$ against this unbiased environment by applying \cref{eq:local_consistency} with $\mathbf{J}$ replaced by $\mathbf{J}_{\text{env}}^{(i)}$, yielding the leave-one-out local support $E_{\text{local}}^{\text{LOO}}(\mathbf{x},\mathbf{v}_i(\mathbf{x});L_i)$.
The final trajectory score used throughout the remainder of the paper is the leave-one-out corrected path integral:

\begin{equation}
    \Phi(L_i) = \frac{1}{|L_i|}\int_{L_i} E_{\text{local}}^{\text{LOO}}(\mathbf{x}(s),\mathbf{v}_i(s);L_i)\,ds
    \label{eq:global_fidelity}
\end{equation}
The length normalization makes $\Phi(L_i)$ a trajectory-fidelity score, which is comparable across paths of different extents rather than a quantity dominated by sample count. Hereafter, $\Phi(L_i)$ refers to this leave-one-out corrected form unless stated otherwise.

\subsection{Spatial Aggregation}
\label{sec:method_vis}

While $\Phi(L_i)$ characterizes an entire trajectory, analysts must also determine \emph{where} support at a queried pixel remains stable or degrades as additional path context is incorporated. The local tensor captures whether a pixel neighborhood exhibits anisotropy, and $\Phi(L_i)$ reflects global trajectory support; however, neither establishes whether support at a specific pixel is consistent across the set of trajectory passages that traverse it. Purely local descriptors are therefore spatially myopic, as they operate on fixed neighborhoods, whereas $\Phi(L_i)$ aggregates over the full trajectory without localizing where support is accumulated or lost.

\begin{figure}[t]
    \centering
    \hfill
    \begin{subfigure}[b]{0.43\linewidth}
        \centering
        \includegraphics[width=\linewidth]{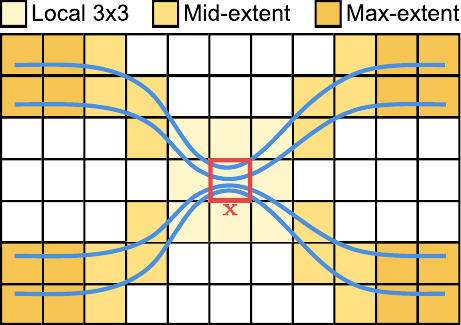}
        \caption{Coherent Corridor}
        \label{fig:arf_stable}
    \end{subfigure}
    \hfill
    \begin{subfigure}[b]{0.43\linewidth}
        \centering
        \includegraphics[width=\linewidth]{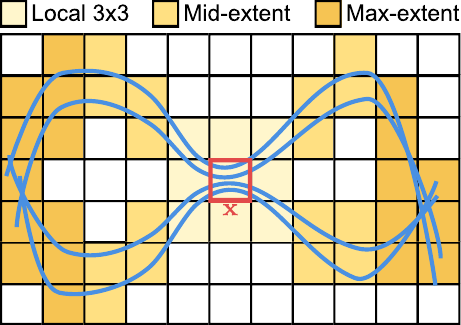}
        \caption{Chaotic Knot}
        \label{fig:arf_mess}
    \end{subfigure}
    \hfill
    \vspace{-2mm}
    \caption{\textbf{Comparison of local and path-aligned support geometries.} The red box marks the queried pixel $\mathbf{x}$. The colored cells illustrate the corresponding support geometries as the context grows from a local $3\times 3$ neighborhood to a mid-sized path extent and finally to the maximal path extent.}
    \label{fig:arf_concept}
    \vspace{-4mm}
\end{figure}

To address this limitation, for a trajectory passing through the queried pixel $(\mathbf{x})$, we follow the trajectory in both directions from $\mathbf{x}$ up to an arc length $(\ell)$, collecting all pixels along the way. Tracing stops on any side that reaches a trajectory endpoint first. This defines the segment over which the passage is evaluated, with a maximum total arc length of $2\ell$, centered at the arc-length position $s_\mathbf{x}$ of the passage.
The scenarios in \cref{fig:arf_concept} illustrate the necessity of this additional spatial quantity. In the vicinity of the queried pixel $\mathbf{x}$, the local $3\times 3$ neighborhoods in (a) and (b) appear nearly indistinguishable, and even intermediate path extents reveal only minor differences. Clear separation emerges only when the centered path extent expands to the maximal support region: in (a) it remains within a coherent corridor, whereas in (b) it incorporates conflicting path context from a chaotic knot. Accordingly, the role of spatial aggregation is not merely to enlarge the neighborhood, but to extend it along trajectory-supported path contexts. This is our motivation to aggregate passage-centered support back into image space.

\begin{figure}[t]
    \centering
    \hfill
    \begin{subfigure}[b]{0.45\linewidth}
        \centering
        \includegraphics[width=\linewidth]{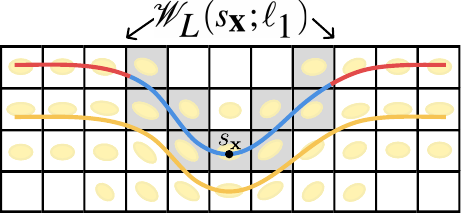}
        \caption{Short path extent $\mathcal{W}_L(s_\mathbf{x};\ell_1)$}
        \label{fig:window_short}
    \end{subfigure}
    \hfill
    \begin{subfigure}[b]{0.45\linewidth}
        \centering
        \includegraphics[width=\linewidth]{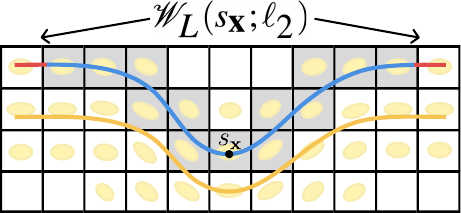}
        \caption{Long path extent $\mathcal{W}_L(s_\mathbf{x};\ell_2)$}
        \label{fig:window_long}
    \end{subfigure}
    \hfill
    \vspace{-2mm}
    \caption{
    \textbf{Passage-centered path extent.} (a) For a representative passage through $\mathbf{x}$, the queried trajectory is evaluated over a symmetric path interval of half-length $\ell_1$ around $\mathbf{x}$. The blue region denotes the continuous extent, and the gray pixels denote the sampled pixels intersected by the discretized trajectory within it. (b) Increasing the extent to $\ell_2>\ell_1$ evaluates the same passage over a longer trajectory segment.
    }
    \label{fig:window_concept}
    \vspace{-4mm}
\end{figure}

In \cref{fig:window_short,fig:window_long}, the yellow trajectories define the leave-one-out environment field, while the queried trajectory $L$ (red) is evaluated against it. The blue region indicates the continuous path extent obtained by moving a distance $\ell$ along $L$ in both directions from $\mathbf{x}$, and the gray pixels indicate the corresponding rasterized trajectory samples. As $\ell$ increases from $\ell_1$ to $\ell_2$, the same passage is evaluated over a longer segment of $L$. Conceptually, we aggregate path-aligned support along the trajectories that traverse $\mathbf{x}$ rather than just accumulating evidence from a fixed neighborhood.
Formally, we denote a passage of $L$ through $\mathbf{x}$ by the pair $(L,s_\mathbf{x})$, where $s_\mathbf{x} \in [0, |L|]$ is the arc-length position at which $L$ traverses $\mathbf{x}$ (in the discrete implementation, the midpoint arc length of the traversal; see \cref{appendix:method_impl}). Also, we denote the corresponding continuous path extent by $\mathcal{W}_L(s_\mathbf{x};\ell)$. The parameter $\ell$ controls how far the query extends from $s_\mathbf{x}$ along the trajectory in each direction, clipped to the trajectory endpoints if necessary. The resulting quantity $\phi(L,s_\mathbf{x};\ell)$ represents the support of one passage averaged over that path extent (formally defined in \cref{eq:passage_support}), not yet the final pixel-level value.

Each trajectory $L : [0, |L|] \rightarrow \Omega$ is parameterized by arc length. $\mathcal{P}_\mathbf{x}$ denotes the set of all passages $(L,s_\mathbf{x})$ passing through pixel $\mathbf{x}$. If a trajectory revisits the same pixel at different arc-length positions, these revisits are treated as distinct passages because they induce different path contexts.
Accordingly, $\phi(L,s_\mathbf{x};\ell)$ and $\Phi(L)$ rely on the same local support primitive but address different scales: $\Phi(L)$ is trajectory-global, whereas $\phi(L,s_\mathbf{x};\ell)$ is passage-centered and conditioned on location.

The corresponding \textbf{Structural Inconsistency Field (SIF)} $M(\mathbf{x};\ell)$ at pixel $\mathbf{x}$ is then defined as one minus the mean passage-centered support across all passages traversing that pixel:
\begin{equation}
    M(\mathbf{x};\ell)
    =
    1 - \frac{1}{|\mathcal{P}_\mathbf{x}|}
    \sum_{(L,s_\mathbf{x}) \in \mathcal{P}_\mathbf{x}}
    \phi(L,s_\mathbf{x};\ell),
    \qquad \mathbf{x}\in\Omega_{\text{supp}}
\label{eq:arf_metric}
\end{equation}

We define $M(\mathbf{x};\ell)$ only on the support region \mbox{$\Omega_{\text{supp}}=\{\mathbf{x}\,:\,|\mathcal{P}_\mathbf{x}|>0\}$}; empty pixels are masked in the rendering and excluded from summation. For pixels with only one or two passages, $M$ remains defined but should be interpreted together with density or passage count as low-confidence evidence. Because $\phi(L,s_\mathbf{x};\ell)\in[0,1]$, it follows that $M(\mathbf{x};\ell)\in[0,1]$ on $\Omega_{\text{supp}}$. Parameter $\ell$ controls the transition from a purely local measurement to increasingly larger path context: $\ell = 0$ is pointwise, intermediate values correspond to intermediate path extents, and larger values extend the evaluation toward the maximal path geometry illustrated in \cref{fig:arf_concept}. In the discrete implementation, the parameter $\ell$ controls how many rasterized trajectory samples are taken before and after the sample corresponding to $\mathbf{x}$ along the trajectory. These sampled values are efficiently accumulated using prefix sums. This allows us to distinguish two cases in \cref{fig:arf_concept}: in the coherent corridor, expanding $\ell$ keeps the passage-centered supports high and thus $M(\mathbf{x};\ell)$ small, whereas in the chaotic knot, larger path extents expose conflicting path context and raise $M(\mathbf{x};\ell)$ substantially.

\subsection{Visual Encoding and Interactive Analytics}
\label{sec:system}

Translating this tensor-guided formulation into an effective analysis tool requires bridging the gap between abstract tensor metrics and visual interpretation. Our system is organized as a progressive analysis workflow. It first provides an overview of the full dataset through density and structural inconsistency views, allowing analysts to identify potentially interesting regions. It then supports region-based exploration through direct selection and TimeBox~\cite{hochheiser2004dynamic} queries, extracts the selected trajectories as a temporary subset, recomputes the trajectory-fidelity scores on that subset alone, and finally enables iterative peeling of the resulting line groups (see supplementary video 
and the static workflow illustrations of the peeling in \cref{appendix:workflow,fig:app_peeling_workflow}).

\noindent\textbf{Decoupling Density from Dissonance.}
To prevent visual interference between scalar volume and structural inconsistency, the overview stage offers three complementary visual encodings. The trajectory density $D(\mathbf{x})$ is from \cref{eq:discrete_cde}; it is mapped to a standard sequential colormap (e.g., Magma or Viridis).
The system can also display the inconsistency field $M(\mathbf{x};\ell)$ as a structural overview, or combine the two by driving an image-space desaturation filter on $D(\mathbf{x})$ with $M(\mathbf{x};\ell)$ so that high-inconsistency pixels lose hue while still retaining density magnitude through luminance. This decoupled encoding lets analysts inspect density alone, inconsistency alone, or both together in a single focus+context view, which serves as the overview layer for locating regions of interest.

\noindent\textbf{Perceptual Focus+Context.} 
After an interesting region has been identified in the overview, analysts can select it either by direct spatial brushing or by TimeBox~\cite{hochheiser2004dynamic} interaction. The selected trajectories are treated as a separate subset, and the tensor-guided pipeline is run again using only those trajectories, i.e., all unselected lines are excluded from the recomputed tensor field and path integrals. This produces trajectory-level fidelity scores $\Phi(L)$ for the selected subset only, turning the overview into a focused, local structural analysis of the region under inspection.

\noindent\textbf{Optional Contrast Enhancement.}
When inconsistency values occupy a narrow display range, the system optionally applies histogram equalization to the rendered inconsistency field as a display-only post-processing step. This monotonic remapping improves visual contrast in the case-studies below, but it does not modify the underlying inconsistency values used for scoring, filtering, interaction, or quantitative evaluation.

\noindent\textbf{Iterative Peeling via Consensus Filter}.
To facilitate exploratory analysis, the interface features a \textit{Consensus Filter}---a selection-conditioned percentile filter over the trajectory-fidelity scores $\Phi(L)$ computed for the current region of interest. This filter acts on the selected trajectory subset, not directly on the pixel-level inconsistency field $M(\mathbf{x};\ell)$, and can be used to isolate either highly supported structures or lower-support anomalies. The filtered subset can then be peeled out as a separate line group, after which the field is recomputed on the remaining trajectories and the same workflow can be repeated. In this way, a complex dataset is progressively decomposed into a collection of interpretable trend subsets, each revealed by the same trajectory-fidelity scoring mechanism. Furthermore, because field recomputation is dominated by fixed-grid filtering and path-extent evaluation uses prefix sums, users can remove the filtered subset and trigger a rapid recomputation of the tensor field. This allows the system to tighten its estimate of the dominant structure around the remaining coherent bundles, enabling iterative peeling until the main structural skeleton is cleanly isolated.

\section{Evaluation}
\label{sec:evaluation}

We evaluate our method in three ways: controlled synthetic benchmarks, a fixed-grid performance evaluation, and real-world case studies. Supplementary path-extent analyses and additional visual comparisons are deferred to the appendix.

\subsection{Synthetic Benchmarks}
\label{sec:eval_synthetic}

\begin{figure}[!t]
    \centering
    \begin{subfigure}[b]{0.22\linewidth}
        \centering
        \fbox{\includegraphics[width=\linewidth]{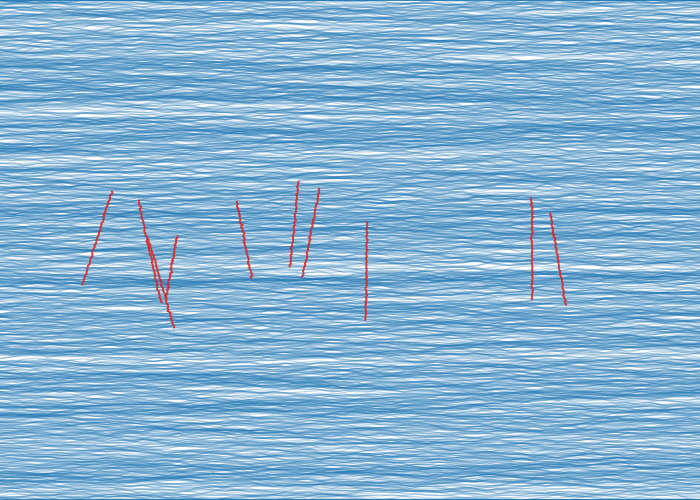}}
        \caption{D1}
        \label{fig:D1_raw}
    \end{subfigure}
    \hfill
    \begin{subfigure}[b]{0.22\linewidth}
        \centering
        \fbox{\includegraphics[width=\linewidth]{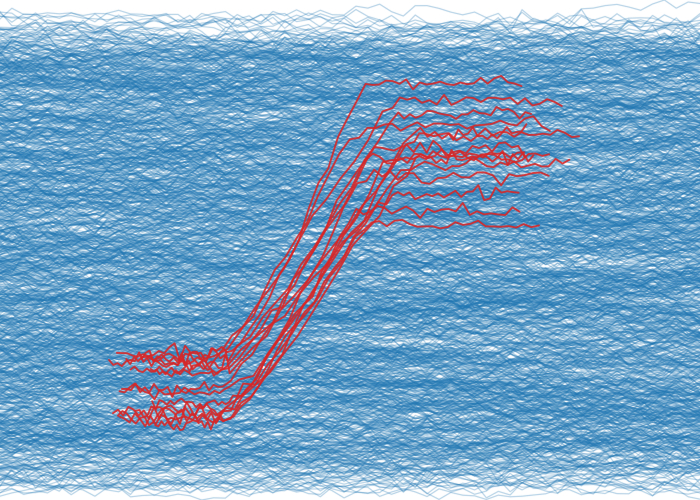}}
        \caption{D2}
        \label{fig:D2_raw}
    \end{subfigure}
    \hfill
    \begin{subfigure}[b]{0.22\linewidth}
        \centering
        \fbox{\includegraphics[width=\linewidth]{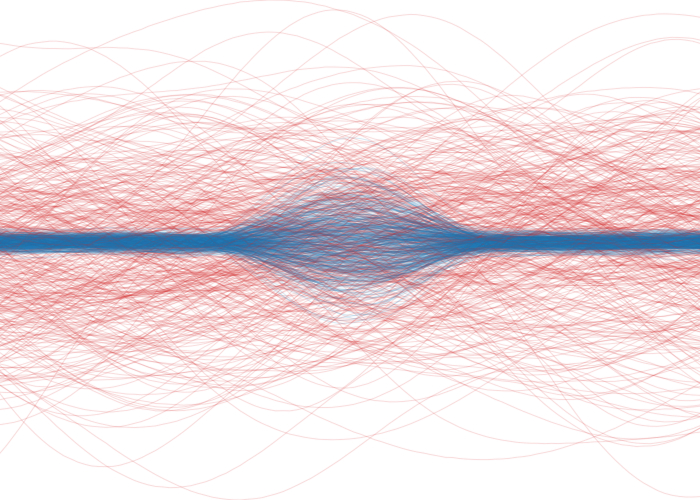}}
        \caption{{D3-C}}
        \label{fig:D3_raw_conn}
    \end{subfigure}
    \hfill
    \begin{subfigure}[b]{0.22\linewidth}
        \centering
        \fbox{\includegraphics[width=\linewidth]{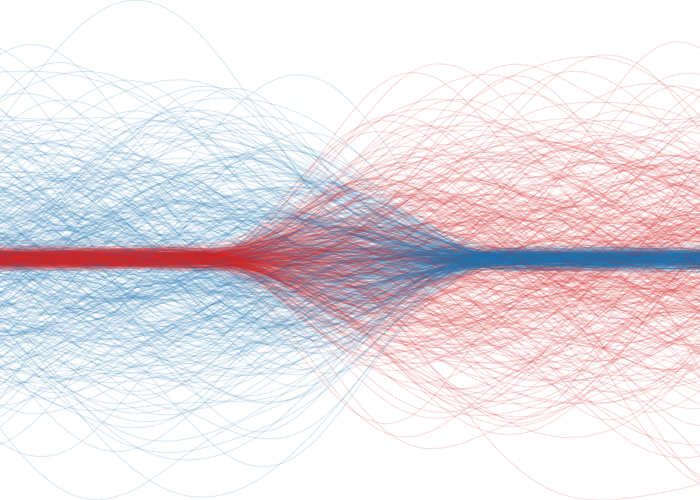}}
        \caption{{D3-D}}
        \label{fig:D3_raw_disconn}
    \end{subfigure}
    \vspace{-2mm}
    \caption{\textbf{Synthetic datasets.} The colors show the generated trajectory categories for cases D1, D2, D3-C, and D3-D in \cref{fig:synthetic_gallery}.}
    \label{fig:synthetic_raw}
    \vspace{-4mm}
\end{figure}

\begin{figure}[!t]
    \centering
    \setlength{\tabcolsep}{2pt}
    \renewcommand{\arraystretch}{1.0}
    \newcommand{\syntheticcell}[1]{\fbox{\includegraphics[width=0.27\linewidth]{#1}}}
    \newcommand{\syntheticlegend}[1]{%
        \makebox[0.27\linewidth][c]{%
            \raisebox{-0.55ex}{\rotatebox{90}{\tiny\textsf{High}}}%
            \hspace{0.4mm}%
            \raisebox{-0.45ex}{\rotatebox{90}{\includegraphics[height=0.26\linewidth,width=0.033\linewidth,trim=12bp 82bp 38bp 82bp,clip]{#1}}}%
            \hspace{0.4mm}%
            \raisebox{-0.55ex}{\rotatebox{90}{\tiny\textsf{Low}}}%
        }%
    }
    \begin{tabular*}{\linewidth}{@{\extracolsep{\fill}}lccc@{}}
        & \begin{tabular}{c}\small\textsf{Scalar Density}\\[-0.5mm]\syntheticlegend{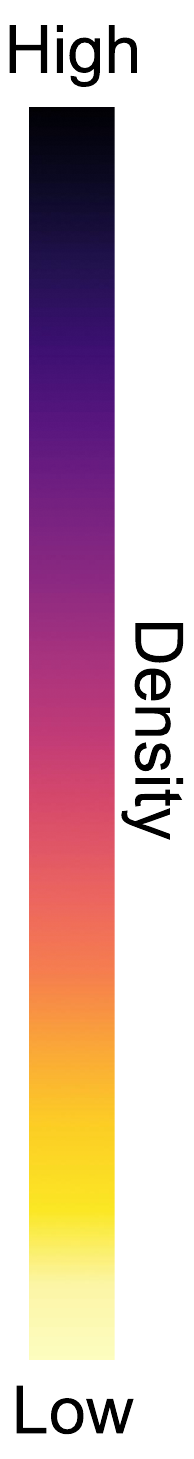}\end{tabular}
        & \begin{tabular}{c}\small\textsf{Structural Inconsistency}\\[-0.5mm]\syntheticlegend{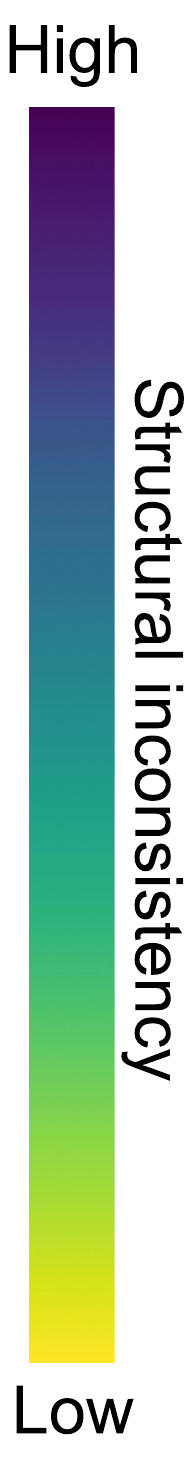}\end{tabular}
        & \begin{tabular}{c}\small\textsf{Cluster Coloring}\\[-0.5mm]\scriptsize\textsf{categorical groups}\end{tabular} \\
        \small\textsf{D1} &
        \syntheticcell{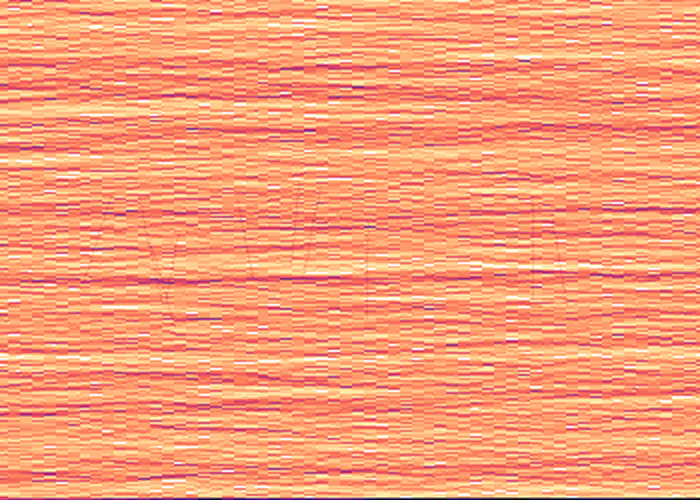} &
        \syntheticcell{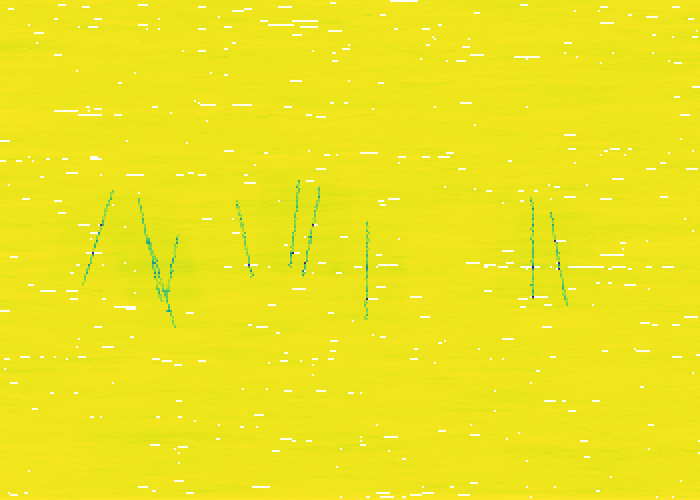} &
        \syntheticcell{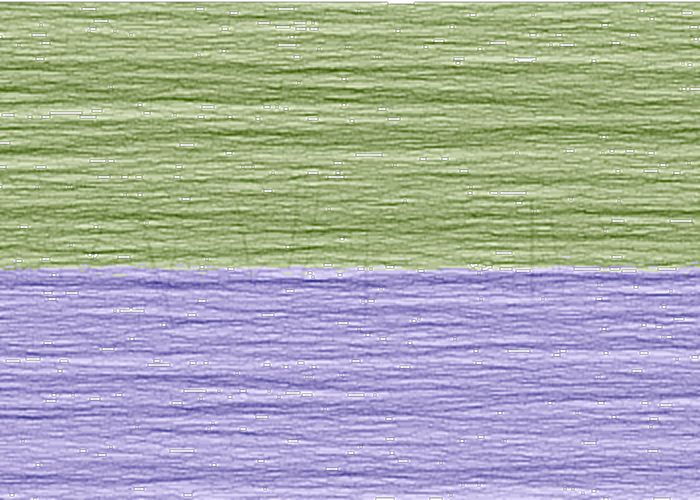} \\[1mm]
        \small\textsf{D2} &
        \syntheticcell{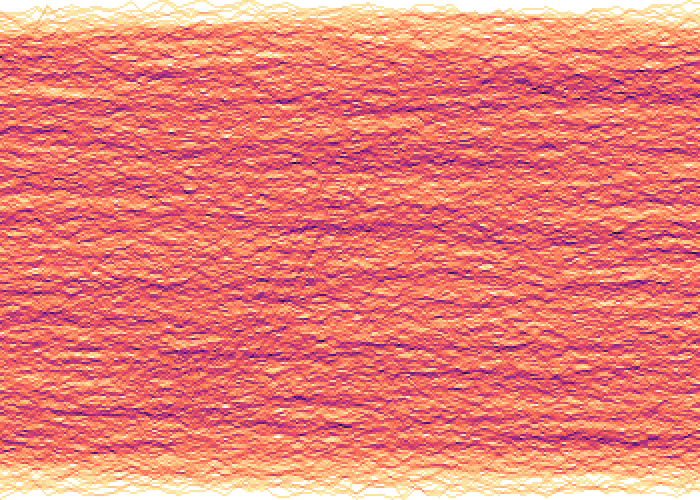} &
        \syntheticcell{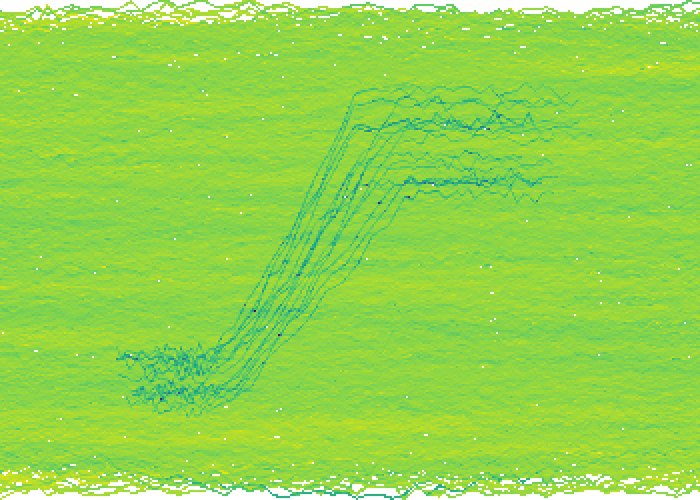} &
        \syntheticcell{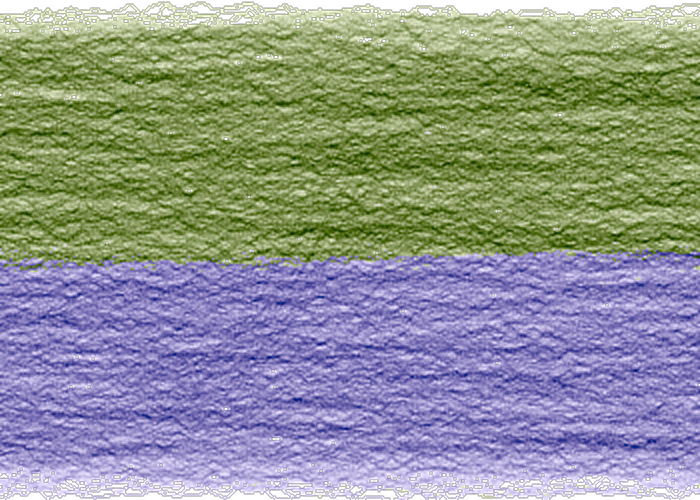} \\[1mm]
        \small\textsf{{D3-C}} &
        \syntheticcell{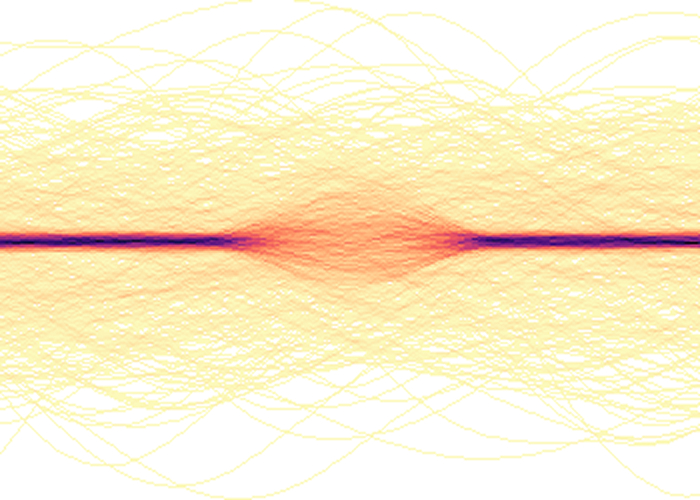} &
        \syntheticcell{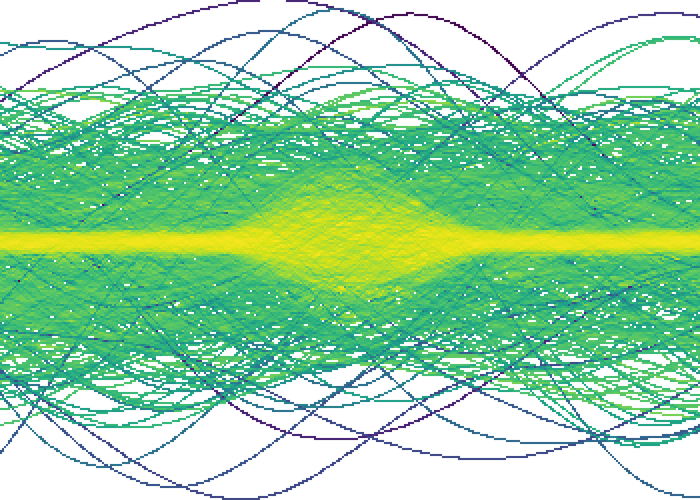} &
        \syntheticcell{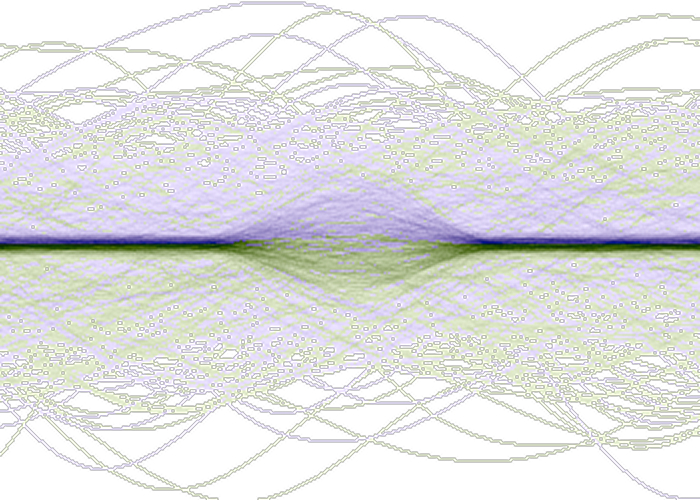} \\[1mm]
        \small\textsf{{D3-D}} &
        \syntheticcell{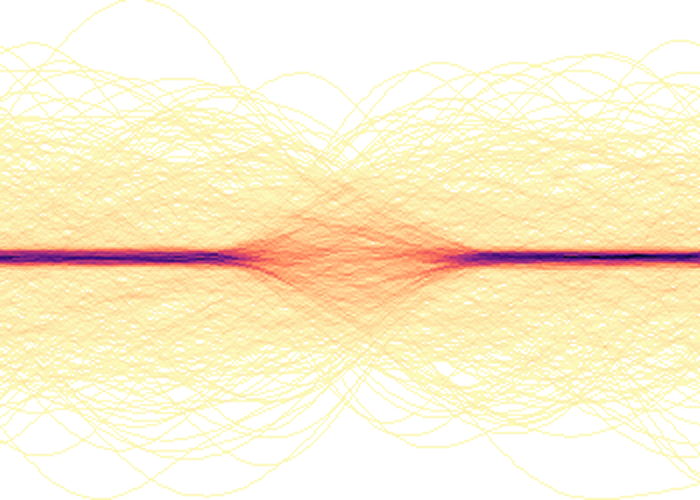} &
        \syntheticcell{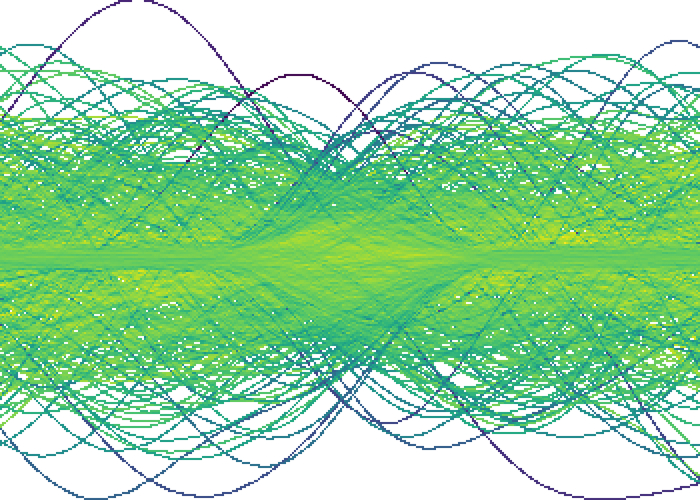} &
        \syntheticcell{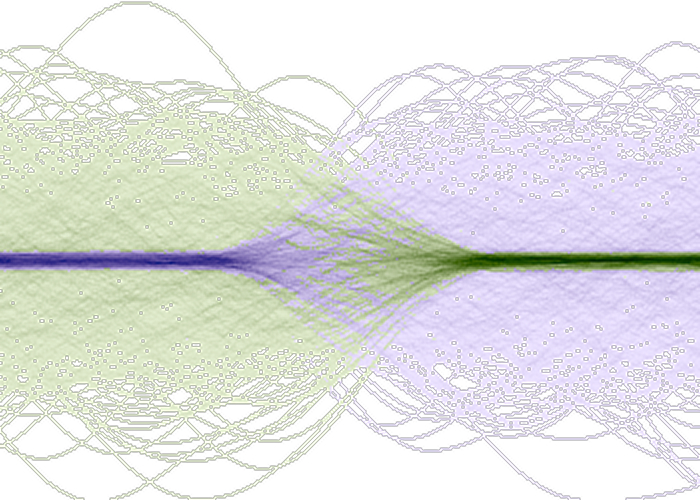} \\
    \end{tabular*}
    \caption{\textbf{Benchmark for synthetic datasets.} The columns show scalar density, our structural inconsistency field, and the qualitative image-space cluster coloring by Xue \textit{et al.}~\cite{xue2024reducing}. Density and inconsistency are continuous scalar fields, whereas the cluster-coloring column uses categorical colors for image-space groups rather than ground-truth trajectory classes.}
    \label{fig:synthetic_gallery}
    \vspace{-7mm}
\end{figure}

We constructed our benchmark datasets D1–D3 according to the rules below. For numerical evaluation, results are averaged over 10 datasets each, while the figures present a single representative example. Our three controlled synthetic datasets are shown in \cref{fig:synthetic_raw} and compared in \cref{fig:synthetic_gallery}. D1 (\cref{fig:D1_raw}) consists of 1000 horizontal lines in blue and additionally 10 slanted outliers in red, which are nearly perpendicular to the inlier direction. We use it to test how subtle hidden orientation conflicts become evident. D2 (\cref{fig:D2_raw}) consists of 1200 horizontal, aligned inliers (blue) and 15 outliers (red). While their beginnings and ends remain locally aligned with the inliers, in their central segment, they bend and thus ``change lanes.''
D3 is an ensemble-level connectivity benchmark adapted from the ambiguity setup of Xue \textit{et al.}~\cite{xue2024reducing} to highlight the visual ambiguity of line-based density plots: as shown in \cref{fig:D3_raw_conn}, in the connected condition D3-C, 400 trajectories converge at both endpoints and diverge through the middle (blue) and are intermingled with 400 background trajectories (red).
In contrast, in the disconnected condition D3-D, two separate groups of 400 trajectories are bundled at opposite sides and diverge across the middle. Each group fans out with no connecting bridge.
Both D3 conditions contain 800 trajectories. 

The first column of \cref{fig:synthetic_gallery} shows the trajectory-based density plot, the second our structural inconsistency field, and the third the image-space cluster coloring by Xue \textit{et al.}~\cite{xue2024reducing}. This baseline hierarchically clusters occupied image bins according to the similarity of the trajectory sets passing through them and assigns a categorical hue to each resulting group. The number of clusters is user-specified. We set it to $k=2$ for all four benchmarks because each dataset was generated from two ground-truth trajectory groups. Thus, the two colors reflect a controlled parameter setting rather than an automatically inferred cluster count or a trajectory-level classification. For D1 and D2, a dominant horizontal pattern obscures subtle trajectory-level outliers in density. The two D3 conditions also remain visually similar in density, despite their different connectivity.
Image-space coloring primarily separates broad rendered regions, whereas our field exposes different structural evidence: it separates outliers from inliers in D1 and D2, and makes \mbox{D3-C} and \mbox{D3-D} visually distinguishable by indicating whether the dense bridge has sustained path support. \mbox{D3-D} also marks a boundary of the method: SIF diagnoses whether a dense bridge is structurally supported, but it is not intended as a standalone categorical clustering method.

To underpin this visual evaluation, we conducted a numerical evaluation using a trajectory-level separation task.
Since the method of Xue \textit{et al.}~\cite{xue2024reducing} partitions rendered image patterns rather than assigning one fidelity score per trajectory, we restrict the quantitative evaluation to methods that natively yield one score per line.
As only D1 and D2 separate inliers from outliers, we only include these datasets in our evaluation. For each trajectory, we average the sampled density and structural inconsistency field values along its rasterized path, yielding one score per line. The scores are oriented so that larger values indicate stronger outlier evidence: the inconsistency readout is used directly, while the density baseline is negated before evaluation. 

To determine whether we measure the right outliers, we compare these oriented line scores against the ground-truth outliers (positive class) using AUROC and AUPRC, two measures for binary classifiers (see \cite{fawcett2006roc}). 
AUROC measures threshold-independent ranking quality across all decision thresholds, where 0.5 corresponds to chance; AUPRC summarizes the precision--recall tradeoff and is more informative under strong class imbalance~\cite{saito2015precision}, with a random baseline equal to the positive prevalence.
For the line-level tasks, AUROC and AUPRC lie in $[0,1]$, with larger values indicating better separation.

\begin{table}[!t]
    \caption{\textbf{Quantitative evaluation for D1 and D2.} The path-extent structural inconsistency field yields substantially better task-aligned line-level separation than density in both cases.}
    \vspace{-2mm}
    \label{tab:eval_synthetic_summary}
    \centering
    \small
    \setlength{\tabcolsep}{4pt}
    \begin{tabular}{@{}llcccc@{}}
        \toprule
        \textbf{Task} & \textbf{Sample Unit} & \multicolumn{2}{c}{\textbf{Density}} & \multicolumn{2}{c}{\textbf{Inconsistency}} \\
        \cmidrule(lr){3-4}\cmidrule(l){5-6}
         &  & \textbf{AUROC} & \textbf{AUPRC} & \textbf{AUROC} & \textbf{AUPRC} \\
        \midrule
        D1 Hidden Cut & line & {0.3902} & {0.0089} & \textbf{{0.9908}} & \textbf{{0.3679}} \\
        D2 Lane Change & line & {0.2741} & {0.0085} & \textbf{{0.9975}} & \textbf{{0.6889}} \\
        \bottomrule
    \end{tabular}
    \vspace{-2mm}
\end{table}
D1 and D2 evaluate line-level separation, whereas D3 evaluates ensemble-level ambiguity resolution.
We therefore report numerical line-level comparisons only for D1 and D2.
Because both cases are strongly class-imbalanced, AUPRC provides an especially informative complement to AUROC. Our method strongly outperforms the density baseline on both line-level ambiguity tasks, reaching much higher AUROC and AUPRC values (\cref{tab:eval_synthetic_summary}).
In the reported discrete implementation, we set $\ell=1$ for D1 and $\ell=150$ for D2.

\subsection{Performance and Scalability}
\label{sec:eval_performance}

\begin{figure}[tb]
    \centering
    \includegraphics[width=0.85\linewidth]{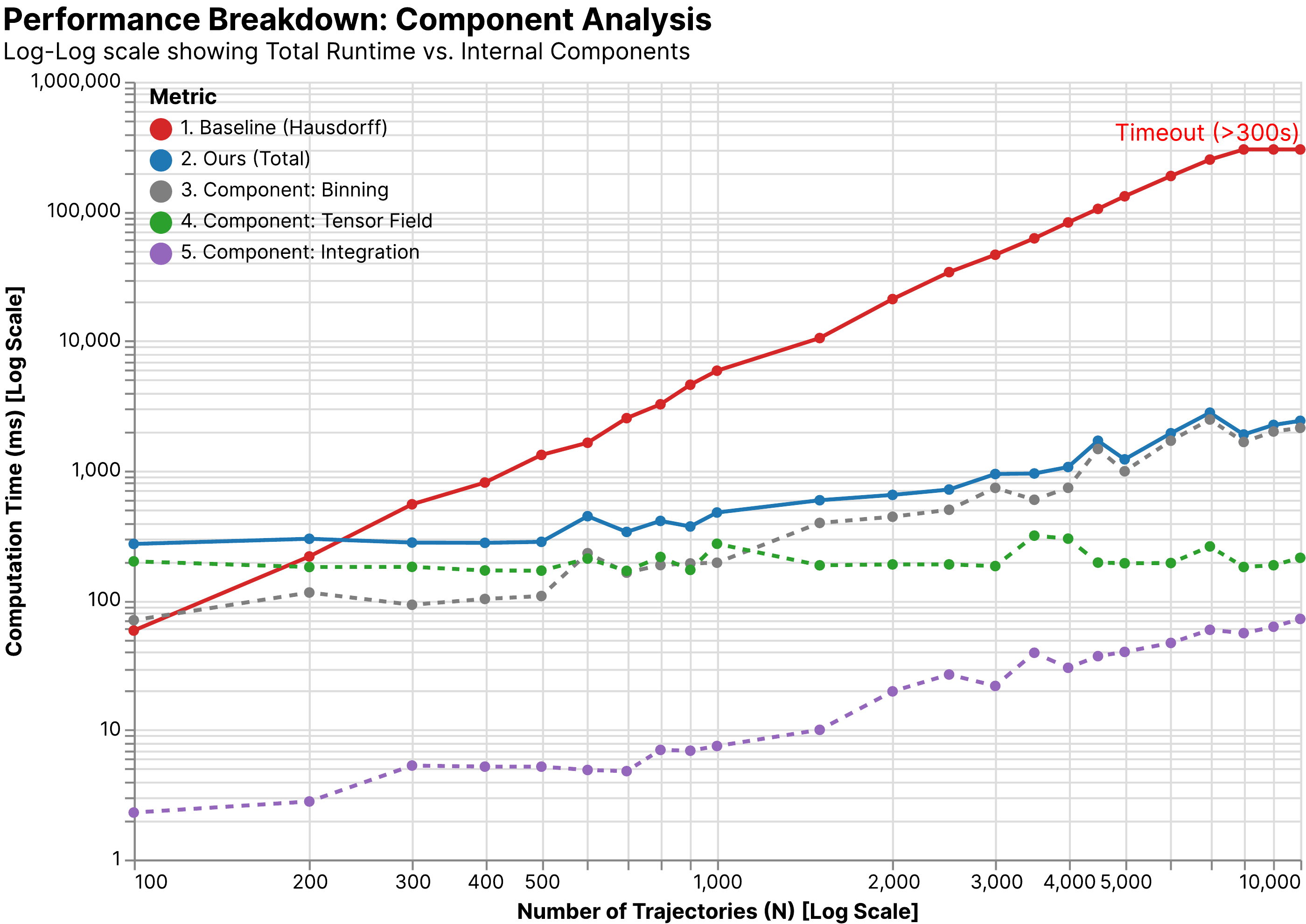}
    \vspace{-2mm}
    \caption{\textbf{Scalability analysis.} Binning scales linearly with the number of trajectories, tensor-field construction is nearly constant with respect to $N$ on a fixed grid, and prefix-sum path-extent evaluation keeps path integration lightweight.}
    \label{fig:eval_performance}
    \vspace{-6mm}
\end{figure}

\begin{figure*}[t]
    \centering
    \begin{subfigure}{0.20\textwidth}
        \centering
        \includegraphics[width=\linewidth]{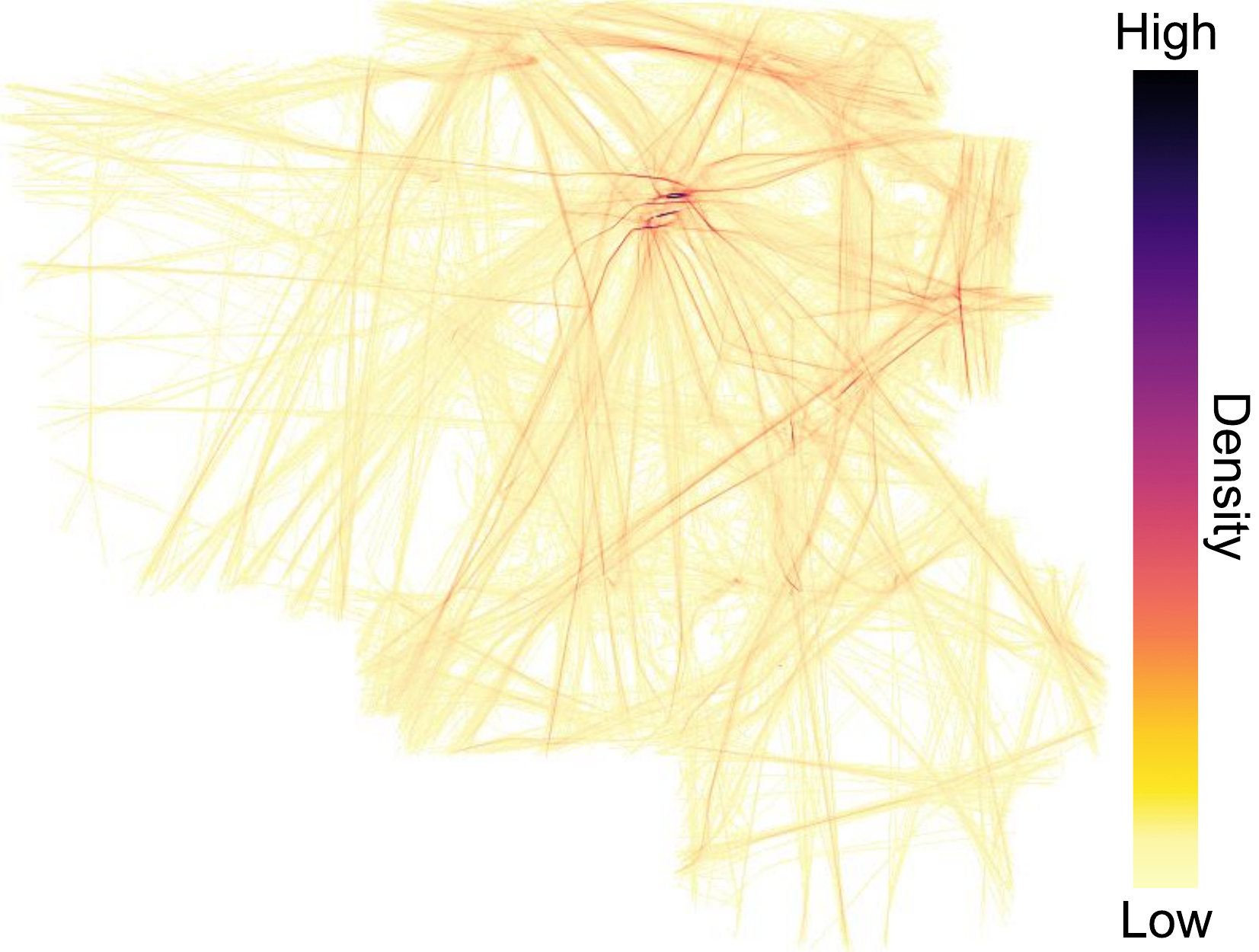}
        \caption{Density}
        \label{fig:france_density}
    \end{subfigure}
    \hfill
    \begin{subfigure}{0.20\textwidth}
        \centering
        \includegraphics[width=\linewidth]{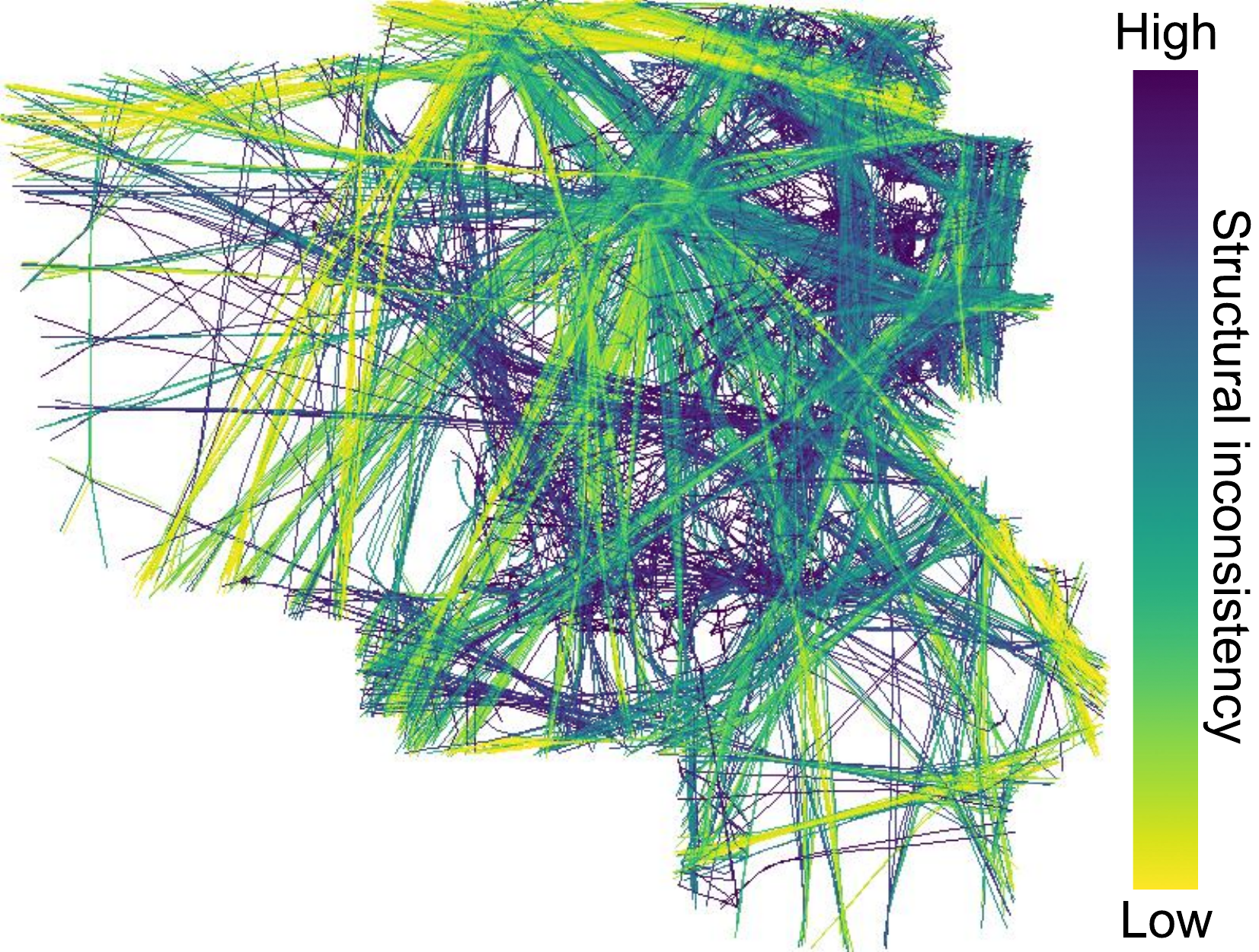}
        \caption{Structural Inconsistency Field}
        \label{fig:france_friction}
    \end{subfigure}
    \hfill
    \begin{subfigure}{0.20\textwidth}
        \centering
        \includegraphics[width=\linewidth]{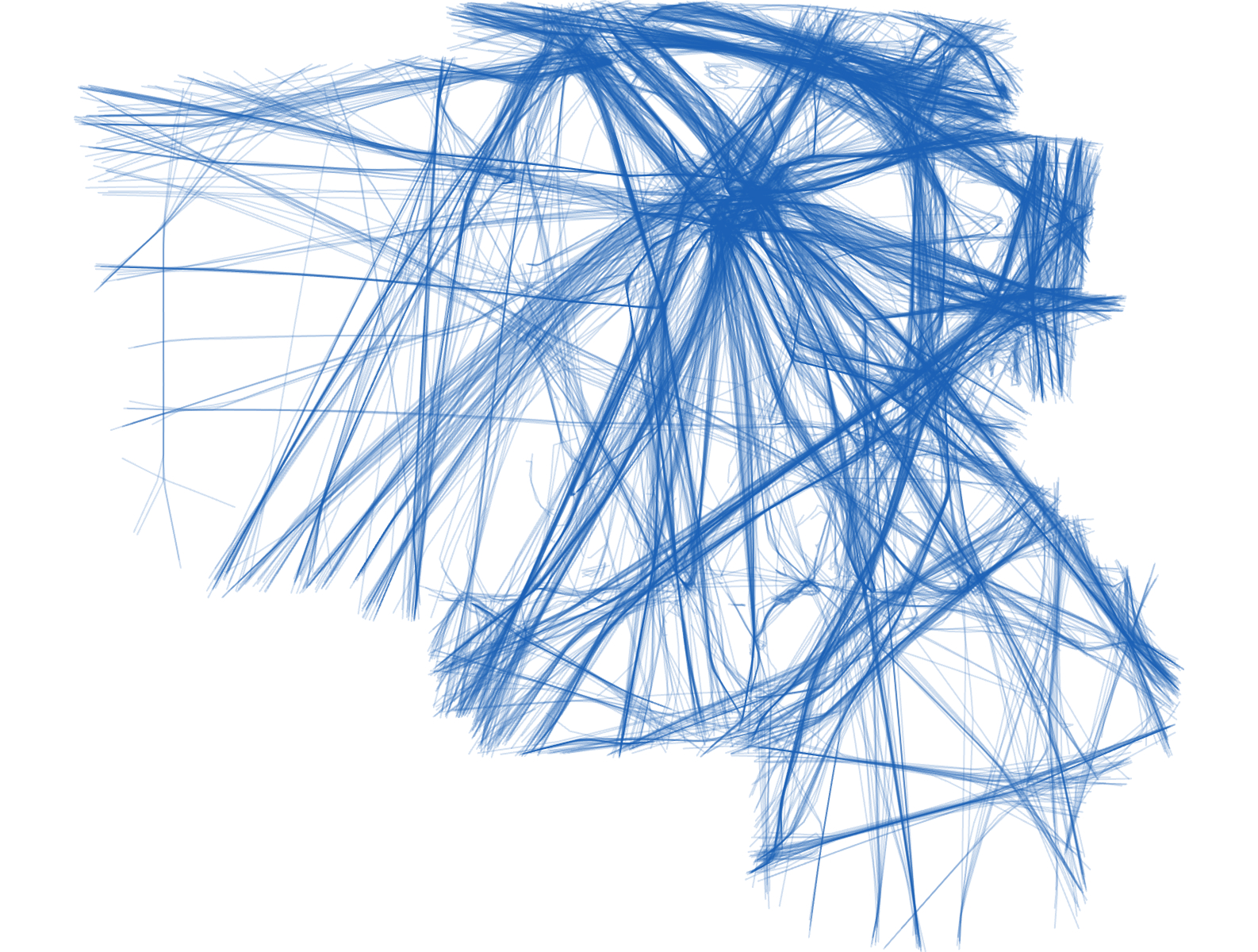}
        \caption{Top 80\% high-score flights}
        \label{fig:france_result_inliers}
    \end{subfigure}
    \hfill
    \begin{subfigure}{0.20\textwidth}
        \centering
        \includegraphics[width=\linewidth]{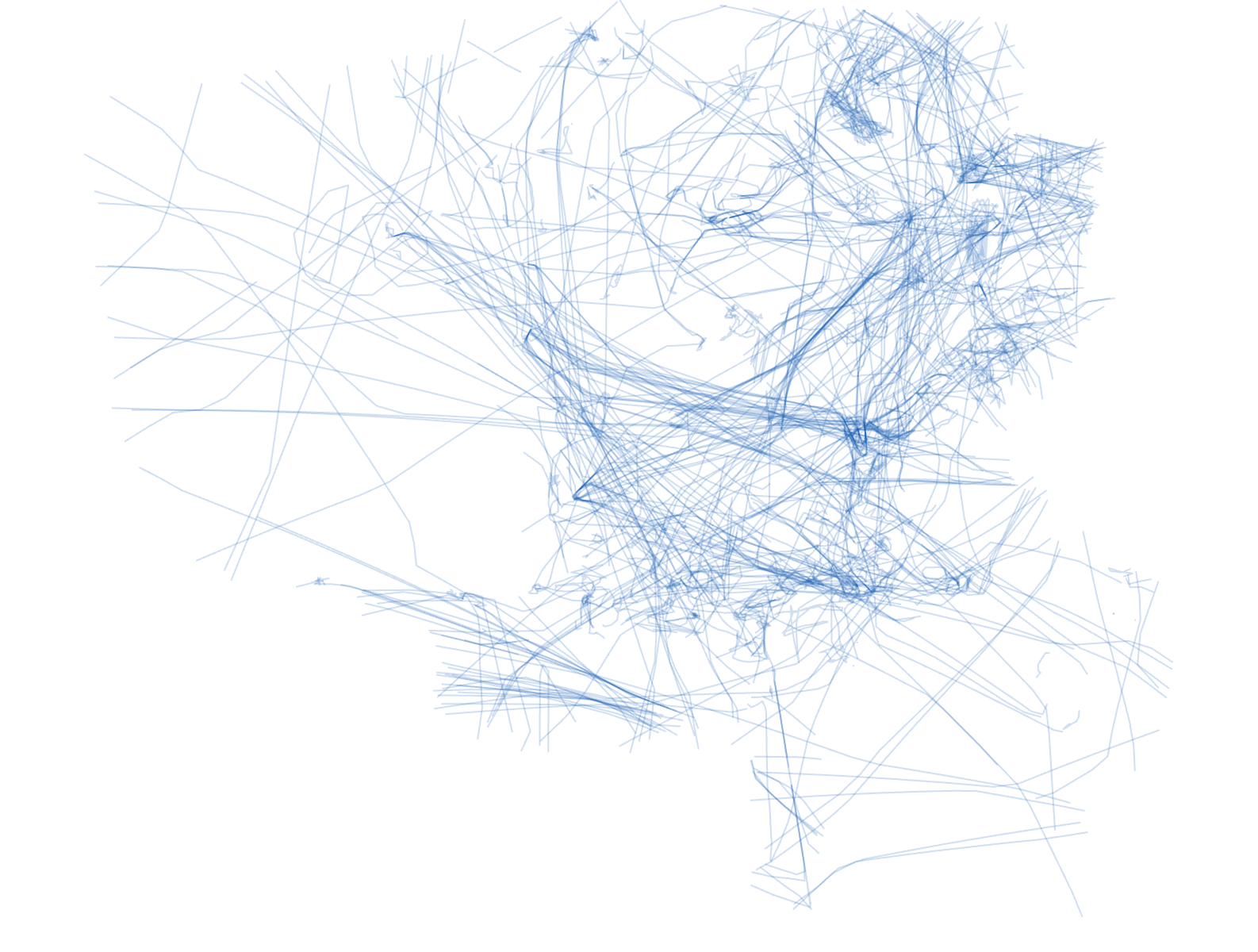}
        \caption{Bottom 20\% low-score flights}
        \label{fig:france_result}
    \end{subfigure}
    \vspace{-2mm}

    \caption{\textbf{France aviation case study.} The full dataset contains 5,360 trajectories. The density map emphasizes the dominant traffic corridors but leaves minority behaviors visually entangled with them. The structural inconsistency field localizes where structural support weakens. Ranking trajectories by $\Phi(L)$ then separates the dataset into a top 80\% high-score subset (4,288 trajectories), which captures the main long-range, route-regular commercial traffic pattern, and a bottom 20\% low-score subset (1,072 trajectories), which analysts informally interpreted as including many smaller-aircraft trajectories, such as private light aircraft, together with looping or holding-like paths and trajectories affected by recording errors.}
    \label{fig:eval_france_case}
    \vspace{-3mm}
\end{figure*}

\begin{figure*}[t]
    \centering
    \begin{subfigure}{0.20\textwidth}
        \centering
        \includegraphics[width=\linewidth]{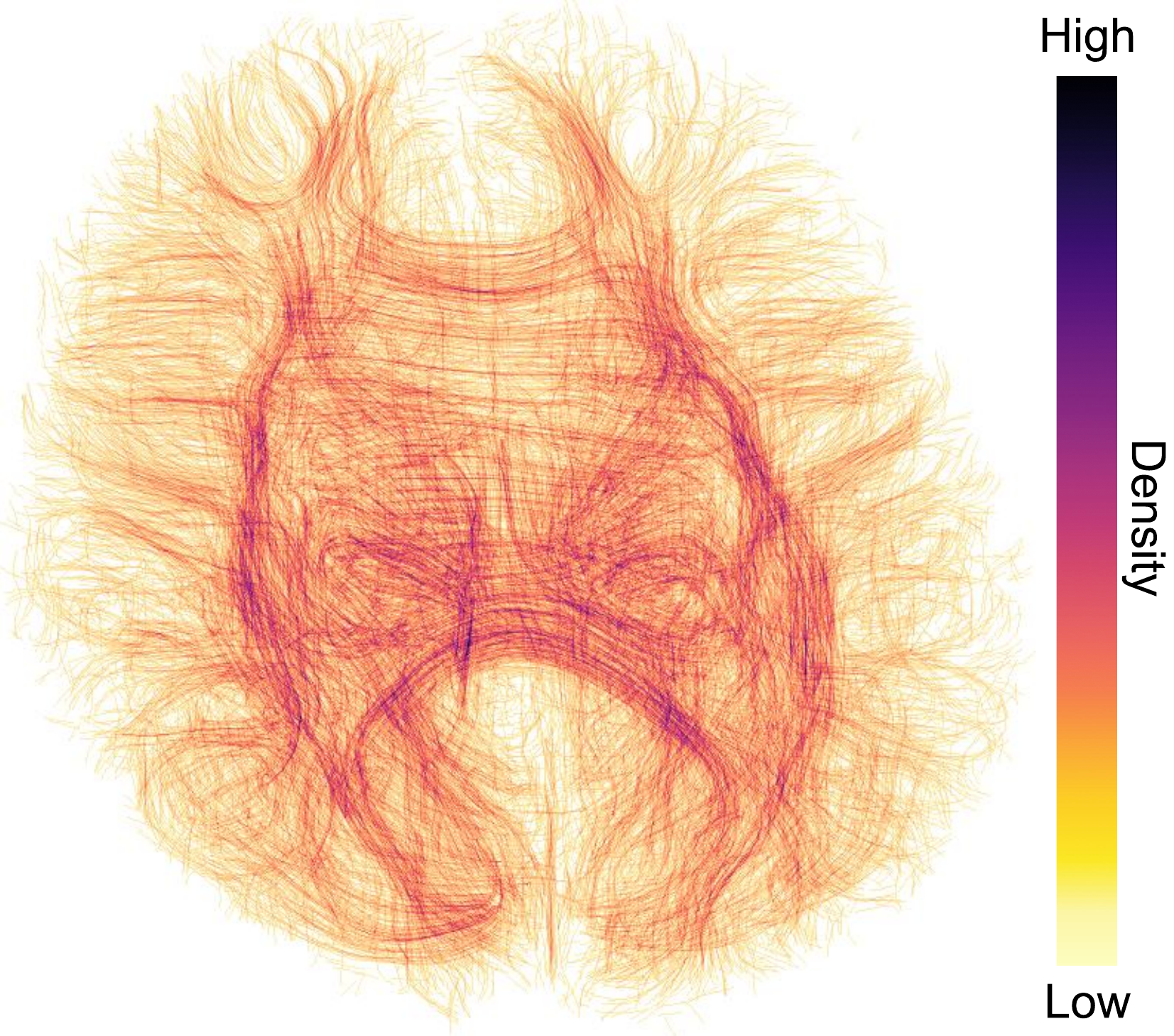}
        \caption{Density}
        \label{fig:case_raw}
    \end{subfigure}
    \hfill
    \begin{subfigure}{0.20\textwidth}
        \centering
        \includegraphics[width=\linewidth]{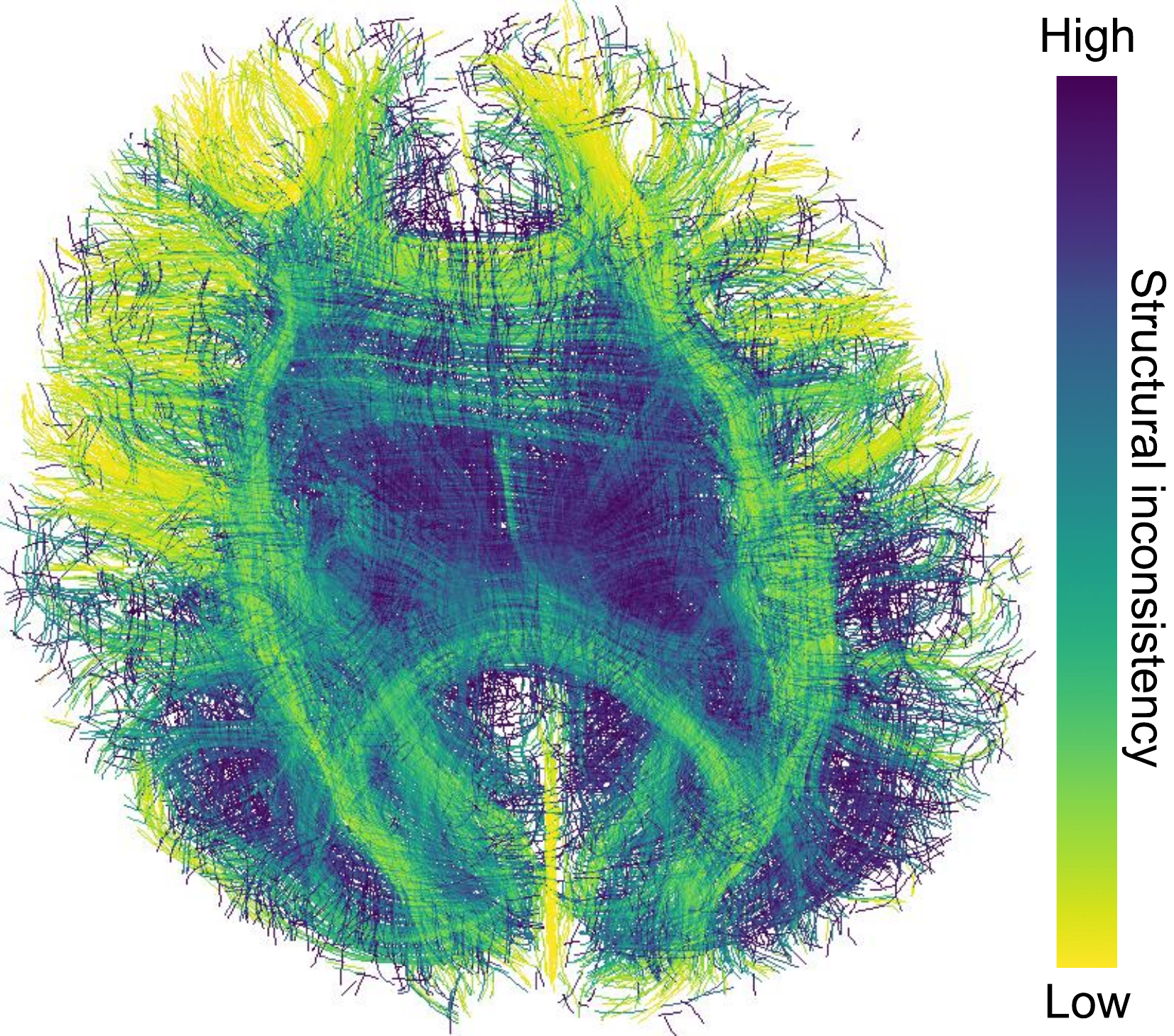}
        \caption{Structural Inconsistency Field}
        \label{fig:case_friction}
    \end{subfigure}
    \hfill
    \begin{subfigure}{0.20\textwidth}
        \centering
        \includegraphics[width=\linewidth]{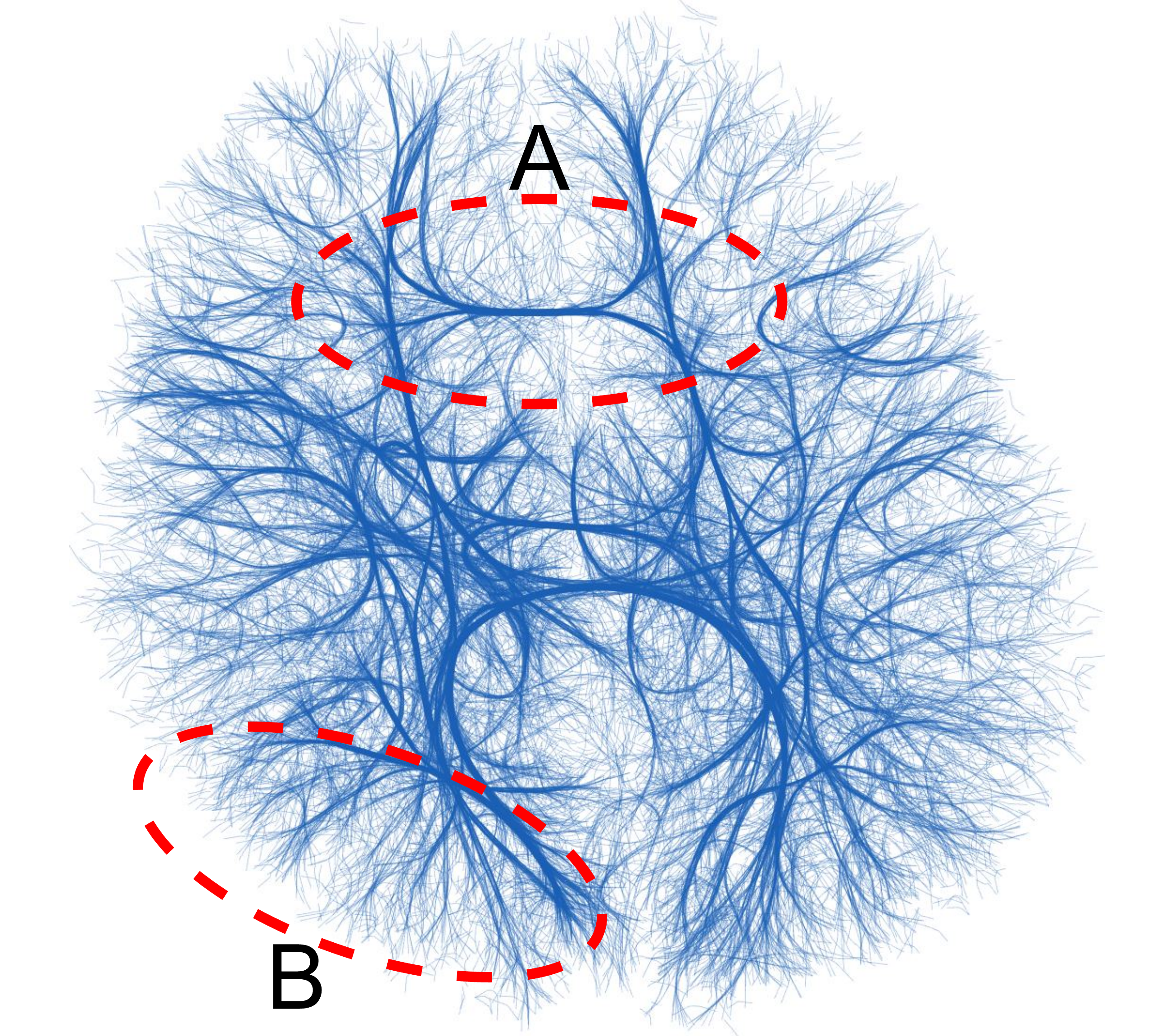}
        \caption{Line Bundling Reference}
        \label{fig:case_bundling}
    \end{subfigure}
    \hfill
    \begin{subfigure}{0.20\textwidth}
        \centering
        \includegraphics[width=\linewidth]{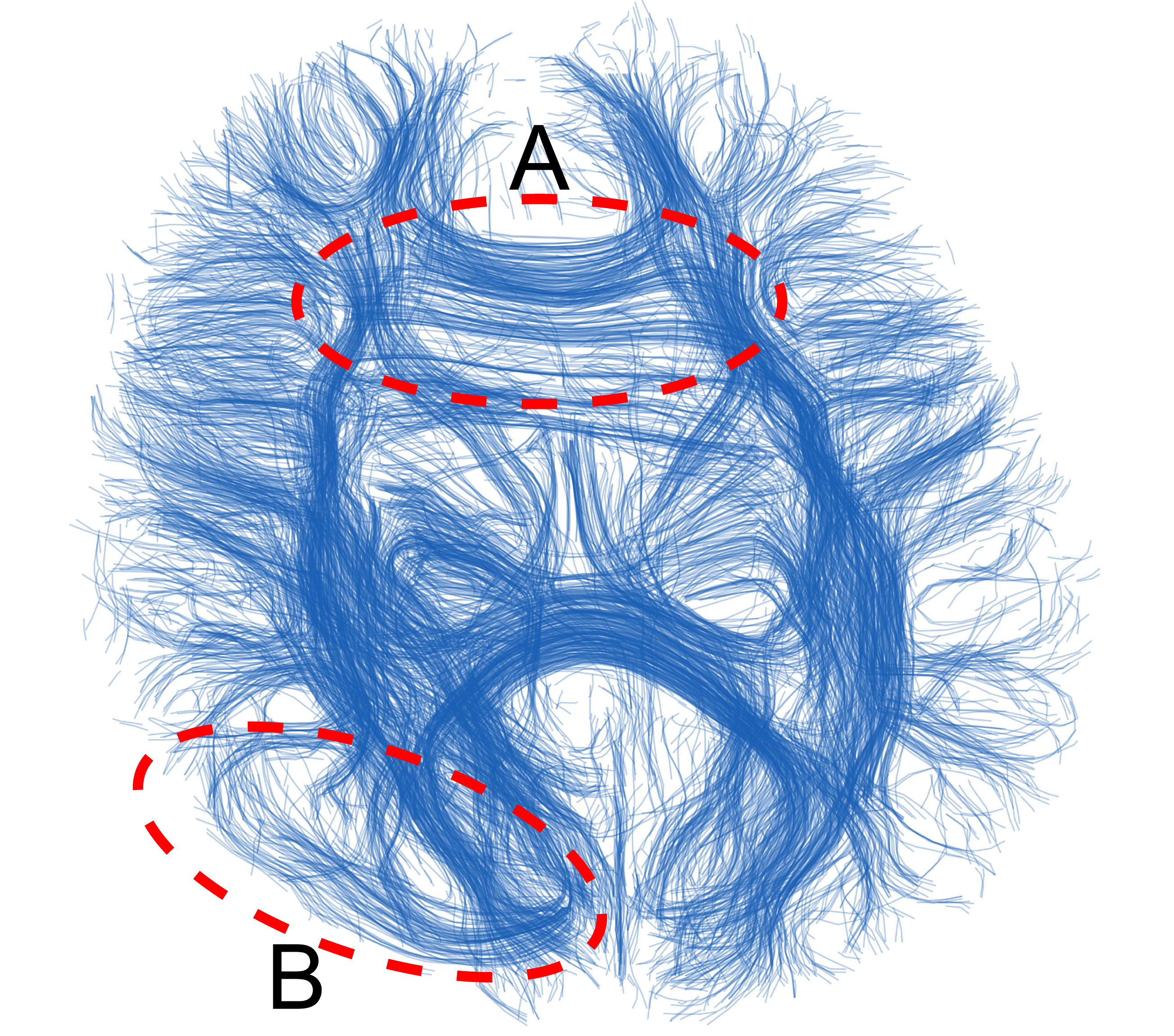}
        \caption{Top 50\% high-score fibers}
        \label{fig:case_result}
    \end{subfigure}
    \vspace{-2mm}
    \caption{\textbf{DTI case study.} This orthographic brain projection uses a 12,059-fiber subset sampled from the 120,593-fiber tractography dataset used in FiberClay~\cite{hurter19fiberclay}. Density (a) collapses overlapping anatomical structures into a hairball, while the structural inconsistency field (b) reveals weakly supported passages. Functional-decomposition bundling~\cite{hurter2018functional} produces a clean tree-like representation (c), but regions A and B show distorted or suppressed bundle geometry. In contrast, the top 50\% of fibers ranked by trajectory fidelity (d) recover a readable scaffold without deformation.}
    \label{fig:eval_dti_case}
    \vspace{-4mm}
\end{figure*}

\cref{fig:eval_performance} reports timings measured on a single randomly sampled subset of the full 120,593-fiber DTI dataset for each tested value of $N$. Measurements use our JavaScript implementation in Google Chrome 144 on an Apple Silicon M1 with 16GB RAM and a fixed $1000 \times 500$ grid of $1 \times 1$ pixel bins. Binning scales linearly and dominates the cost; fixed-grid Gaussian smoothing keeps tensor construction near 200ms, while prefix sums make each centered path-extent query $O(1)$ after preprocessing. Memory comprises $O(W \cdot H)$ fixed-grid buffers plus rasterized passages, with no pairwise $O(N^2)$ term or additional buffers when the path extent changes. Because each plotted timing value is a single run, minor fluctuations should be interpreted as JavaScript runtime variability, such as JIT optimization and garbage collection, rather than as estimates of statistical uncertainty. The full pipeline reaches about 2.4 seconds at $N=10{,}000$ on a single thread, whereas the quadratic Hausdorff reference reaches a computational wall beyond $N \approx 8{,}000$. This supports practical browser-side preprocessing followed by cached interactive queries.

\subsection{Real-World Case Studies}
\label{sec:eval_corroboration}

Real-world case studies complement our three synthetic datasets by focusing on distinct aspects. 
To demonstrate the versatility of our method, we use real-world datasets from diverse domains, including time series (\emph{ACIS}, \cref{fig:teaser}), geolocated trajectories (\emph{France aviation}, \cref{fig:eval_france_case}), and medical data (\emph{DTI}, \cref{fig:eval_dti_case}). In contrast to the case in \cref{fig:teaser}, where we examined local windows, in \cref{fig:eval_france_case} and \cref{fig:eval_dti_case} we do a global peeling based on trajectory-level fidelity scores.
These figures show the line-based density plot, the corresponding inconsistency field where structural support is sustained (or breaks down), and the extracted subset that allows inspection once the metric guides filtering.

\noindent\textbf{ACIS Data.}
\emph{ACIS Temperature Data}~\cite{acis} is a line-based time-series ensemble with 293,175 weekly maximum temperature values from 6,187 U.S. weather stations. Its pronounced seasonal envelope provides a controlled setting for comparing query specificity: although many stations follow the same annual pattern, density-guided querying can still return broad and structurally mixed subsets. In \cref{fig:teaser}, the density hotspot A returns 3,776 trajectories and a diffuse geographic footprint, whereas the SIF makes the low-inconsistency corridors B1 and B2 salient; querying them extracts 1,183 and 593 trajectories with a clearer north--south climate separation. This contrast is important because the dense annual envelope is visually plausible but analytically nonspecific: it reflects the dominant seasonal trend rather than a coherent subgroup. The structure-guided windows instead target portions of the plot where similar local temperatures are sustained by compatible path context, which makes the selected station sets more geographically interpretable. Thus, the case illustrates how structure-guided querying can be more selective than density-guided querying even when the dominant seasonal structure is already visually apparent.

\noindent\textbf{France Aviation Data.}
The \emph{France aviation} dataset is a 2D trajectory dataset of 5,360 flights in a geo-location context, based on the \emph{Anonymized Aircraft trajectories} dataset released by Hurter~\cite{hurter2023anonymizedaircraft} and previously used in \emph{FromDaDy}~\cite{hurter5290707}. Most flights in this dataset follow relatively regular, standardized routes, so the task is both to recover the dominant traffic pattern and to examine the trajectories that deviate from it. In the raw density view, overlapping long-range connections and hub-centered crossings make both aspects difficult to distinguish from the surrounding traffic volume. We therefore rank trajectories by the trajectory-fidelity score $\Phi(L)$ and inspect both the highest-scoring 80\% and the lowest-scoring 20\%. The high-score subset in \cref{fig:france_result_inliers} contains 4,288 flights and captures the main long-range, route-regular commercial traffic pattern. The low-score subset in \cref{fig:france_result} contains 1,072 flights and aligns spatially with the high-inconsistency regions in \cref{fig:france_friction}. In the density plot in \cref{fig:france_density}, the same regions are much harder to separate from adjacent low-inconsistency regions.
In informal discussions with civil aviation analysts, this low-score subset was interpreted as containing many smaller-aircraft trajectories, including private light aircraft, together with looping or holding-like patterns and sparse spurious routes caused by erroneous instrument recordings.
This second case, therefore, extends the argument from subgroup discovery to expert-guided anomaly inspection in 2D traffic data, showing that the combination of $\Phi(L)$ and the SIF can direct attention not only to coherent structure, but also to structurally weak or abnormal lines that density leaves visually entangled with the dominant flow.

\noindent\textbf{View-Dependent Exploration of DTI Fibers.}
The third case adds representational complexity through 3D brain fibers analyzed in an orthographic 2D screen-space projection, in a tractography visualization context similar to that of Everts~\textit{et al.}~\cite{everts2015whitematter}. Our full dataset contains 120,593 fibers; to keep the example interactive, we visualize a sampled subset of 12,059 fibers. This case evaluates whether screen-space structural scoring can recover useful projected organization while preserving the original fiber geometry. Accordingly, all interpretations are view-dependent: projection-induced overlaps can create ambiguity that requires 3D-aware analysis for anatomical validation. In \cref{fig:eval_dti_case}, density collapses the C-shaped \textit{Cingulum Bundle}, radiating callosal fibers, and peripheral crossings into a hairball. The SIF separates coherent projected corridors from diffuse or weakly supported passages, and selecting the highest-scoring 50\% yields a readable scaffold without deforming trajectories. The circled regions in \cref{fig:case_bundling,fig:case_result} make this distinction concrete. In region A, retained fibers preserve a parallel projected bundle with visible width, whereas bundling collapses this geometry into an unnaturally tight track. In region B, the retained subset reveals a slanted bundle of substantial length and thickness at roughly $45^\circ$, while bundling largely suppresses this structure. Together, these observations show that the retained geometry preserves structures that line bundling can collapse or suppress; the complementary low-score half is shown in the appendix (\cref{fig:app_dti_outlier}).

\section{Discussion}
\label{sec:discussion}

Visualizing massive line ensembles entails an inherent trade-off between occlusion from overplotting in trajectory-centric representations and semantic loss from density-based aggregation.
Our work introduces a complementary structural view for density-based visualizations in which second-order orientation statistics mediate between field-based aggregation and trajectory-centric evaluation. By combining a trajectory-fidelity score with structural inconsistency, we move beyond merely displaying where trajectories accumulate to characterizing how consistently they align and where structural support breaks down.

\noindent\textbf{From Occupancy to Structural Agreement.}
The central idea of our work is a shift from occupancy-based analysis to agreement-based analysis. Traditional density maps answer the question \textit{``How many trajectories pass here?''} Our SIF and path-integrated fidelity score address a different question: \textit{``How consistently do the trajectories align with a dominant local orientation pattern?''} This shift helps to resolve the semantic blindness of scalar fields. High-density regions are no longer treated as monolithic; they can instead be decomposed into regions of sustained agreement and regions of structural conflict. As the D3 benchmark makes explicit, the limitation of density is not simply that it contains no global cue, but that two ensembles that are structured differently can still appear visually similar under scalar accumulation. The SIF contributes a different form of evidence: It localizes where structural support is maintained and where it collapses.

\noindent\textbf{Trajectory-Aligned Aggregation.}
The use of SIF can be interpreted as a trajectory-aligned aggregation mechanism. Instead of assigning each pixel a fixed Euclidean neighborhood, it aggregates evidence along the subset of trajectories that traverse that pixel. This expands the effective spatial support of the measurement from a local bin neighborhood to a path-centered neighborhood while preserving the original coordinates. In practice, this is the key step that enables our method to separate locally similar yet globally distinct structures.

\noindent\textbf{Outlier and Cluster Exploration.}
The method provides structural evidence for ensemble exploration but does not assign categorical cluster labels. In practice, density locates occupied regions, SIF distinguishes sustained support from structural conflict, and the trajectory-fidelity score $\Phi(L)$ ranks or peels trajectories within a selected region. When categorical assignment is required, image-space grouping or trajectory clustering can provide an initial partition, while SIF indicates where that partition is structurally supported or where its support breaks down. A practical workflow is therefore to use density to define a region of interest, inspect its SIF, rank trajectories by $\Phi(L)$, and compare the resulting high- and low-support subsets with an explicit clustering. The absence of a cluster-shaped region from the SIF, as in D3-D, does not imply that the corresponding trajectories are absent from the ensemble. It rather indicates that the region lacks sustained structural support under the selected path context.

\noindent\textbf{Self-Bias and Counterfactual Verification.}
A subtle yet critical contribution is our treatment of the \textit{Self-Fulfilling Prophecy} via Dynamic Leave-One-Out (LOO) filtering. In dense data analysis, the observer (the candidate trajectory) may distort the observed environment (the local tensor field). A static metric can therefore overestimate the consistency of a strong outlier because the outlier partially defines the field against which it is evaluated. Our LOO formulation removes the candidate trajectory's contribution from every bin within the kernel footprint of its rasterized samples and evaluates the centered path extents by prefix sums. This turns the interaction into a counterfactual test: \textit{``How well would this trajectory align if its own contribution were removed from the surrounding field?''} 
Our spindle ablation (\cref{appendix:loo_interpretation}) confirms that LOO matters even in a 500-line ensemble: it removes the artificial self-support of sparse outer trajectories while leaving the densely supported center nearly unchanged.

\noindent\textbf{Limitations and Boundary Conditions.}
Despite its robustness, our framework operates under several assumptions that warrant discussion. Our method is most appropriate when multiple independent trajectories corroborate local orientation support. Sparsely populated regions, especially one- or two-line neighborhoods after LOO, should be interpreted as low-confidence rather than as a fixed failure threshold. The supplementary LOO ablation (\cref{appendix:loo_interpretation}) quantifies this regime. There is therefore no universal minimum line count: stability depends jointly on kernel bandwidth, grid resolution, local passage count, path extent, and local orientation variance; reliable interpretation requires sufficient independent orientation support after LOO. In practice, SIF should therefore be inspected in conjunction with density or passage count, because high inconsistency in very sparse regions may indicate insufficient independent evidence rather than a stable structural conflict. Several additional boundary conditions follow from the same formulation. The method follows a \textit{majority-rules} logic: if a dataset is dominated by anomalous or adversarial trajectories, the structure tensor will encode that majority as the local consensus, potentially suppressing a valid minority structure. The current formulation is also \textit{orientation-based} rather than \textit{direction-based}. Because local contributions are accumulated as $\mathbf{v}\mathbf{v}^T$, antiparallel trajectories reinforce one another. This is desirable in some domains, such as bidirectional traffic corridors, but it is a limitation in settings where counterflow must be distinguished explicitly. Isotropic regions remain intrinsically ambiguous. When $\lambda_1 \approx \lambda_2$, every trajectory accumulates at most moderate ($\approx 0.5$) local agreement, which correctly marks the region as structurally complex but limits our ability to separate multiple coherent crossings within that region. The present implementation also operates on 2D projections. Extending the tensor field to 3D is mathematically direct, but the corresponding visual analysis becomes substantially more challenging because of occlusion and depth perception. For projected 3D data such as DTI fibers, the resulting scores and fields are view-dependent and intended for screen-space exploration; anatomical inference requires a 3D-aware analysis.

\noindent\textbf{Future Directions.}
The concept of passage-centered structural aggregation opens several avenues for future work, including space-time extensions for time-varying ensembles, direction-aware variants for counterflow or multi-branch flows, and clustering methods that combine geometric distance with trajectory-fidelity or support-derived distances. Another useful extension would be explicit confidence encoding for sparse regions. The current implementation masks empty pixels, but it does not separately encode confidence in one- or two-line neighborhoods after LOO; a future interface could expose the effective support count, tensor anisotropy, and path-extent coverage together, so that analysts can distinguish ``high inconsistency'' from ``insufficient evidence'' more directly. Our results suggest that perceptual disambiguation and structural scoring should be treated as complementary goals.
\section{Conclusion}
\label{sec:conclusion}
In this work, we present a visual analytics framework for structuring massive line ensembles.
We complement density-based visualization in regimes where scalar accumulation reaches its semantic limit: density preserves occupancy, but does not explain whether the observed trajectories are structurally coherent.
Visually similar density patterns can therefore reflect different connectivity or bundle support.
We addressed this limitation with a tensor-guided, path-integrated formulation that evaluates trajectories against an ensemble-derived structure field.

Our method follows a local-to-global progression. The structure tensor encodes local orientation consensus; path integration converts this local evidence into a trajectory-level fidelity measure; dynamic leave-one-out evaluation mitigates self-bias; and passage-centered aggregation maps structural support back into a structural inconsistency field, indicating where coherence is sustained or fails. This pipeline links field-based macroscopic representations with trajectory-centric microscopic analysis without compromising geometric fidelity.

We further showed that the framework is practical for visual analytics. Fixed-grid field construction and prefix-sum path-extent evaluation keep the computation interactive, while the tensor-guided peeling workflow reuses these same scores to iteratively remove low-consensus trajectories and re-evaluate the remaining field, exposing coherent skeletons buried under clutter.

We argue that the ``hairball'' problem is not merely an issue of visibility, but of interpretation. Our framework provides a way to read dense line ensembles in terms of consensus, outliers, and clutter, allowing analysts to move from seeing where trajectories accumulate to understanding why a structure should be trusted.

\acknowledgments{%
	This work was funded in part by Deutsche Forschungsgemeinschaft (DFG) Project 410883423 and Project 251654672 -- TRR 161 ``Quantitative methods for visual computing.'' Yunhai Wang was supported by the grants of NSFC (No.62132017 and No.U2436209), the Shandong Provincial Natural Science Foundation (No.ZQ2022JQ32), the Fundamental Research Funds for the Central Universities, and the Research Funds of Renmin University of China.
}

\bibliographystyle{abbrv-doi-hyperref}

\bibliography{template}

\clearpage
\appendix
\makeatletter
\twocolumn[%
\begin{@twocolumnfalse}
\begin{center}
    {\LARGE\bfseries Structuring Line Ensembles with Path-Integrated Fidelity and \\ Structural Inconsistency Fields\par}
    \vspace{0.6em}
    {\Large\bfseries Appendix\par}
\end{center}

\vspace{0.75em}
\noindent
This appendix provides supplementary background material, implementation details, parameter analyses, ablation results, workflow illustrations, and additional examples that complement the main paper.

\vspace{1.25em}
\end{@twocolumnfalse}
]
\makeatother
\section{Supplementary Background Illustration of CDE}
\label{appendix:cde}

\begin{figure}[ht]
    \centering
    \begin{subfigure}[b]{0.32\linewidth}
        \centering
        \includegraphics[width=\linewidth]{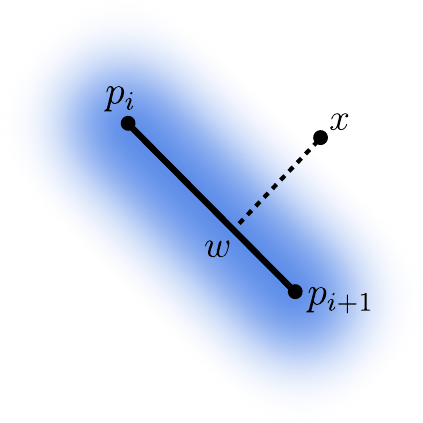}
        \caption{Continuous CDE}
        \label{fig:app_cde_continuous}
    \end{subfigure}
    \hfill
    \begin{subfigure}[b]{0.32\linewidth}
        \centering
        \includegraphics[width=\linewidth]{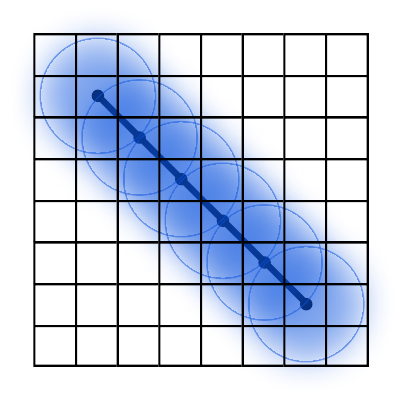}
        \caption{Rasterization Process}
        \label{fig:app_cde_discrete_1}
    \end{subfigure}
    \hfill
    \begin{subfigure}[b]{0.32\linewidth}
        \centering
        \includegraphics[width=\linewidth]{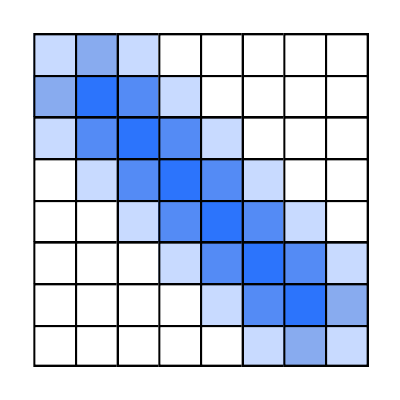}
        \caption{Scalar Field Output}
        \label{fig:app_cde_discrete_2}
    \end{subfigure}
\caption{\textbf{Supplementary illustration of the zero-order moment (CDE).} (a) Continuous kernel integration along curve geometry. (b) Discrete rasterization for fixed-grid approximation. (c) The resulting scalar occupancy field used downstream as the tensor mass term.}
    \label{fig:app_cde}
\end{figure}

This appendix complements \cref{sec:bg_cde} by using \cref{fig:app_cde} to connect the continuous Curve Density Estimate (CDE) view to the fixed-grid approximation used in the main paper. We use $\rho(x)$ to denote the continuous CDE field and $D(x)$ for its rasterized approximation on the analysis grid.

\Cref{fig:app_cde_continuous} illustrates the continuous CDE interpretation. For a line segment with endpoints $p_i$ and $p_{i+1}$, let $w$ denote the perpendicular projection of the query point $x$ onto the segment, clipped to the segment endpoints when necessary. In the segment-based CDE view, the contribution of this segment to $x$ is written as
\[
L_h(x,p_i,p_{i+1}) = L_h^{1D}(w,p_i,p_{i+1})\,K_h(x-w),
\]
where $K_h$ is an isotropic spatial kernel with bandwidth $h$, and the associated one-dimensional line kernel is
\[
L_h^{1D}(w,p_i,p_{i+1})
=
\frac{\int_{p_i}^{p_{i+1}} K_h(w-t)\,dt}{\|p_{i+1}-p_i\|}.
\]
Summing these segment contributions over the full ensemble yields the continuous occupancy field. Equivalently, for a curve $L$ parameterized by arc length $s$, this can be written as
\[
\rho(x)=\int_L K_h(x-y(s))\,ds,
\]
where $y(s)$ denotes the point on the curve at arc-length position $s$. As shown in \cref{fig:app_cde_continuous}, this form corresponds to continuous kernel accumulation along the curve geometry and serves as the ideal zero-order quantity underlying the later tensor construction.

\Cref{fig:app_cde_discrete_1} shows how this continuous construction is approximated in the implementation. For efficient fixed-grid computation, each segment is rasterized into a set of pixels $P_{\text{segment}}$, and the continuous line integral is replaced by a discrete summation over rasterized pixel centers:
\[
L_h^{1D}(w,p_i,p_{i+1})
\approx
\frac{1}{|P_{\text{segment}}|}
\sum_{p\in P_{\text{segment}}} K_h(w-p).
\]
Conceptually, the rasterized pixels act as discrete samples of the segment support, making the continuous CDE compatible with the fixed-grid accumulation used throughout the main paper.

Substituting this discretization into the ensemble accumulation yields the practical approximation shown in \cref{fig:app_cde_discrete_2}:
\[
D(x)\approx \sum_{\text{segments}} \omega_s \sum_{p\in P_{\text{segment}}} K_h(x-p),
\]
where $p$ denotes the center of a rasterized pixel and $\omega_s$ is a compact per-segment constant that absorbs the arc-length normalization in this discrete approximation. In the implementation-oriented notation of \cref{appendix:method_impl}, this compact factor is expanded into sample-to-bin weights $\omega_s(b)$.

The scalar field in \cref{fig:app_cde_discrete_2} is therefore the rasterized counterpart of the continuous CDE in \cref{fig:app_cde_continuous}: $\rho(x)$ denotes the continuous occupancy field, whereas $D(x)$ denotes its fixed-grid approximation on the analysis lattice. This is the zero-order mass field used downstream in the main paper. Because each tensor contribution $v_p v_p^{T}$ has unit trace, the structure tensor field in \cref{sec:bg_structure_tensor} is a strict extension of the same mass field, satisfying
\[
\mathrm{tr}(J(x)) = D(x)
\]
under the fixed-grid discretization adopted in this work.

\section{Supplementary Method Details}
\label{appendix:method_impl}

This section summarizes the implementation-level formulas omitted from the main paper for the sake of compactness and readability. Our main paper focuses only on the core definitions. A more detailed definition of the discrete weighting, the projected leave-one-out realization, and the centered path-extent implementation is provided here for completeness.

\noindent\textbf{Discrete Raster Weighting.}
Each trajectory $L_i$ is rasterized into samples $s$ with spatial position $\mathbf{x}_s$, unit tangent $\mathbf{v}_s$, and arc-length weight $\Delta s_s$. A sample contributes to bin $b$ centered at $\mathbf{x}_b$ through the nonnegative kernel weight
\begin{equation}
    \omega_s(b)=K_h(\mathbf{x}_b-\mathbf{x}_s)\Delta s_s.
    \label{eq:app_discrete_weight}
\end{equation}
Substituting \cref{eq:app_discrete_weight} into \cref{eq:tensor_accumulation,eq:loo_tensor} yields the exact discrete tensor and self-contribution used in our implementation.

\noindent\textbf{Projected Leave-One-Out Evaluation.}
The exact conceptual leave-one-out definitions used in the main paper are
\begin{equation}
    \begin{aligned}
        \mathbf{C}_i(b) &= \sum_{s \in L_i} \omega_s(b) (\mathbf{v}_s \mathbf{v}_s^T), \\
        \mathbf{J}_{\text{env}}^{(i)}(b) &= \mathbf{J}_{\text{total}}(b) - \mathbf{C}_i(b)
    \end{aligned}
    \label{eq:loo_tensor}
\end{equation}
\begin{equation}
    E_{\text{local}}^{\text{LOO}}(\mathbf{x},\mathbf{v}_i(\mathbf{x});L_i) =
    \frac{\mathbf{v}_i(\mathbf{x})^T \mathbf{J}_{\text{env}}^{(i)}(\mathbf{x})\mathbf{v}_i(\mathbf{x})}
    {\operatorname{tr}(\mathbf{J}_{\text{env}}^{(i)}(\mathbf{x})) + \epsilon}
    \label{eq:loo_local_exact}
\end{equation}
These exact expressions define the unbiased environment and the local support primitive conceptually. The implementation below evaluates the same quantity through cached scalar components.
Rather than materializing $\mathbf{J}_{\text{env}}^{(i)}(b)$ at every queried sample, we cache its two sufficient scalar statistics in the queried tangent direction: the projected energy and the trace. These statistics are evaluated per queried sample. For a sample $s$ of $L_i$ lying in bin $b$, we cache:
\begin{equation}
    \begin{aligned}
        q_i &= \mathbf{v}_i(b)^T \mathbf{J}_{\text{total}}(b)\mathbf{v}_i(b), \qquad
        \rho_i = \operatorname{tr}(\mathbf{J}_{\text{total}}(b)), \\
        q_i^{\text{self}} &= \mathbf{v}_i(b)^T \mathbf{C}_i(b)\mathbf{v}_i(b), \qquad
        \rho_i^{\text{self}} = \operatorname{tr}(\mathbf{C}_i(b)).
    \end{aligned}
    \label{eq:app_loo_components}
\end{equation}
\begin{equation}
    \tilde{q}_i = \max(0, q_i-q_i^{\text{self}}), \qquad
    \tilde{\rho}_i = \max(0, \rho_i-\rho_i^{\text{self}})
    \label{eq:app_loo_projected_score}
\end{equation}
\begin{equation}
    E_{\text{local}}^{\text{LOO}}(\mathbf{x},\mathbf{v}_i(\mathbf{x});L_i) = \frac{\tilde{q}_i}{\tilde{\rho}_i+\epsilon}.
    \label{eq:app_loo_projected_eval}
\end{equation}
When $\epsilon=0$ and $\tilde{\rho}_i=0$, which occurs when leave-one-out removes all independent tensor mass at the queried sample, we define the local support to be zero. This convention treats zero remaining mass as absence of independent evidence rather than as an undefined orientation agreement. Under this convention, when the cached terms are accumulated from the same raster samples and kernel weights as $\mathbf{J}_{\text{total}}$, \cref{eq:app_loo_projected_eval} is equivalent to the exact definition in \cref{eq:loo_local_exact}. The clamping in \cref{eq:app_loo_projected_score} acts only as a numerical safeguard.

\noindent\textbf{Passage-Centered Evaluation.}
For a passage $(L,s_\mathbf{x})$ through pixel $\mathbf{x}$ and extent parameter $\ell$, the corresponding centered arc-length interval is
\begin{equation}
    \mathcal{W}_L(s_\mathbf{x};\ell)=
    \left[\max(0,s_\mathbf{x}-\ell),\;\min(|L|,s_\mathbf{x}+\ell)\right].
    \label{eq:app_arf_window}
\end{equation}
The corresponding passage-centered support is then
\begin{equation}
    \phi(L,s_\mathbf{x};\ell)=
    \begin{cases}
        E_{\text{local}}^{\text{LOO}}
        (L(s_\mathbf{x}),\mathbf{v}_L(s_\mathbf{x});L), & \ell=0,\\[1mm]
        \dfrac{1}{|\mathcal{W}_L(s_\mathbf{x};\ell)|}
        \displaystyle\int_{s\in\mathcal{W}_L(s_\mathbf{x};\ell)}
        E_{\text{local}}^{\text{LOO}}
        (L(s),\mathbf{v}_L(s);L)\,ds, & \ell>0.
    \end{cases}
    \label{eq:passage_support}
\end{equation}
Conceptually, $\ell$ controls the extent of the path-centered evaluation: the continuous path interval runs from $s_\mathbf{x}-\ell$ to $s_\mathbf{x}+\ell$, so its total arc-length span is at most $2\ell$, and it is clipped to the trajectory endpoints whenever one side reaches them first. In the discrete implementation, we keep the same symbol $\ell$ for the corresponding raster-sample extent, meaning that we take rasterized trajectory samples before and after the sample at $\mathbf{x}$ along the same passage and evaluate the resulting passage-centered averages with prefix sums.
Conceptually, each element of $\mathcal{P}_\mathbf{x}$ corresponds to one contiguous traversal through pixel $\mathbf{x}$ and uses the midpoint arc-length of that traversal as $s_\mathbf{x}$; repeated visits at different arc-length positions are counted separately. Empty pixels remain outside the support region $\Omega_{\text{supp}}=\{\mathbf{x}:|\mathcal{P}_\mathbf{x}|>0\}$ and are omitted in the final rendering.

\section{Supplementary Peeling Workflow}
\label{appendix:workflow}

\begin{figure*}[t]
    \centering
    \includegraphics[width=0.85\linewidth]{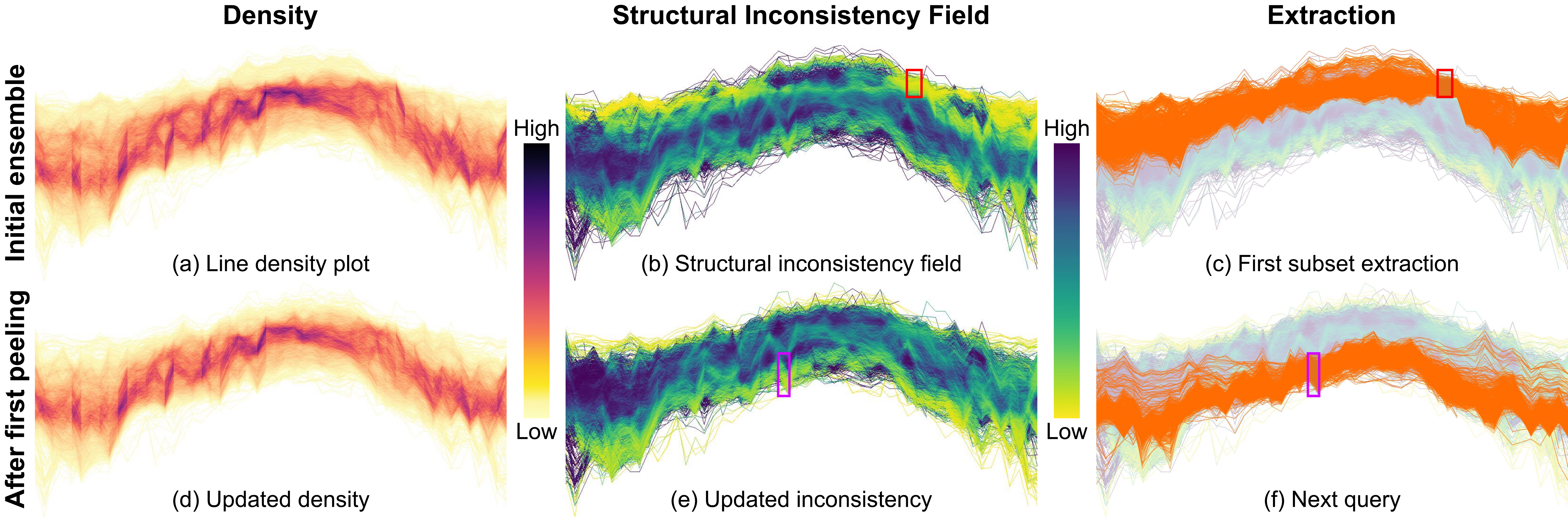}
    \vspace{-2mm}
    \caption{\textbf{Supplementary peeling workflow.} Top row: density, inconsistency, and the first extraction on the full ensemble. Bottom row: the corresponding density, inconsistency, and next extraction after peeling the first subset and recomputing the fields on the remainder.}
    \label{fig:app_peeling_workflow}
    \vspace{-4mm}
\end{figure*}

\cref{fig:app_peeling_workflow} illustrates the iterative peeling workflow introduced in \cref{sec:system} as a static figure. Starting from the full ensemble, analysts inspect density and inconsistency together, extract a first subset, remove it, and recompute the overview fields on the residual ensemble. The updated density and inconsistency views then expose secondary structure that was masked by the dominant set of trajectories removed in the first peeling iteration, enabling a second query on the remainder.

\section{Qualitative Effect of Path Extent}
\label{appendix:path_extent}

\cref{fig:app_d3_path_extent} illustrates the effect of different path extents $\ell$ on our ambiguous synthetic dataset D3. We used a short ($\ell=0$), a medium ($\ell=400$) and a maximal ($\ell=1000$) extent. As these values depend on the discretization resolution, different values must be chosen if the resolution is altered to obtain a similar result.  The key point is that D3 cannot be determined solely from the local geometry. At short and medium extents, both D3-C and D3-D still evaluate structural support mostly within the locally coherent central region. Thus, the middle region remains visually similar for D3-C and D3-D. Once the path extent reaches its maximum, the evaluation of structural support reaches the regions where the underlying structures of D3-C and D3-D differ greatly. In D3-C, the center stays comparatively light because the dense bridge is supported by trajectories that remain coherent between the two bundles at both sides. In D3-D, the same central density pattern loses its distinctive low inconsistency because the longer path extent now includes the divergent, structurally mixed regions towards the margins of the plot. This is why variable path extent is useful: it allows the field to transition from local agreement to global structural support, thereby revealing whether a dense visual pattern corresponds to a genuine bridge or an ambiguous superposition.

\section{Supplementary Leave-One-Out Ablation}
\label{appendix:loo_interpretation}

To isolate self-support bias, we construct nested, noncrossing trajectories $L_i(x)=(x,y_i(x))$, $x\in[0,1]$, where $y_i(x)=0.08\sin(2\pi x)+z_i[0.08+0.22\sin(\pi x)]$. In each trial, 500 coefficients $z_i$ are sampled from jittered strata of a standard normal distribution. The resulting ensemble is concentrated around its centerline and becomes progressively sparser toward its outer envelopes. We define the center-outward depth proxy $d_i=-|z_i|$. For this nested construction, its ordering follows the functional-depth principle that central curves are more representative than curves near the envelope~\cite{mirzargar2014curve,lopez2014simplicial}; we do not use functional band depth itself as an input to our method. The lowest-depth 5\% of trajectories constitute the positive outlier class. For the tabulated quantitative comparison, all fidelities are computed from 100 samples per trajectory on a $420 \times 260$ coordinate domain, using Gaussian bandwidth $\sigma=1$ pixel, density regularization $\epsilon=0$ with the zero-support convention above, and path extent 51 in both conditions. Under this convention, samples with no remaining LOO tensor mass are assigned zero local support rather than evaluated as an undefined ratio. Without LOO, a trajectory that is the sole local contributor induces a rank-one tensor aligned with its own tangent and can consequently receive near-unit local support. LOO removes this deterministic self-contribution, so the remaining fidelity measures support supplied by other trajectories.

\begin{table}[tb]
    \centering
    \small
    \setlength{\tabcolsep}{2.8pt}
    \caption{\textbf{Controlled LOO ablation on spindle ensembles.} Values are means over 20 independently generated trials with 500 trajectories per trial. AUROC and AUPRC are computed using lower trajectory fidelity as the outlier evidence, equivalently ranked by $1-\Phi$, with the lowest-depth 5\% as the positive class. The center-tail gap is the mean fidelity of the highest-depth 5\% minus that of the lowest-depth 5\%; a positive value indicates lower support in the tail. Gaps are computed before rounding.}
    \label{tab:app_loo_spindle}
    \begin{tabular}{@{}lccccc@{}}
        \toprule
        \textbf{Method} &
        \textbf{Tail AUROC} &
        \textbf{Tail AUPRC} &
        \textbf{Center $\Phi$} &
        \textbf{Tail $\Phi$} &
        \textbf{Gap} \\
        \midrule
        Without LOO & 0.009 & 0.027 & 0.9999 & 1.0000 & -0.000025 \\
        With LOO    & \textbf{0.831} & \textbf{0.785} & 0.9999 & 0.8514 & \textbf{0.1486} \\
        \bottomrule
    \end{tabular}
\end{table}

\cref{tab:app_loo_spindle} quantifies the resulting bias. Without LOO, mean fidelity in the lowest-depth tail (1.0000) slightly exceeds that in the center (0.9999), yielding a negative center-tail gap and reversing the intended outlier ranking. The corresponding AUROC and AUPRC are 0.009 and 0.027, respectively. With LOO, the central mean remains 0.9999, whereas the tail mean decreases to 0.8514; the center-tail gap consequently increases to 0.1486, with AUROC 0.831 and AUPRC 0.785. Based on 2,000 bootstrap resamples of the 20 trial-level AUROC values, the 95\% confidence interval for the mean AUROC is [0.007, 0.012] without LOO and [0.804, 0.859] with LOO.

\begin{figure*}[tb]
    \centering
    \begin{minipage}[t]{0.045\textwidth}
        \vspace{0pt}
        \centering
        \includegraphics[height=0.16\textheight]{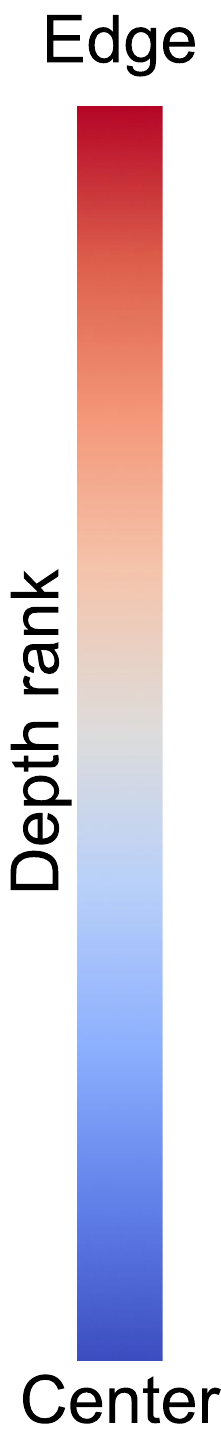}
    \end{minipage}
    \hfill
    \begin{subfigure}[t]{0.285\textwidth}
        \vspace{0pt}
        \centering
        \includegraphics[width=\linewidth]{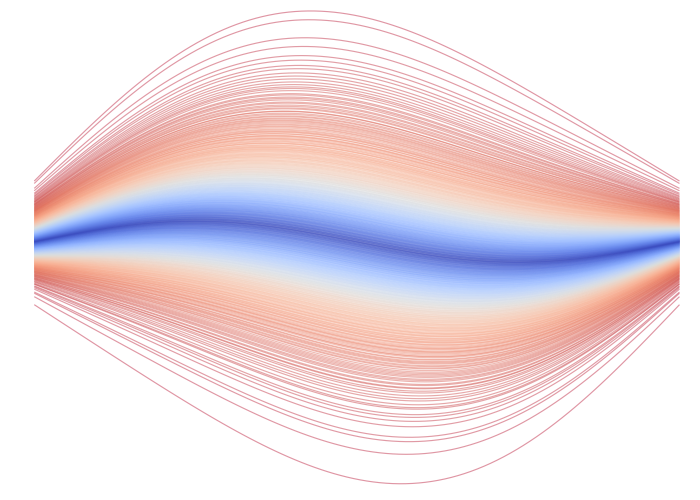}
        \caption{Depth proxy}
        \label{fig:app_loo_spindle_depth}
    \end{subfigure}
    \hfill
    \begin{subfigure}[t]{0.285\textwidth}
        \vspace{0pt}
        \centering
        \includegraphics[width=\linewidth]{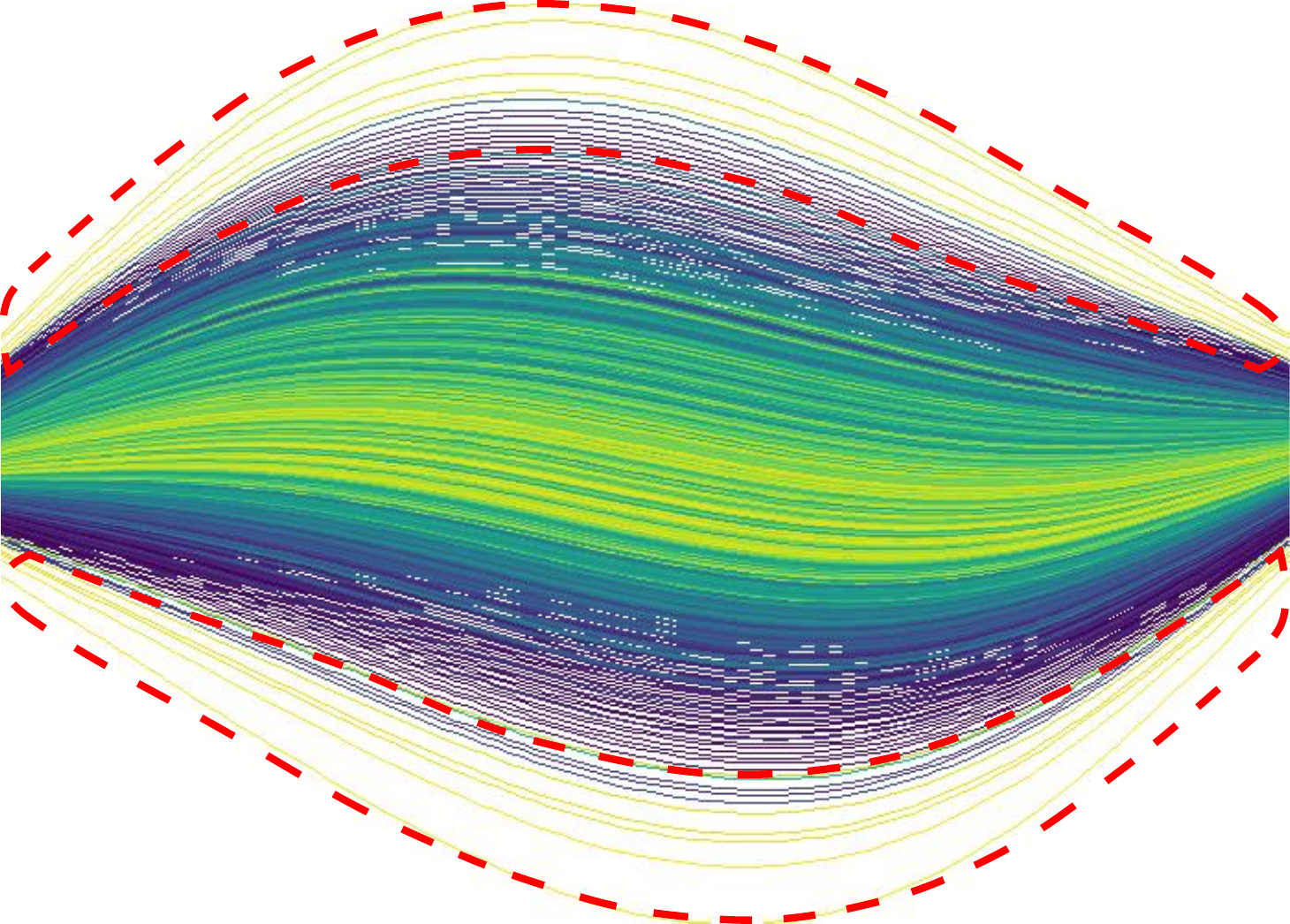}
        \caption{SIF without LOO}
        \label{fig:app_loo_spindle_sif_without}
    \end{subfigure}
    \hfill
    \begin{subfigure}[t]{0.285\textwidth}
        \vspace{0pt}
        \centering
        \includegraphics[width=\linewidth]{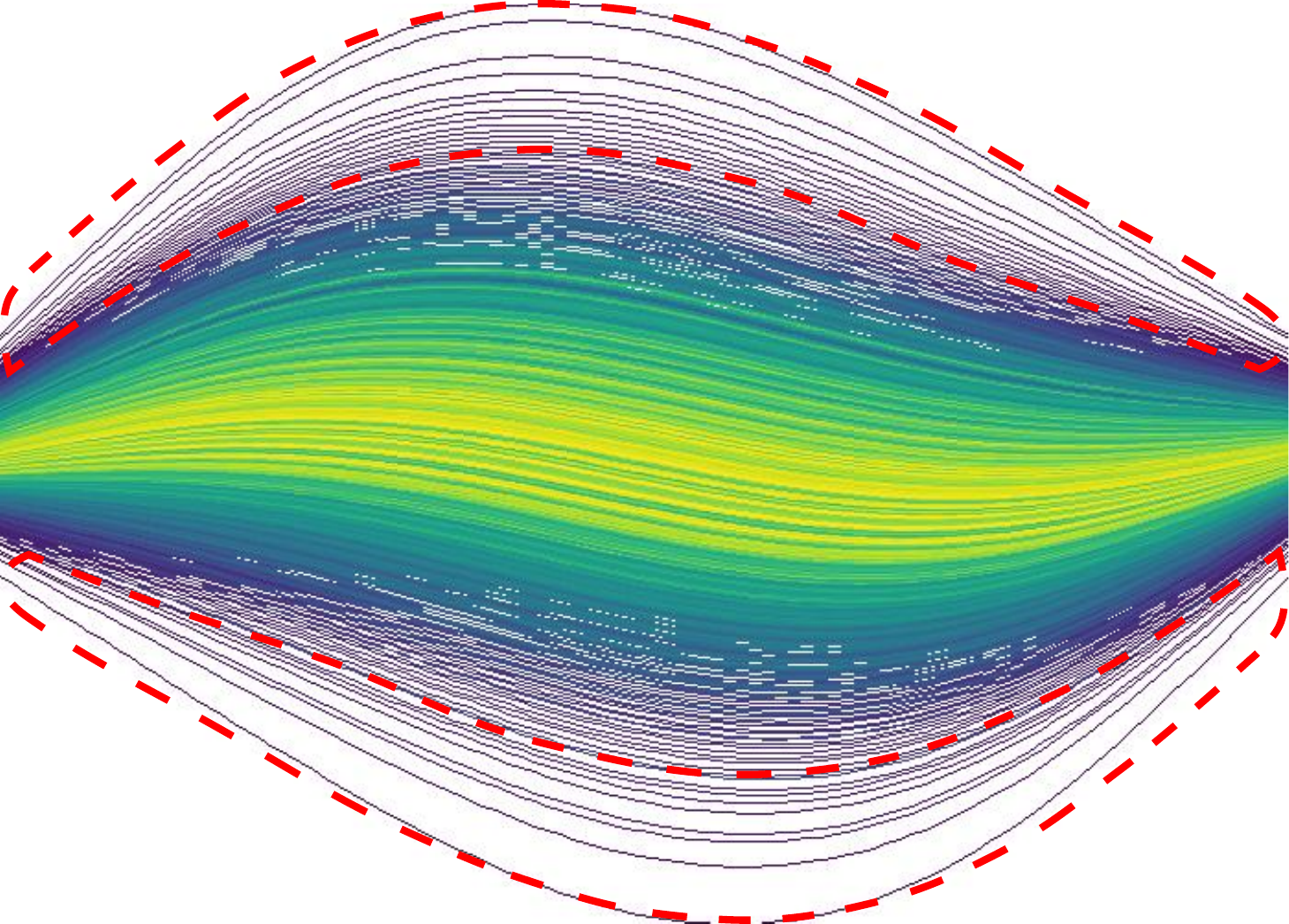}
        \caption{SIF with LOO}
        \label{fig:app_loo_spindle_sif_with}
    \end{subfigure}
    \hfill
    \begin{minipage}[t]{0.05\textwidth}
        \vspace{0pt}
        \centering
        \includegraphics[height=0.16\textheight]{figs/evaluation/inconsistency_colormap.pdf}
    \end{minipage}
    \caption{\textbf{Controlled spindle ablation and spatial effect of LOO.} (a) Nested noncrossing trajectories colored by the center-outward depth proxy $d_i=-|z_i|$ (blue: central; red: outer). The SIF views are qualitative renderings of the same representative seed on a $700 \times 500$ canvas; they use a common scale, are vertically aligned with (a), and use the maximal path extent in both conditions. Red dashed boxes mark sparse outer regions where the visual difference is most apparent. Without LOO (b), these trajectories exhibit artificially low inconsistency because their own tangent contributions dominate the local tensors. With LOO (c), the same trajectories exhibit high inconsistency because little independent support remains, whereas the densely supported center changes minimally.}
    \label{fig:app_loo_spindle}
\end{figure*}

We further characterize local sparsity in the representative trial shown in \cref{fig:app_loo_spindle_depth} (seed 9100) on the same $420 \times 260$ quantitative grid. For each rasterized trajectory sample, we define rounded-pixel occupancy as the number of distinct trajectories represented in the same rounded pixel. Samples from the highest-depth 5\% have a median occupancy of eight, and none has occupancy at most two. In contrast, samples from the lowest-depth 5\% have a median occupancy of one: 80.6\% occur in pixels containing only the queried trajectory, and 96.0\% occur in pixels containing at most two trajectories. Every sample from the extreme 1\% tail occurs in a single-trajectory pixel under this diagnostic. The representative ensemble therefore contains both densely corroborated and locally unsupported trajectories despite its fixed global size of 500.

\cref{fig:app_loo_spindle_sif_without,fig:app_loo_spindle_sif_with} show the spatial consequence of this occupancy difference under the same maximal path extent. The red dashed boxes highlight the sparse outer regions where the discrepancy between the non-LOO and LOO fields is most pronounced. Without LOO, isolated outer trajectories provide their own orientation evidence and therefore appear artificially low in inconsistency. Removing their contributions changes the same trajectories to high inconsistency, while the densely corroborated center remains nearly unchanged. The experiment does not imply a universal minimum trajectory count. Reliability is governed by the independent tensor mass remaining after removal of the queried trajectory and therefore depends jointly on kernel bandwidth, grid resolution, local passage count, path extent, and local directional variance. A single isolated trajectory leaves no independent tensor support after LOO, and one or two weakly overlapping trajectories may likewise provide insufficient evidence. Such locations should be interpreted as weakly supported rather than as reliable evidence of a particular orientation pattern, in conjunction with density and passage count.

\section{Further Examples}
\label{appendix:further_examples}

This section collects additional examples that complement the main case studies without changing the method or evaluation protocol. It first provides the complementary low-score view of the DTI case and the native planar Rayleigh--Bénard pathline example, followed by four real-world examples spanning two time-series datasets and two trajectory datasets. Across these cases, density provides the occupancy overview, the Structural Inconsistency Field (SIF) indicates where structural support is maintained or breaks down, and trajectory-level querying, filtering, or peeling exposes the corresponding line subsets.

\newcommand{\appendixfieldlegend}[1]{%
    \makebox[\linewidth][c]{%
        \raisebox{-0.55ex}{\rotatebox{90}{\tiny\textsf{High}}}%
        \hspace{0.7mm}%
        \raisebox{-0.45ex}{\rotatebox{90}{\includegraphics[height=0.70\linewidth,width=0.026\linewidth,trim=12bp 82bp 38bp 82bp,clip]{#1}}}%
        \hspace{0.7mm}%
        \raisebox{-0.55ex}{\rotatebox{90}{\tiny\textsf{Low}}}%
    }%
}

\newcommand{\appendixfieldlegendrow}{%
    \begin{minipage}[t]{0.48\textwidth}
        \centering
        \appendixfieldlegend{figs/evaluation/density_colormap.pdf}
    \end{minipage}%
    \hfill
    \begin{minipage}[t]{0.48\textwidth}
        \centering
        \appendixfieldlegend{figs/evaluation/inconsistency_colormap.pdf}
    \end{minipage}%
}

\subsection{Supplementary DTI Filtering View}

\begin{figure}[tb]
    \centering
    \includegraphics[width=0.8\linewidth]{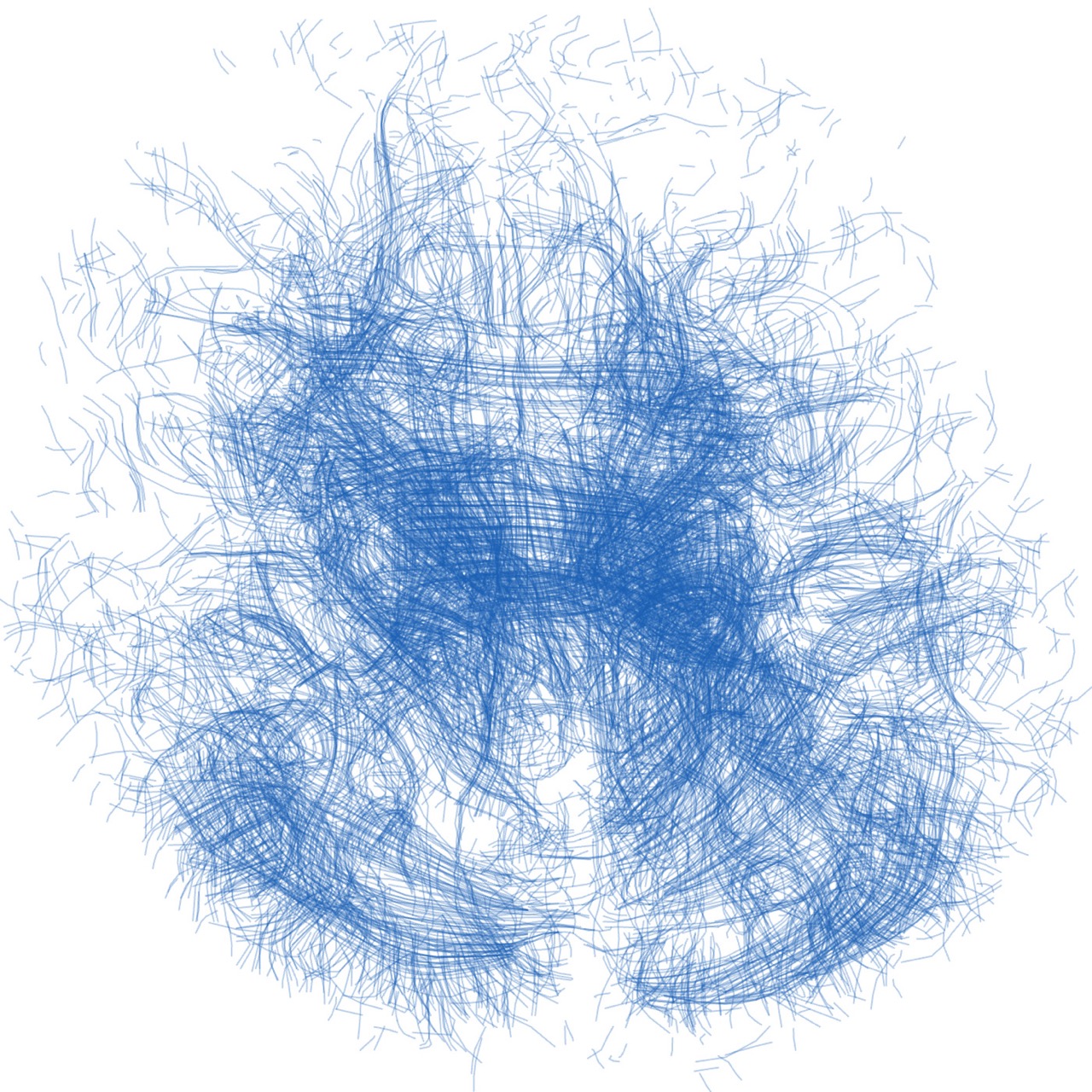}
    \caption{\textbf{Supplementary low-score half of the DTI example.} The lowest-scoring 50\% of fibers removed from \cref{fig:case_result}.}
    \label{fig:app_dti_outlier}
    \vspace{-4mm}
\end{figure}

This supplementary view complements the DTI case study in the main paper by showing what remains outside the highest-scoring half. The discarded fibers are not just minor side branches of the retained scaffold. Instead, they contain many short, peripheral, and directionally inconsistent trajectories around the boundary and crossing regions, so they remain visually diffuse and structurally mixed. The circled regions in \cref{fig:eval_dti_case} further clarify the difference to line bundling: in region A, the retained fibers preserve a parallel bundle with visible width, whereas bundling collapses it into an unnaturally tight track; in region B, the retained subset reveals a slanted bundle at roughly $45^\circ$ that disappears in the bundled view. Together, \cref{fig:case_result,fig:app_dti_outlier} make the screen-space filtering effect easier to interpret: the retained set forms a coherent projected scaffold, while the removed half captures competing or weakly supported projected fibers that stay entangled in the raw density view.

\subsection{Rayleigh--Bénard Pathlines}
\label{appendix:rayleigh_benard}

We evaluate the method on pathlines derived from the \emph{Rayleigh--Bénard convection} dataset in The Well~\cite{ohana2024well}. The dataset describes a two-dimensional fluid layer heated from below and cooled from above, producing convection cells, vortices, plume-like transport, and boundary layers. The simulations were generated with Dedalus~\cite{burns2020dedalus}; the released fields comprise buoyancy, pressure, and velocity over 200 time steps on a $512\times128$ grid. The physical domain is $[0,4]\times[0,1]$, with periodic horizontal boundaries and wall boundaries in the vertical direction. Because the velocity field and derived pathlines are intrinsically two-dimensional, this case permits assessment of the screen-space representation without introducing a separate 3D-to-2D projection.

Trajectory extraction uses simulation index 4 from the data with Rayleigh number $10^8$ and Prandtl number $1$. We initialize 12,059 particles (matching the DTI subset size) on a deterministic jittered grid and integrate them over $t\in[20.0,25.0]$ using fourth-order Runge--Kutta integration, bilinear spatial interpolation, and linear temporal interpolation. Each source pathline contains 101 output samples, and the coordinates are normalized to $[0,1]^2$. In this normalized domain, a periodic wrap produces an apparent horizontal displacement greater than 0.5 between consecutive samples. We split each pathline at such displacements before visualization and analysis, thereby preventing nonphysical full-width chords in the non-periodic screen-space rendering. This preprocessing produces 12,717 output polylines while preserving the sampled geometry within each contiguous segment.

\begin{figure*}[t]
    \centering
    \begin{subfigure}[b]{0.48\textwidth}
        \centering
        \includegraphics[width=\linewidth]{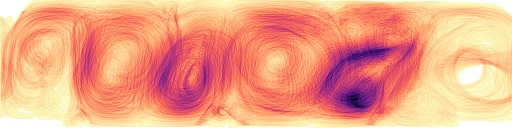}
        \caption{Density}
        \label{fig:app_rayleigh_density}
    \end{subfigure}
    \hfill
    \begin{subfigure}[b]{0.48\textwidth}
        \centering
        \includegraphics[width=\linewidth]{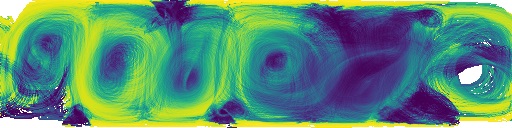}
        \caption{Structural inconsistency}
        \label{fig:app_rayleigh_sif}
    \end{subfigure}
    \par\vspace{-1mm}
    \appendixfieldlegendrow
    \vspace{1mm}
    \begin{subfigure}[b]{0.48\textwidth}
        \centering
        \includegraphics[width=\linewidth]{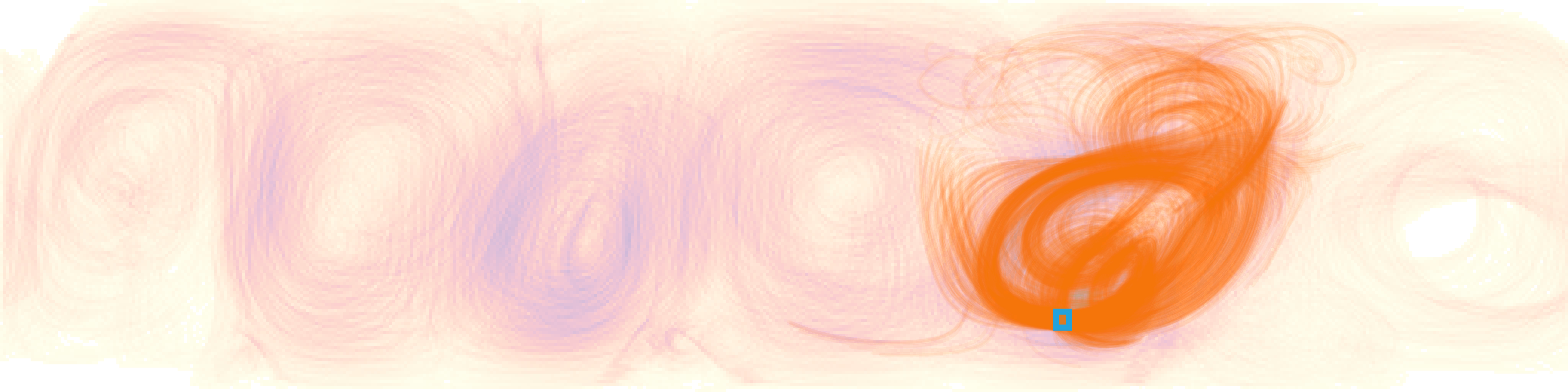}
        \caption{Density-guided query}
        \label{fig:app_rayleigh_density_query}
    \end{subfigure}
    \hfill
    \begin{subfigure}[b]{0.48\textwidth}
        \centering
        \includegraphics[width=\linewidth]{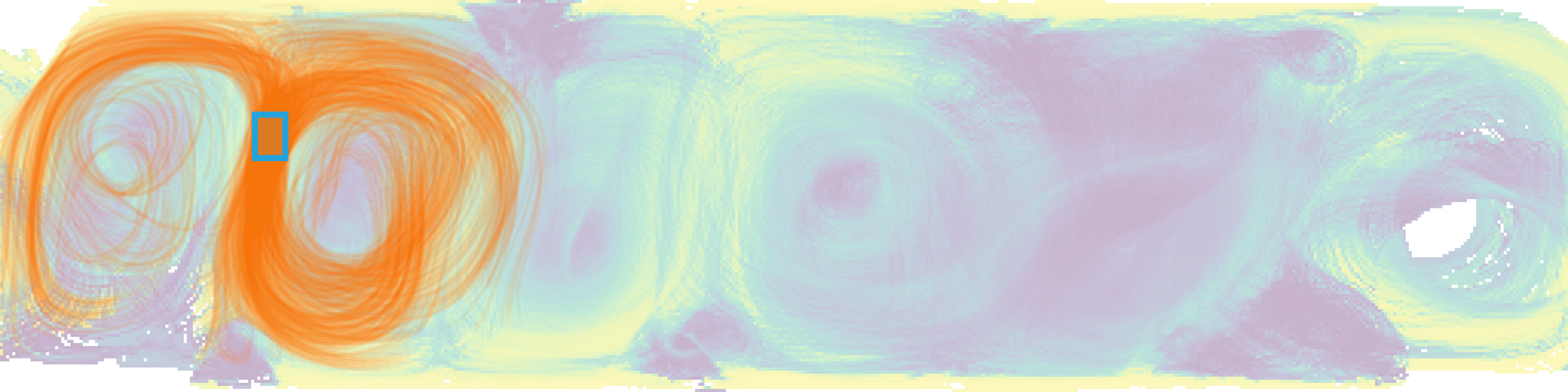}
        \caption{Structure-guided query}
        \label{fig:app_rayleigh_sif_query}
    \end{subfigure}
    \caption{\textbf{Supplementary Rayleigh--Bénard pathline case study.} Pathlines derived from the native 2D unsteady-flow fields in The Well Rayleigh--Bénard convection dataset. Density (a) emphasizes frequently visited recirculation regions. The SIF (b) assigns lower inconsistency to many recirculating bands and higher inconsistency to interfaces and transition regions containing overlapping orientations. From the 12,717 output polylines, a high-density query (c) selects 739 trajectories across several loop families and connecting arcs, whereas a low-inconsistency query (d) selects 787 trajectories concentrated around one recirculation cell. The queries target distinct regions and provide a qualitative comparison.}
    \label{fig:app_rayleigh_benard}
\end{figure*}

\cref{fig:app_rayleigh_benard} compares occupancy and structural support in this planar flow. The density view in \cref{fig:app_rayleigh_density} assigns high intensity to frequently visited portions of the recirculation cells, but it does not indicate whether coincident passages follow one locally supported orientation or several competing orientations. The SIF in \cref{fig:app_rayleigh_sif} assigns relatively low inconsistency to many smoothly recirculating bands and higher inconsistency to regions where differing local orientations overlap, including interfaces, cell boundaries, and transition corridors between neighboring rolls.

The two queries illustrate how the views support different exploration targets; they are qualitative examples selected from distinct regions rather than a matched-region quantitative comparison. In \cref{fig:app_rayleigh_density_query}, querying a high-density region selects 739 trajectories spanning the right-center recirculation pattern, several adjacent loop families, and connecting arcs. In \cref{fig:app_rayleigh_sif_query}, querying a low-inconsistency region selects 787 trajectories visually concentrated around one recirculation cell, with fewer passages from neighboring cells and transition corridors. Density therefore provides an occupancy-based overview, whereas the SIF identifies locations at which the path-integrated orientation support is sustained or weakened.

\subsection{Additional Real-World Examples}
\label{appendix:additional_real_examples}

\begin{figure*}[t]
    \centering
    \begin{subfigure}[t]{0.48\textwidth}
        \centering
        \includegraphics[width=\linewidth]{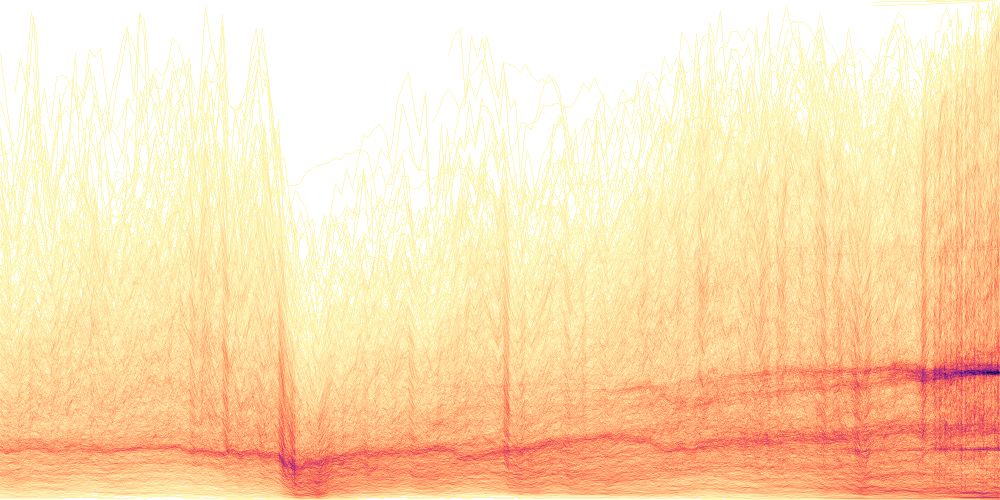}
        \caption{Density}
    \end{subfigure}
    \hfill
    \begin{subfigure}[t]{0.48\textwidth}
        \centering
        \includegraphics[width=\linewidth]{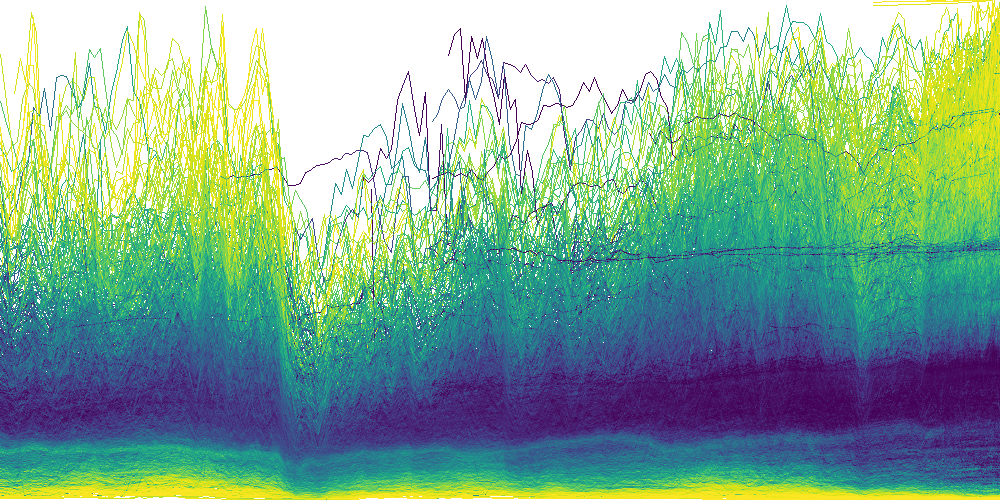}
        \caption{SIF}
    \end{subfigure}
    \par\vspace{-1mm}
    \appendixfieldlegendrow
    \vspace{4pt}
    \begin{subfigure}[t]{0.48\textwidth}
        \centering
        \includegraphics[width=\linewidth]{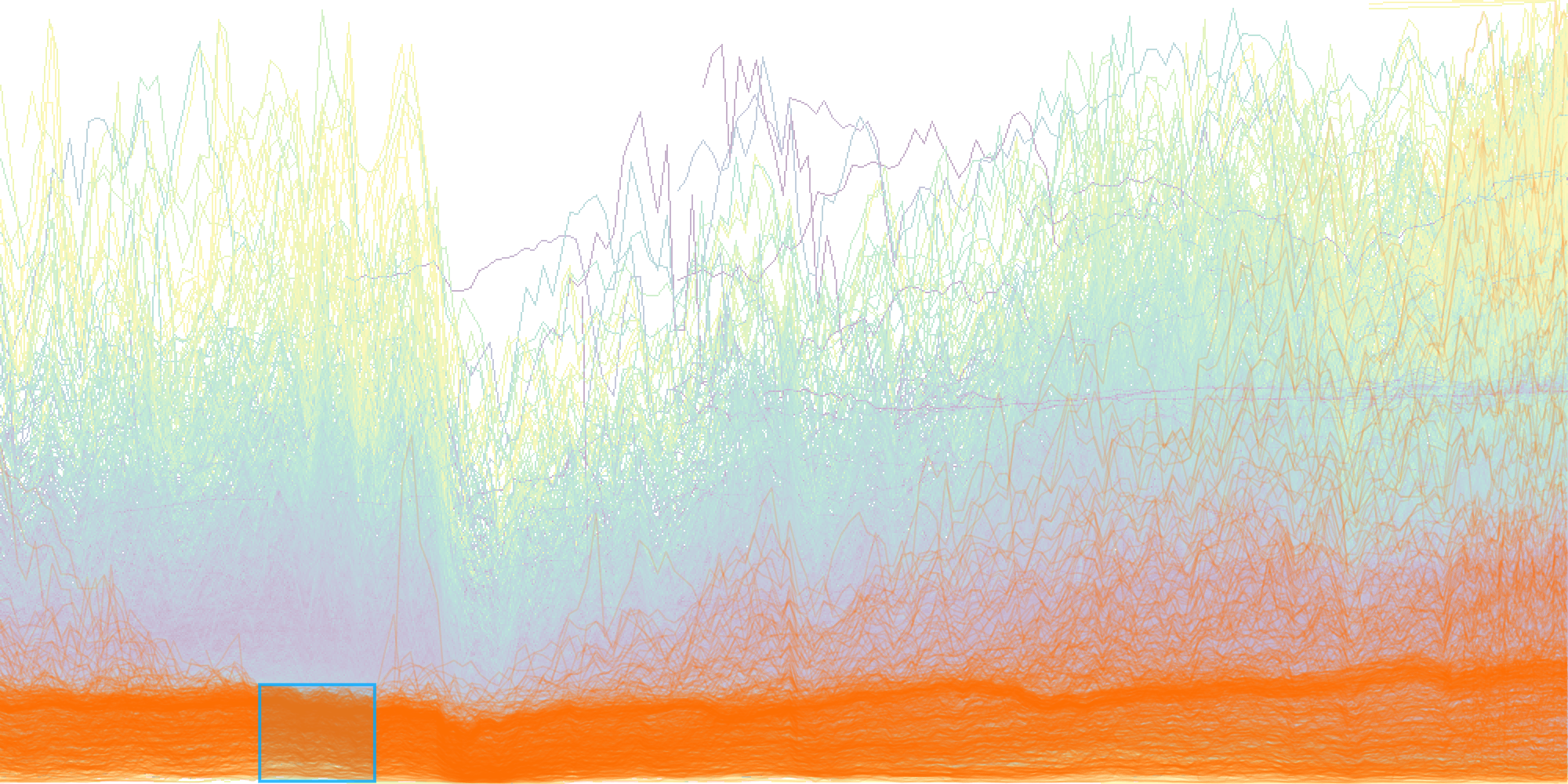}
        \caption{TimeBox Query (658)}
    \end{subfigure}
    \hfill
    \begin{subfigure}[t]{0.48\textwidth}
        \centering
        \includegraphics[width=\linewidth]{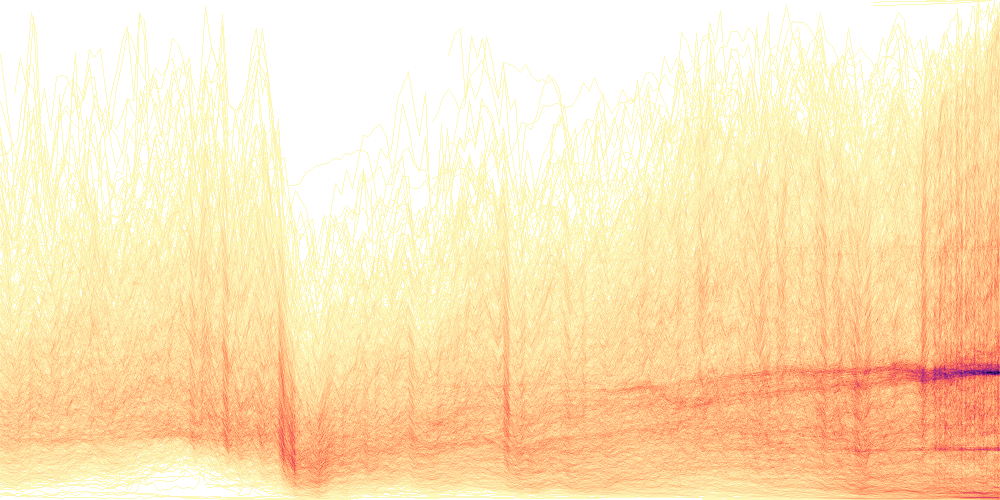}
        \caption{After Peeling}
    \end{subfigure}
    \caption{\textbf{Stock price trajectories.} The long bottom band is simultaneously prominent in the density view and low in structural inconsistency in the SIF. A TimeBox query retrieves 658 lines from this band, and peeling them removes it from the residual density view, confirming that it corresponds to a coherent subset of the stock ensemble.}
    \label{fig:supp_real_example_stock}
\end{figure*}

\begin{figure*}[t]
    \centering
    \begin{subfigure}[t]{0.48\textwidth}
        \centering
        \includegraphics[width=\linewidth]{figs/further_examples/stock/density.png}
        \caption{Density}
    \end{subfigure}
    \hfill
    \begin{subfigure}[t]{0.48\textwidth}
        \centering
        \includegraphics[width=\linewidth]{figs/further_examples/stock/sif.png}
        \caption{SIF}
    \end{subfigure}
    \par\vspace{-1mm}
    \appendixfieldlegendrow
    \vspace{4pt}
    \begin{subfigure}[t]{0.48\textwidth}
        \centering
        \includegraphics[width=\linewidth]{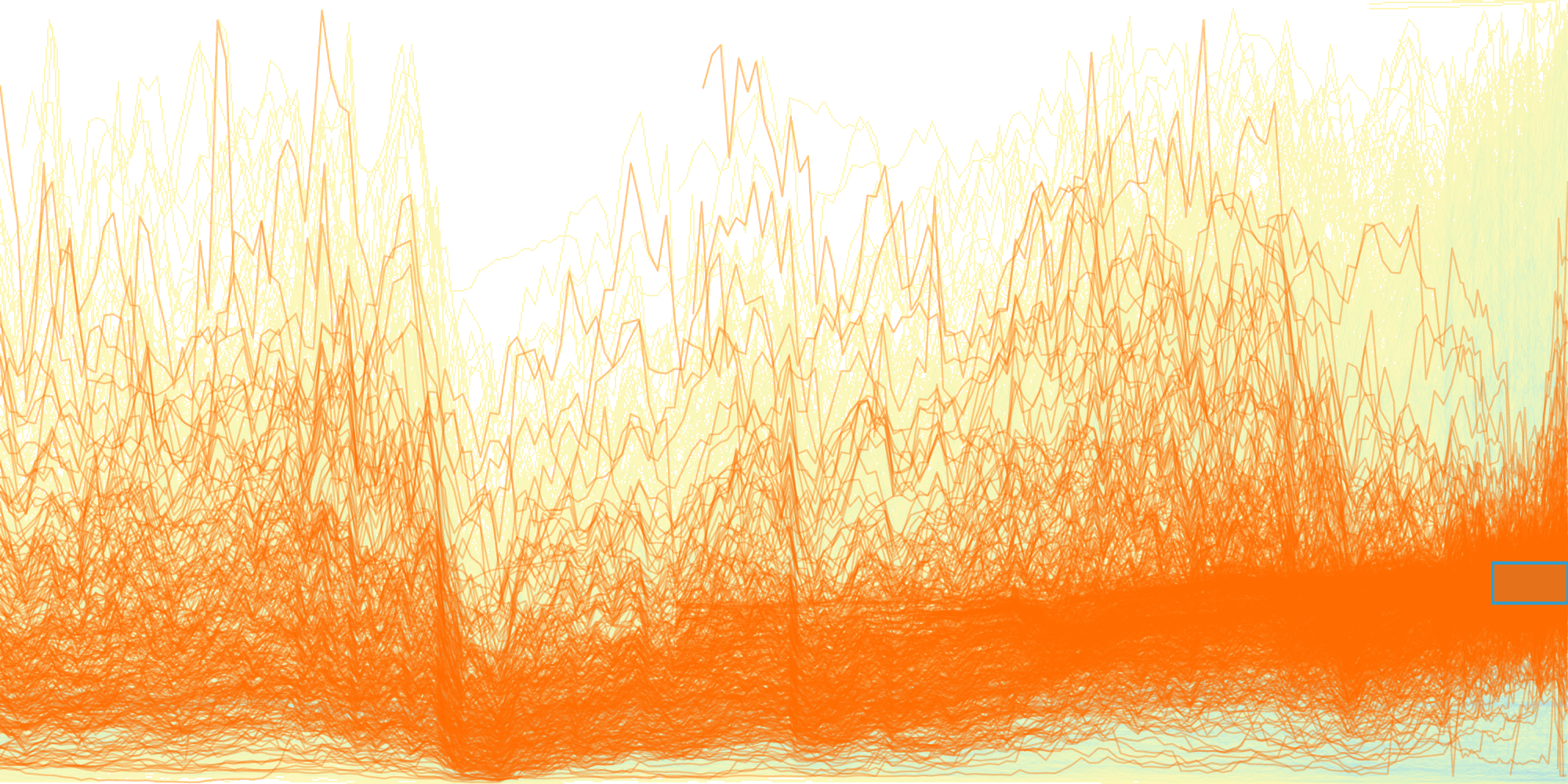}
        \caption{Window Query (1,232)}
    \end{subfigure}
    \hfill
    \begin{subfigure}[t]{0.48\textwidth}
        \centering
        \includegraphics[width=\linewidth]{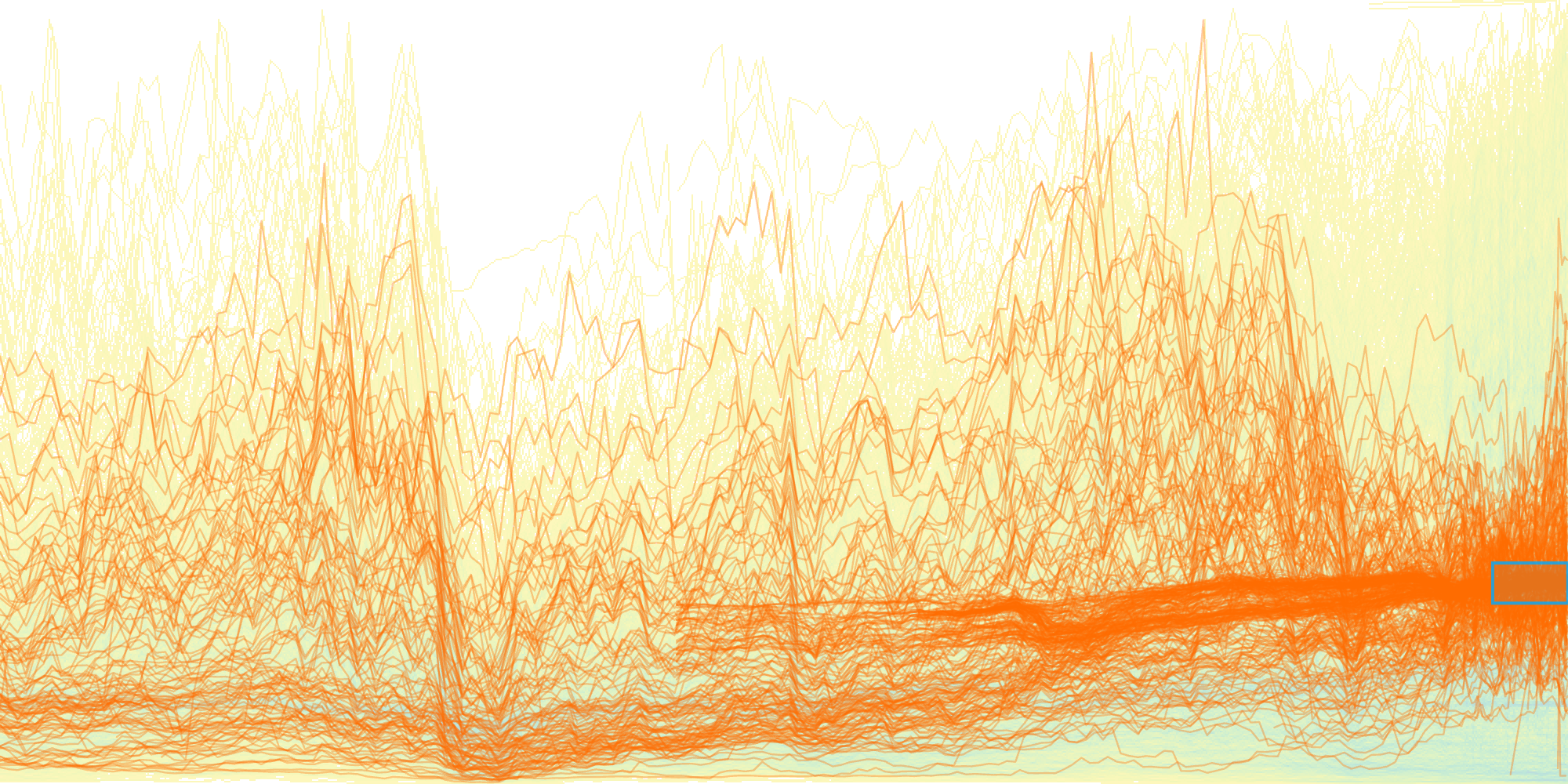}
        \caption{After Filtering (370)}
    \end{subfigure}
    \caption{\textbf{Stock price trajectories, second pattern.} The second prominent stock pattern is visually strong in the density view but is marked as high structural inconsistency in the SIF. We therefore use a window query instead of a TimeBox because the lines entering from above or below are part of the phenomenon rather than a distraction. The 1,232-line selection is heavily cluttered, while the filtered 370-line subset reveals a coherent late-starting minority around or after 2012. The example matches the majority-rules limitation discussed in the main paper: a valid minority structure can be suppressed by a dominant mixed background and remain marked as highly inconsistent in the SIF.}
    \label{fig:supp_real_example_stock_second}
\end{figure*}

\begin{figure*}[t]
    \centering
    \begin{subfigure}[t]{0.48\textwidth}
        \centering
        \includegraphics[width=\linewidth]{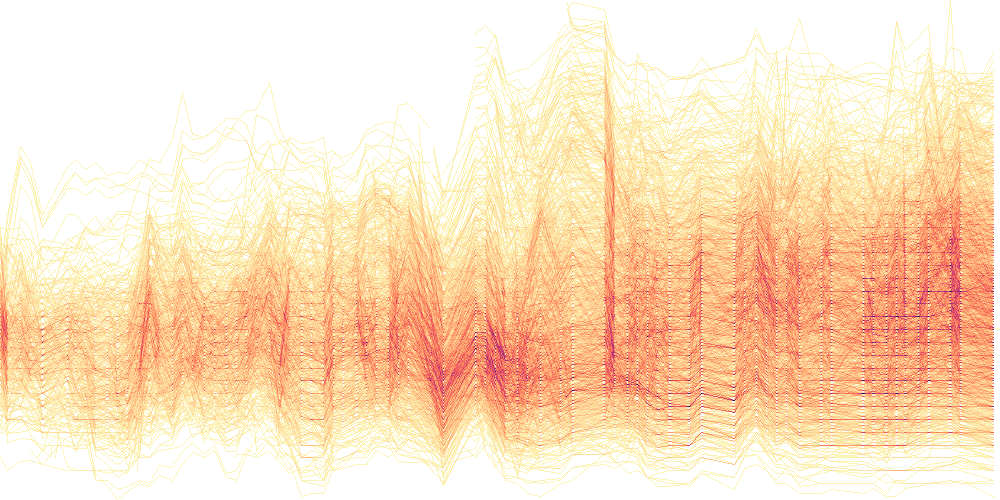}
        \caption{Density}
    \end{subfigure}
    \hfill
    \begin{subfigure}[t]{0.48\textwidth}
        \centering
        \includegraphics[width=\linewidth]{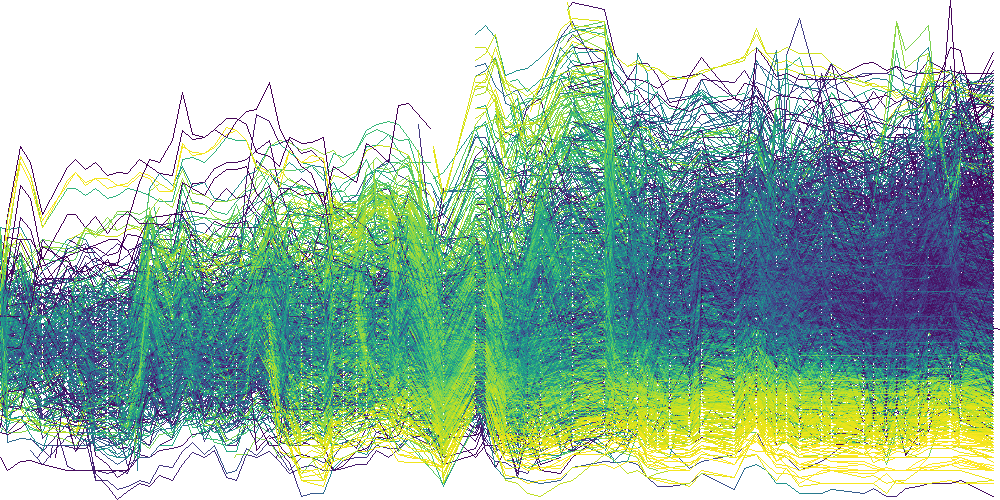}
        \caption{SIF}
    \end{subfigure}
    \par\vspace{-1mm}
    \appendixfieldlegendrow
    \vspace{4pt}
    \begin{subfigure}[t]{0.48\textwidth}
        \centering
        \includegraphics[width=\linewidth]{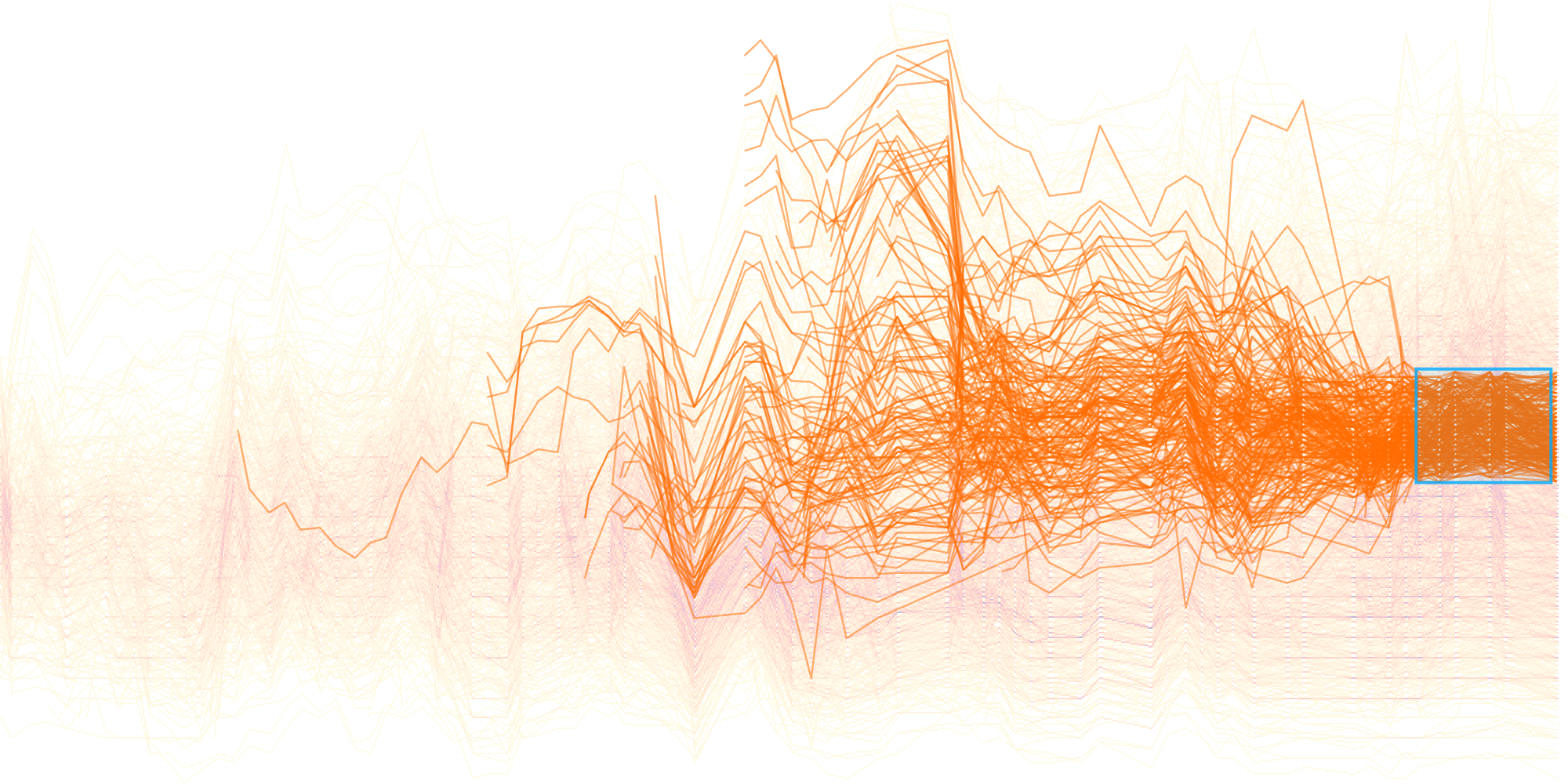}
        \caption{Density-Guided TimeBox (223)}
    \end{subfigure}
    \hfill
    \begin{subfigure}[t]{0.48\textwidth}
        \centering
        \includegraphics[width=\linewidth]{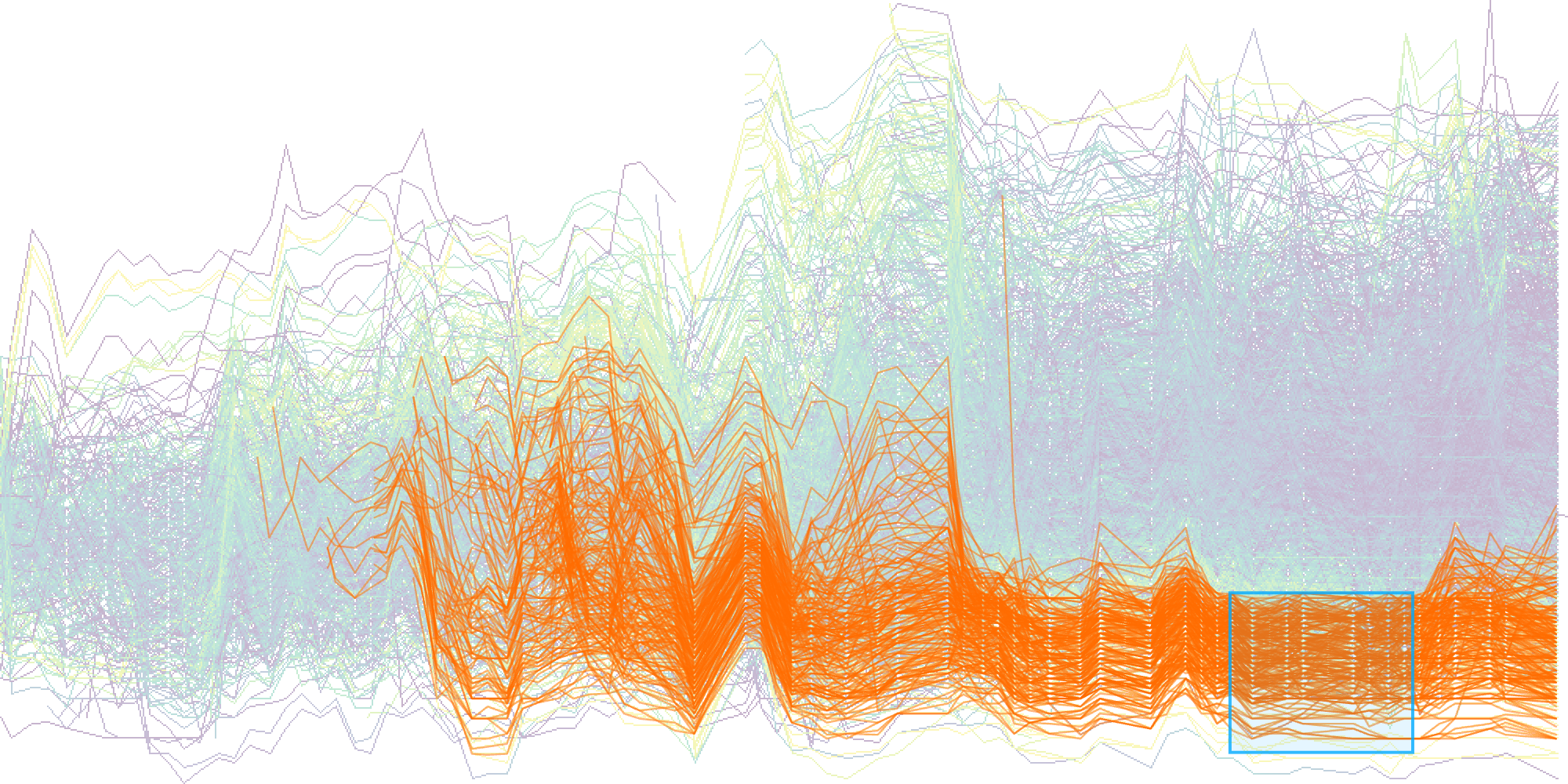}
        \caption{SIF-Guided TimeBox (229)}
    \end{subfigure}
    \caption{\textbf{Hard-drive temperature series.} The density-guided and SIF-guided TimeBox choices emphasize different parts of the same data. A denser region yields a 223-line selection that is still strongly affected by fluctuations, whereas the SIF-guided lower-right query returns 229 lines that are visually more regular. This reversal arises because steep oscillations can inflate density through frequent top/bottom crossings, while the TimeBox ignores those crossings.}
    \label{fig:supp_real_example_harddrive}
\end{figure*}

\begin{figure*}[t]
    \centering
    \begin{subfigure}[t]{0.48\textwidth}
        \centering
        \includegraphics[width=\linewidth]{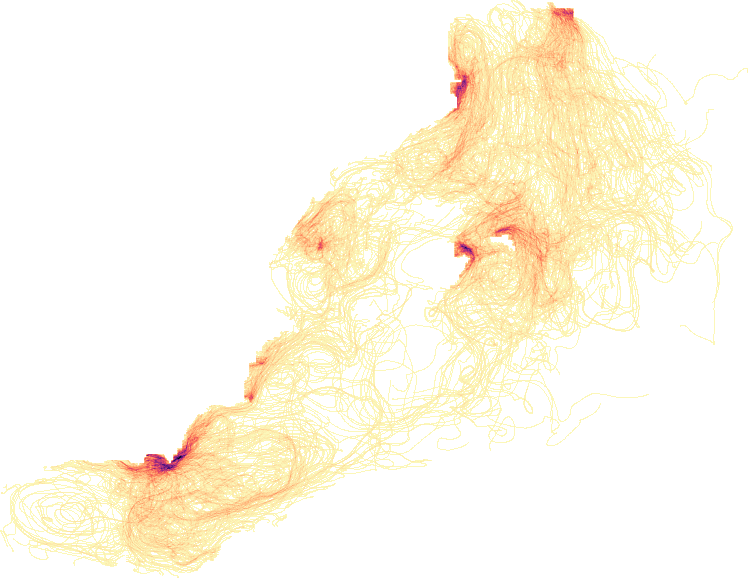}
        \caption{Density}
    \end{subfigure}
    \hfill
    \begin{subfigure}[t]{0.48\textwidth}
        \centering
        \includegraphics[width=\linewidth]{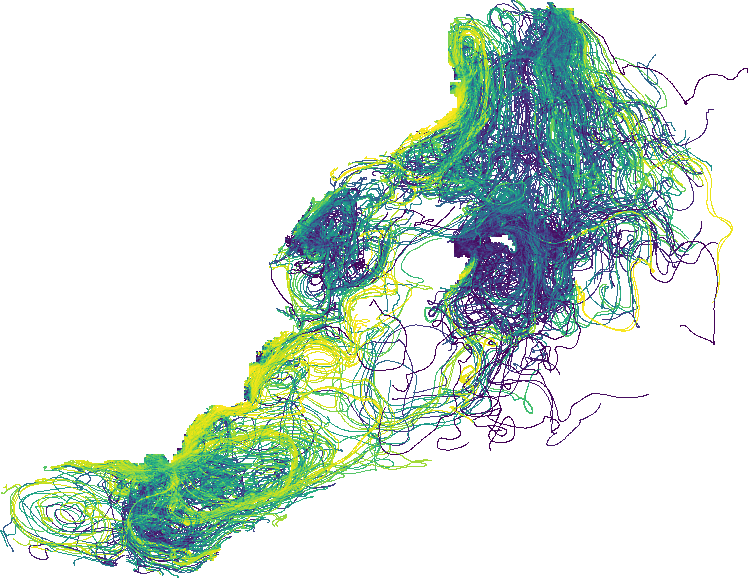}
        \caption{SIF}
    \end{subfigure}
    \par\vspace{-1mm}
    \appendixfieldlegendrow
    \vspace{4pt}
    \begin{subfigure}[t]{0.48\textwidth}
        \centering
        \includegraphics[width=\linewidth]{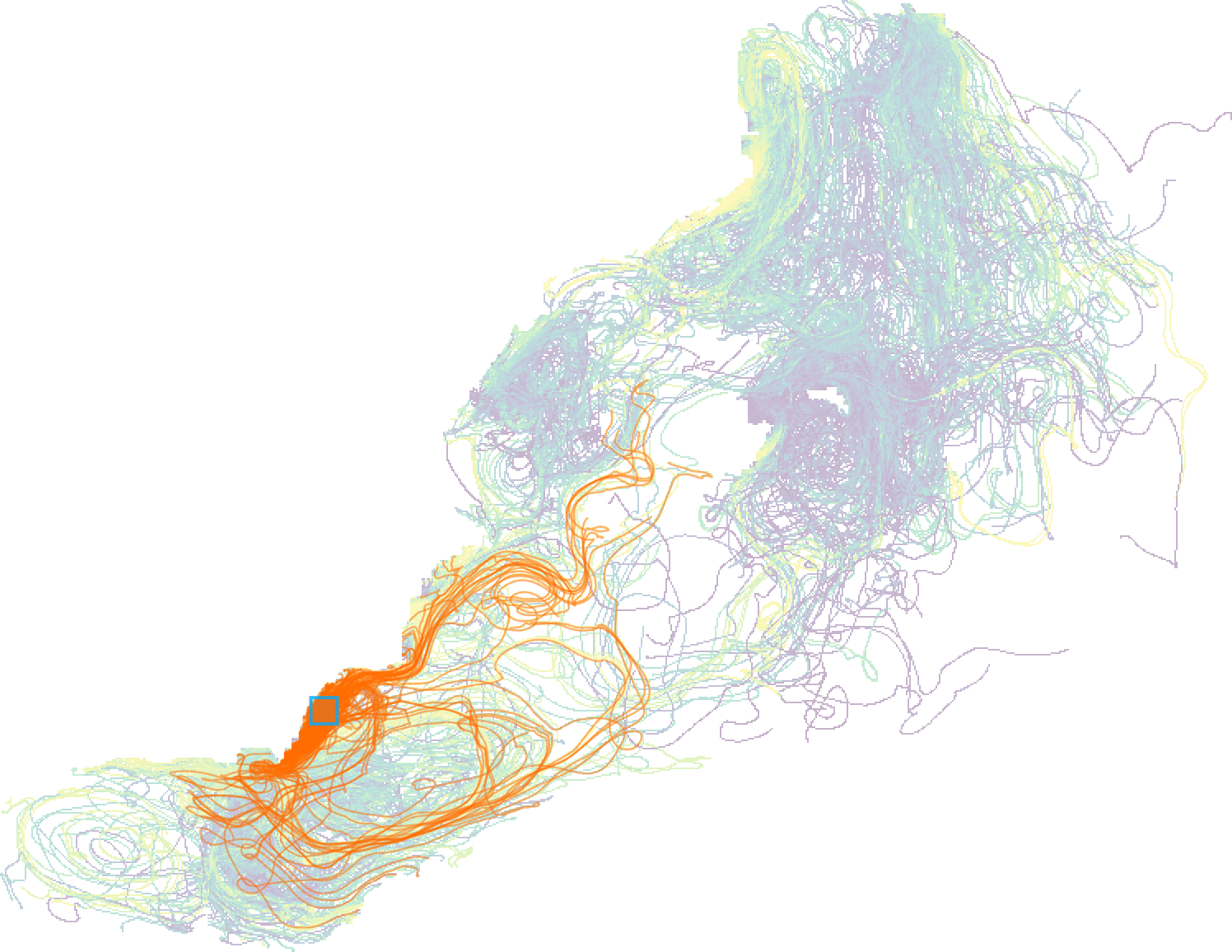}
        \caption{Low-Inconsistency Selection (144)}
    \end{subfigure}
    \hfill
    \begin{subfigure}[t]{0.48\textwidth}
        \centering
        \includegraphics[width=\linewidth]{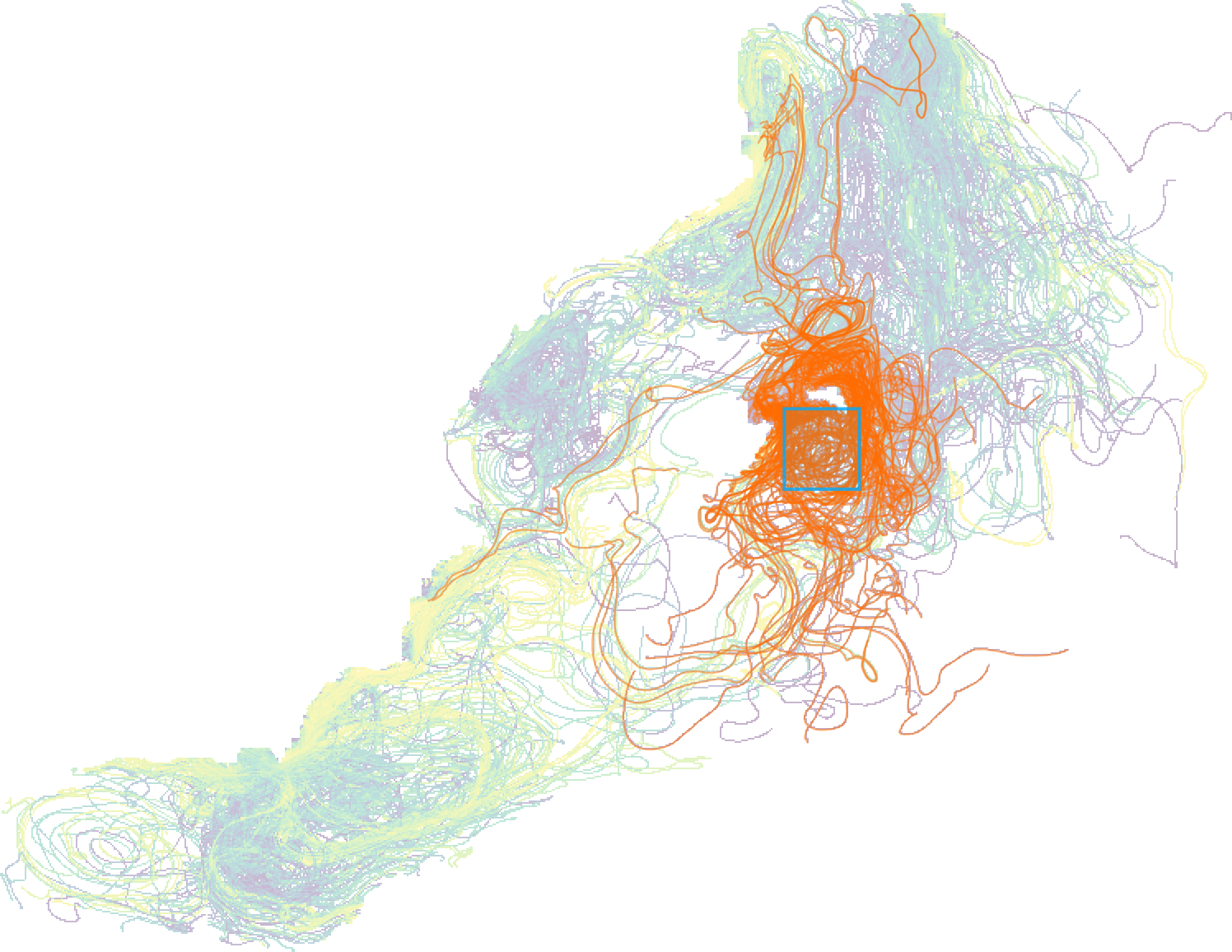}
        \caption{High-Inconsistency Selection (138)}
    \end{subfigure}
    \caption{\textbf{Mediterranean Sea trajectories.} The low-inconsistency selection contains 144 lines yet stays concentrated in a compact, coherent corridor. The high-inconsistency selection contains a similar number of lines, 138, but occupies a much larger visual region and exhibits substantially weaker structural agreement.}
    \label{fig:supp_real_example_mediterranean}
\end{figure*}

\begin{figure*}[t]
    \centering
    \begin{subfigure}[t]{0.48\textwidth}
        \centering
        \includegraphics[width=\linewidth]{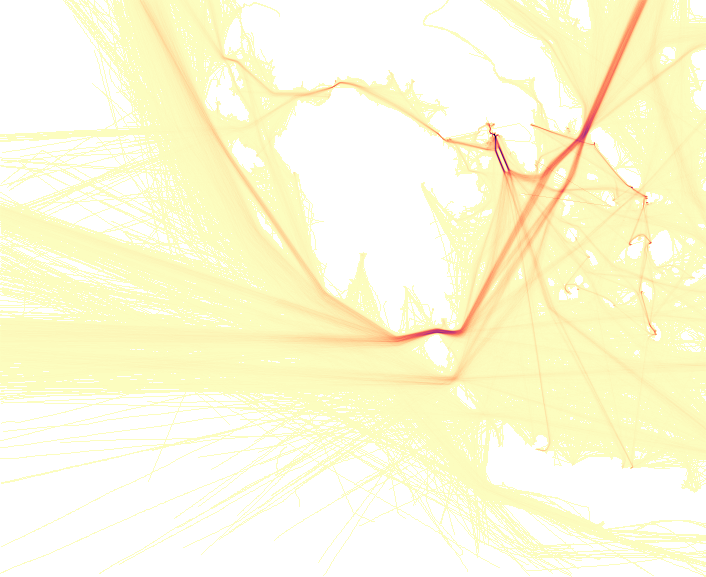}
        \caption{Density}
    \end{subfigure}
    \hfill
    \begin{subfigure}[t]{0.48\textwidth}
        \centering
        \includegraphics[width=\linewidth]{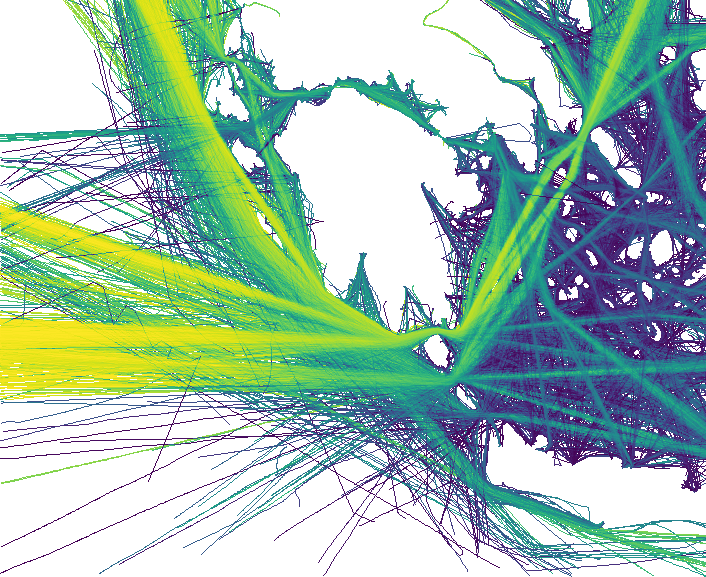}
        \caption{SIF}
    \end{subfigure}
    \par\vspace{-1mm}
    \appendixfieldlegendrow
    \vspace{4pt}
    \begin{subfigure}[t]{0.48\textwidth}
        \centering
        \includegraphics[width=\linewidth]{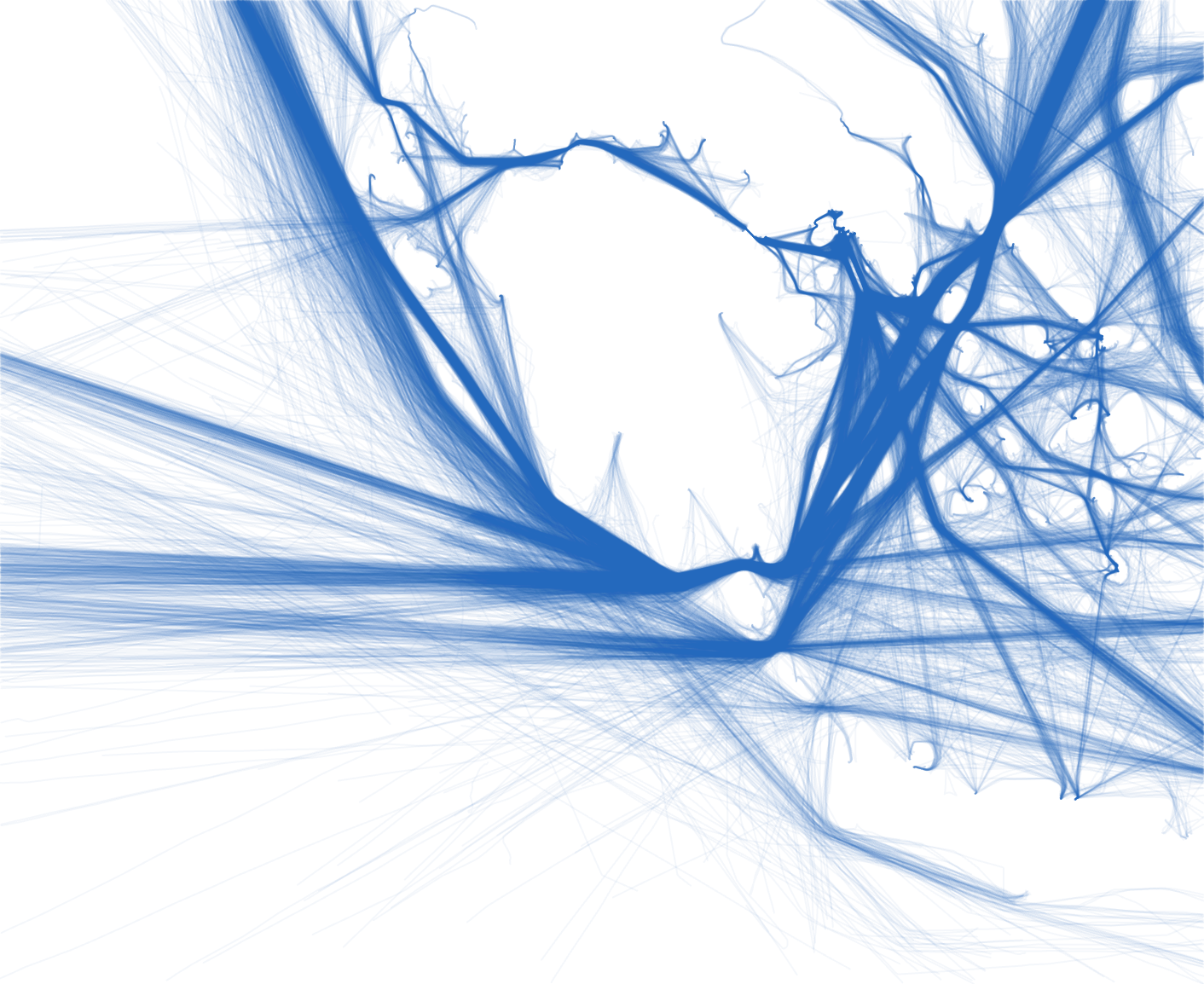}
        \caption{Filtered 97\%}
    \end{subfigure}
    \hfill
    \begin{subfigure}[t]{0.48\textwidth}
        \centering
        \includegraphics[width=\linewidth]{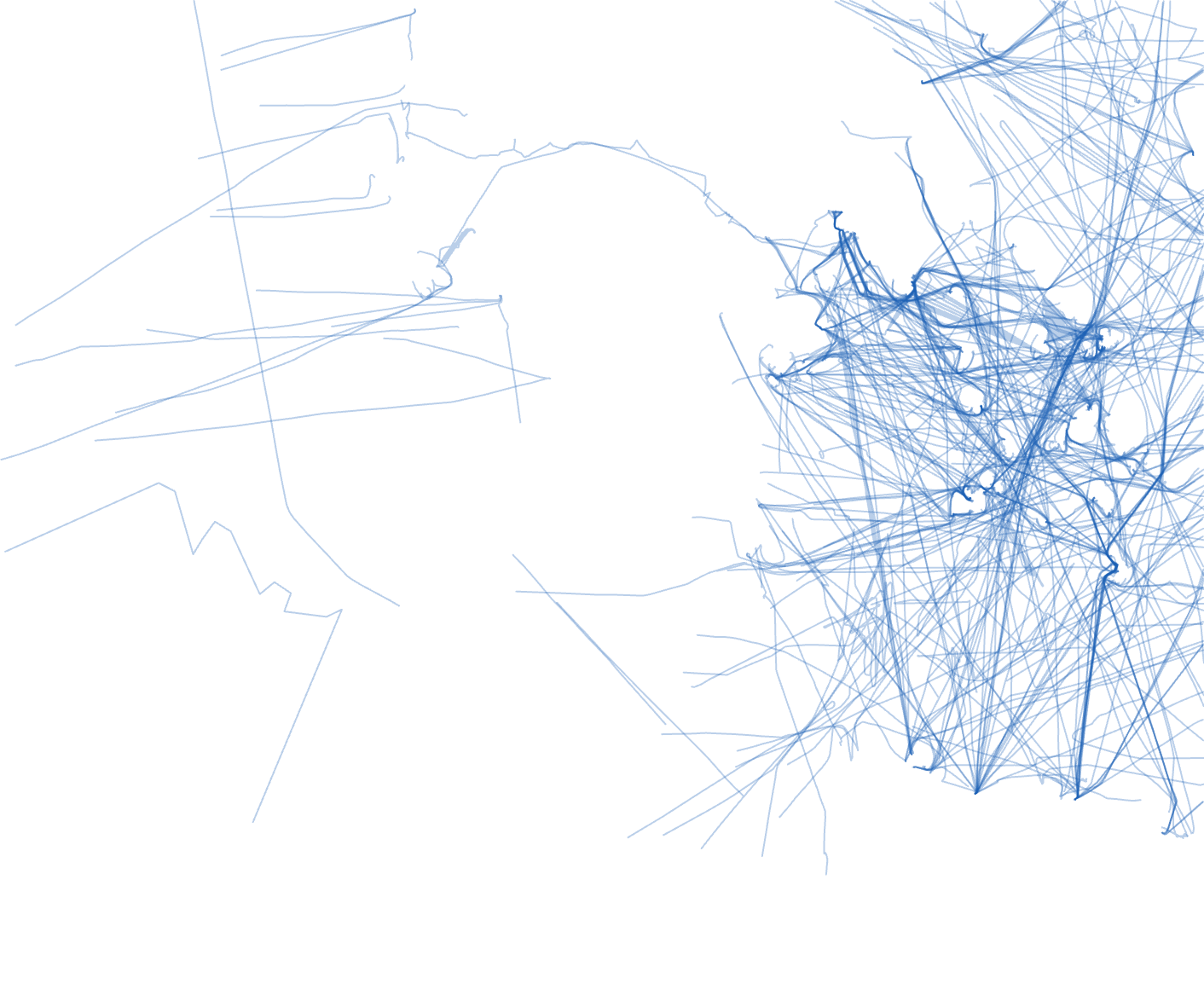}
        \caption{3\% Low-Fidelity Outliers}
    \end{subfigure}
    \caption{\textbf{Ship trajectories.} Because this dataset is already dominated by regular shipping lanes, we use the SIF primarily as confirmation that the main routes have low structural inconsistency. We then apply a trajectory-fidelity split at 3\%/97\%: the retained 97\% form visually regular, mostly straight point-to-point routes, while the lowest 3\% contain visually irregular, low-fidelity trajectories, including unusual routes in the lower-left part of the figure.}
    \label{fig:supp_real_example_ship}
\end{figure*}

The following examples are balanced across two time-series datasets and two trajectory datasets. Following the data sources used in the supplementary material of Xue \textit{et al.}~\cite{xue2025enhancing}, we include stock trajectories, hard-drive temperature series, Mediterranean Sea trajectories, and Hellenic Trench ship trajectories. For the time-series cases, we use either TimeBox or window queries depending on whether top/bottom crossings should be excluded. We first summarize the four analyses and then present their full-width visual panels in \cref{fig:supp_real_example_stock,fig:supp_real_example_stock_second,fig:supp_real_example_harddrive,fig:supp_real_example_mediterranean,fig:supp_real_example_ship}. All four examples can also be explored in our interactive online system\footnote{https://structuring-line-ensembles.vercel.app/}.

\subsubsection{Stock Price Trajectories}

This example uses 4,393 stock price trajectories derived from New York Stock Exchange historical data~\cite{nyse}. Two patterns stand out. The long band near the bottom is salient in the density view and also appears as a low-structural-inconsistency region in the SIF, indicating sustained structural support. A TimeBox query retrieves 658 lines, and peeling them removes this band from the residual density, confirming that it corresponds to a coherent subset of the ensemble. A second pattern, slightly higher and further to the right, tells the opposite story: it is visually strong in the density view but marked as highly inconsistent in the SIF. Here we use a window query rather than a TimeBox because the lines entering from above and below are part of the phenomenon itself. The query returns 1,232 lines, and filtering reveals a coherent 370-line minority whose series begin around or after 2012 instead of spanning the full timeline from the earliest records in 2005. Most queried lines, however, are long and strongly fluctuating. This case therefore matches the majority-rules limitation discussed in the main paper: a valid minority structure can be suppressed by a dominant mixed background.

\subsubsection{Hard-Drive Temperature Series}

This example uses 2,996 hard-drive temperature series~\cite{backblaze}. The two selected TimeBox regions are informative because their salience is reversed between density and SIF. A density-guided choice returns 223 lines, but they remain strongly fluctuation-dominated. The SIF instead emphasizes the lower-right region, whose TimeBox returns 229 lines that are visually more regular. This reversal follows directly from TimeBox semantics: lines entering the box from above or below are ignored, while steep oscillations can still inflate density through repeated top/bottom crossings.

\subsubsection{Mediterranean Sea Trajectories}

This example uses a 2,000-trajectory sample of the Mediterranean Sea Trajectory dataset~\cite{rath2021mediterranean}. It illustrates how similar line counts can correspond to very different structure. The low-inconsistency selection contains 144 lines and remains concentrated in a narrow transport corridor, whereas the high-inconsistency selection contains 138 lines but spreads over a much larger region with substantially weaker directional agreement.

\subsubsection{Ship Trajectories}

This example uses 10,000 trajectories from the Hellenic Trench AIS dataset~\cite{frantzis2018hellenic}. Because most vessels already follow established shipping lanes, the SIF mainly confirms that the dominant corridors have low structural inconsistency. We therefore analyze the full dataset with a simple 3\%/97\% trajectory-fidelity split: the retained 97\% form regular, mostly point-to-point routes, whereas the lowest 3\% contain visually irregular, low-fidelity trajectories, including unusual routes in the lower-left part of the view.

\subsection{Real-Data Comparison with Image-Space Coloring}

\begin{figure*}[t]
    \centering
    \setlength{\tabcolsep}{2pt}
    \renewcommand{\arraystretch}{1.0}
    \newcommand{\realcolorcell}[2]{\fbox{\parbox[c][#1][c]{0.23\textwidth}{\centering\includegraphics[width=0.23\textwidth,height=#1,keepaspectratio]{#2}}}}
    \begin{tabular*}{\textwidth}{@{\extracolsep{\fill}}lccc@{}}
        & \small\textsf{Image-space coloring} & \small\textsf{High-score subset} & \small\textsf{Low-score subset} \\
        \small\textsf{France} &
        \realcolorcell{0.185\textwidth}{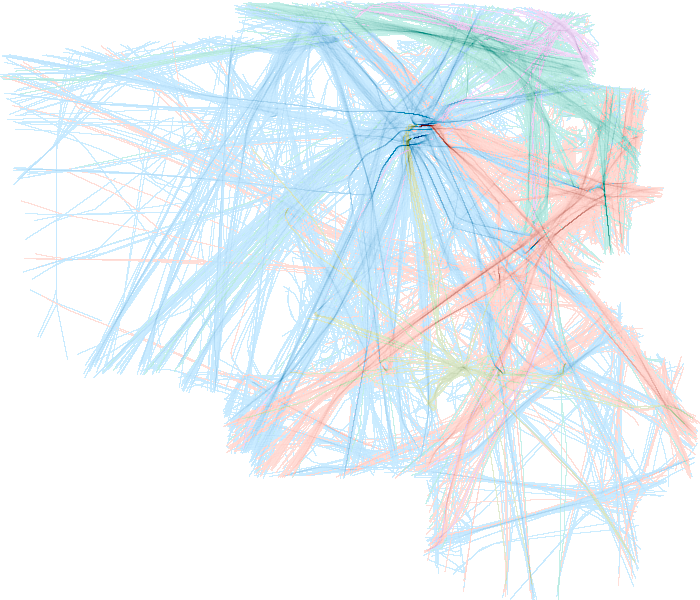} &
        \realcolorcell{0.185\textwidth}{figs/evaluation/france_inliers4288.jpg} &
        \realcolorcell{0.185\textwidth}{figs/evaluation/france_outliers1072.jpg} \\
        \small\textsf{DTI} &
        \realcolorcell{0.225\textwidth}{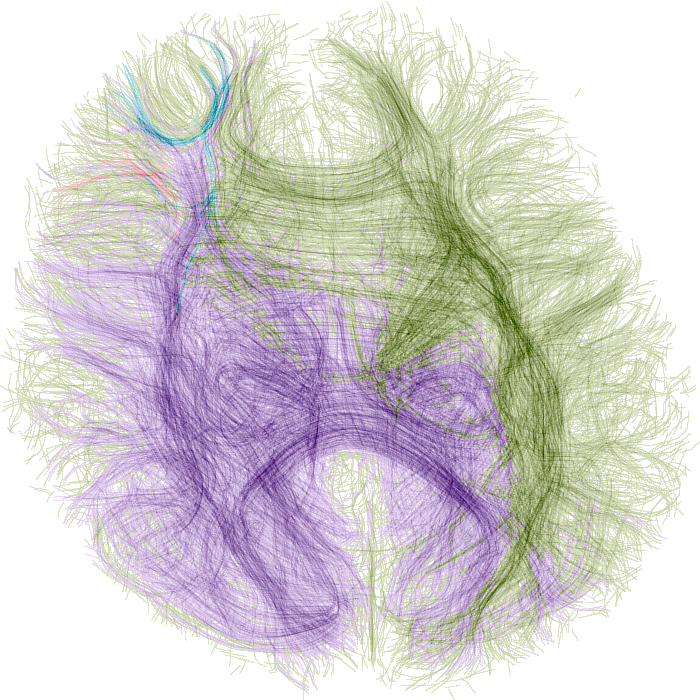} &
        \realcolorcell{0.225\textwidth}{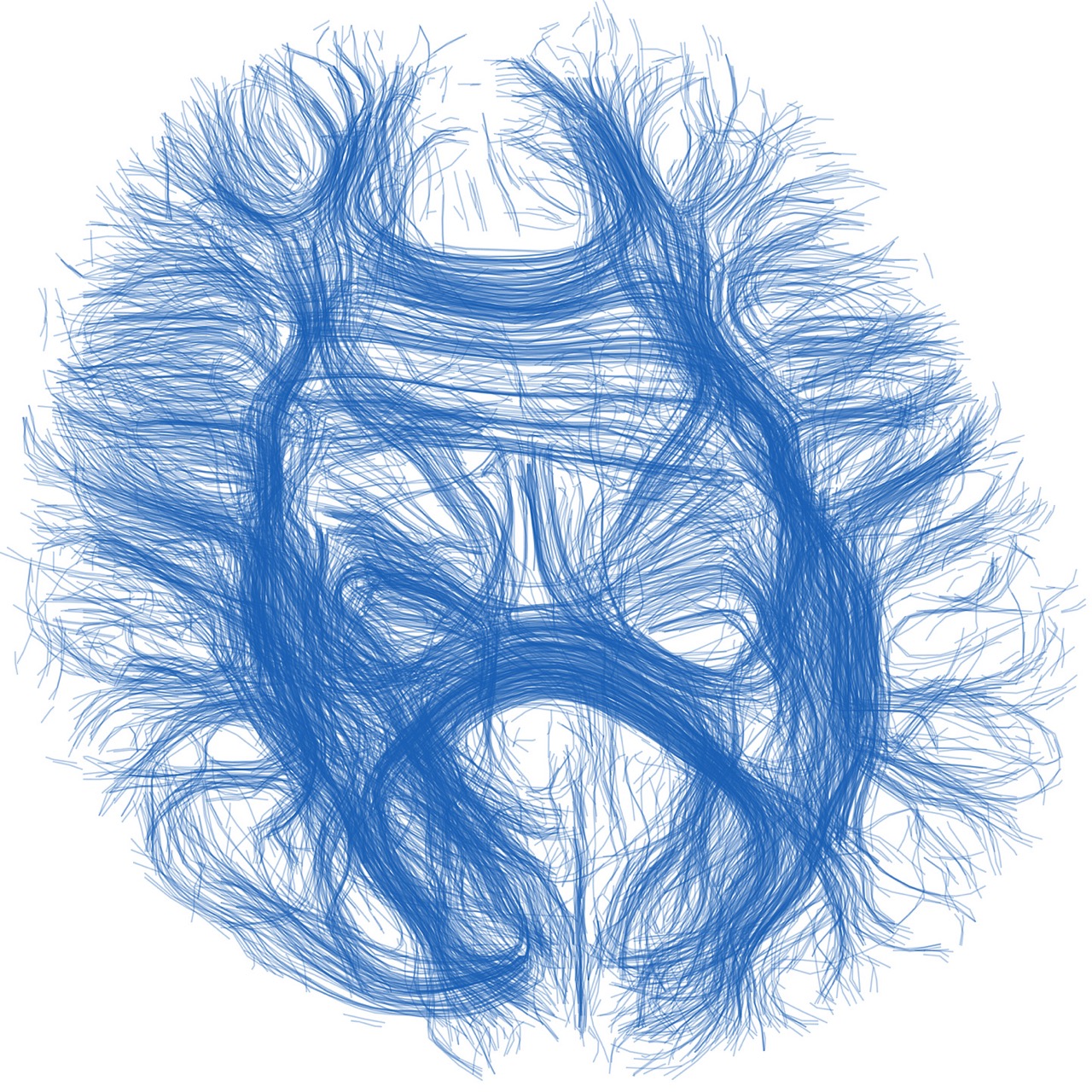} &
        \realcolorcell{0.225\textwidth}{figs/evaluation/DTI_outlier.jpg} \\
    \end{tabular*}
    \vspace{-2mm}
    \caption{\textbf{Real-data comparison with image-space coloring.} Left: the colorization of Xue \textit{et al.}~\cite{xue2024reducing}. Middle and right: the highest- and lowest-scoring subsets from our method.}
    \label{fig:app_real_coloring}
    \vspace{-4mm}
\end{figure*}

\cref{fig:app_real_coloring} highlights the same limitation on both real datasets. The image-space coloring of Xue \textit{et al.}~\cite{xue2024reducing} partitions the rendered view into coarse perceptual regions, but it does not isolate which trajectories actually support the main structure and which ones deviate from it. In the French aviation data, route-regular long-range traffic and structurally weaker trajectories around hubs or sparse side routes still appear in the same color groups when they occupy similar screen-space sectors. Thus, this method is not able to visually separate these different patterns.
In the DTI projection, major scaffold fibers and peripheral or crossing fibers are likewise merged into broad color regions because they overlap in the rendered image. Our method answers a different question: by scoring each trajectory against structural support from the surrounding ensemble, it separates the coherent main scaffold from its weakly supported complement while keeping the original trajectories unchanged. This makes an explicit distinction that an image-space coloring method cannot provide on its own, namely, which lines are structurally supported and which lines remain competing or inconsistent with that support.

\begin{figure*}[t]
    \centering
    \begin{minipage}{0.5\textwidth}
        \centering
        {\scriptsize\textsf{Structural Inconsistency}}\\[-0.8mm]
        \appendixfieldlegend{figs/evaluation/inconsistency_colormap.pdf}
    \end{minipage}
    \par\vspace{1mm}
    \setlength{\tabcolsep}{2pt}
    \renewcommand{\arraystretch}{1.0}
    \newcommand{\extentcell}[1]{\fbox{\includegraphics[width=0.23\textwidth]{#1}}}
    \begin{tabular*}{\textwidth}{@{\extracolsep{\fill}}lccc@{}}
        & \small\textsf{Short extent} & \small\textsf{Mid extent} & \small\textsf{Maximal extent} \\
        \small\textsf{D3-C} &
        \extentcell{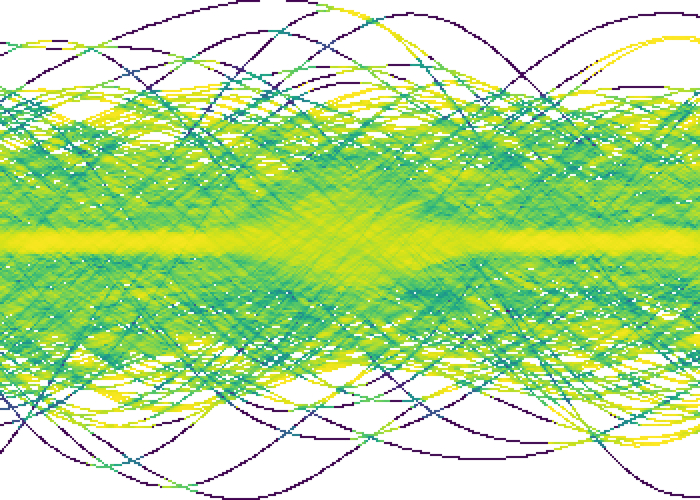} &
        \extentcell{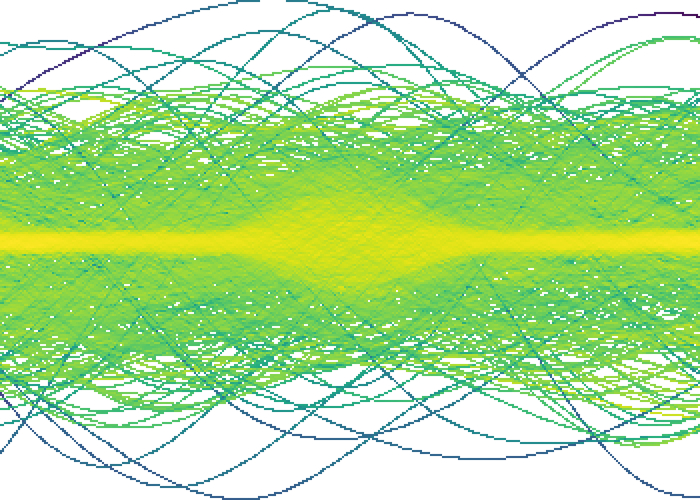} &
        \extentcell{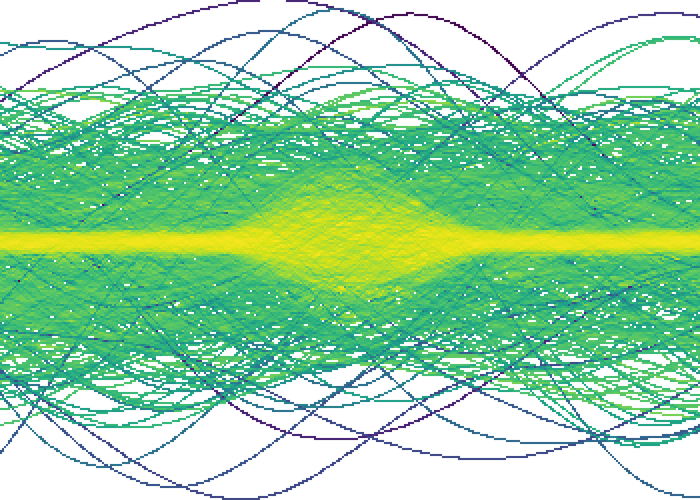} \\
        \small\textsf{D3-D} &
        \extentcell{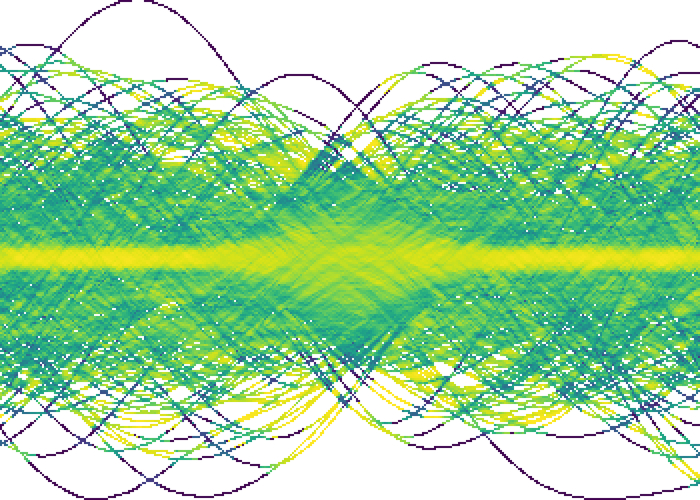} &
        \extentcell{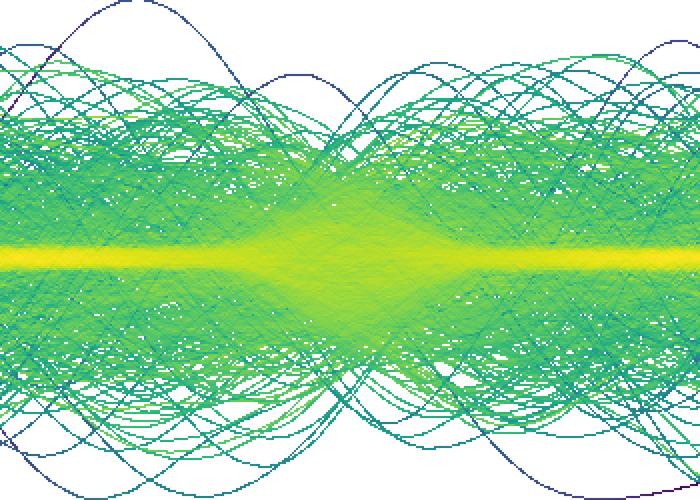} &
        \extentcell{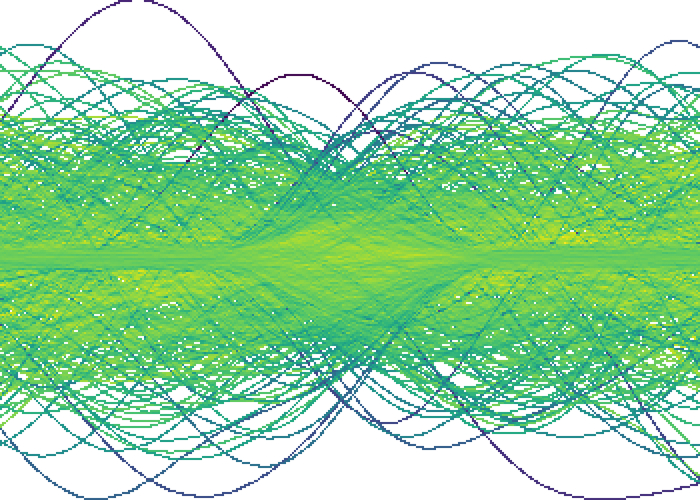} \\
    \end{tabular*}
    \vspace{-2mm}
    \caption{\textbf{Qualitative effect of increasing path extent on D3.} Top row: D3-C. Bottom row: D3-D. Left to right: short, intermediate, and maximal path extent. D3-C keeps a lighter central bridge across all three settings. D3-D looks similar at short and intermediate extents because the evaluation has not yet reached the divergent side regions; only the maximal extent exposes the ambiguous support by making the center visually similar to its surroundings.}
    \label{fig:app_d3_path_extent}
    \vspace{-2mm}
\end{figure*}

\FloatBarrier

\end{document}